\documentclass[letterpaper]{IEEEtran}
\usepackage[numbers,sort&compress]{natbib}

\makeatletter

\usepackage{etex}

\usepackage{zref-savepos,varioref}

\newcounter{mnote}

\def\xmarginnote{%
	\xymarginnote{\hskip -\marginparsep \hskip -\marginparwidth}}

\def\ymarginnote{%
	\xymarginnote{\hskip\columnwidth \hskip\marginparsep}}

\long\def\xymarginnote#1#2{%
	\vadjust{#1%
		\smash{\hbox{{%
					\hsize\marginparwidth
					\@parboxrestore
					\@marginparreset
					\footnotesize #2}}}}}

\def\mnoteson{%
	\gdef\mnote##1{\refstepcounter{mnote}\label{##1}%
		\zsavepos{##1}%
		\ifnum20432158>\number\zposx{##1}%
		\xmarginnote{{\color{blue}\bf $\langle$\arabic{mnote}$\rangle$}}%
		\else
		\ymarginnote{{\color{blue}\bf $\langle$\arabic{mnote}$\rangle$}}%
		\fi%
	}
}
\gdef\mnotesoff{\gdef\mnote##1{}}
\mnoteson
\mnotesoff

\newcommand{\figref}[1]{Fig.~\ref{#1}}

\usepackage{framed}

\usepackage{subcaption}
\usepackage{comment}

\usepackage[svgnames,dvipsnames]{xcolor}

\usepackage[all]{xy}

\usepackage{tikz}
\usetikzlibrary{positioning,matrix,through,calc,arrows,fit,shapes,decorations.pathreplacing,decorations.markings,}

\tikzstyle{block} = [draw,fill=blue!20,minimum size=2em]

\usepackage{qsymbols,amssymb,mathrsfs}
\usepackage[standard,thmmarks]{ntheorem}
\theoremstyle{plain}

\theoremsymbol{\ensuremath{_\vartriangleleft}}
\theorembodyfont{\itshape}
\theoremheaderfont{\normalfont\bfseries}
\theoremseparator{}

\theoremstyle{nonumberplain}
\theoremheaderfont{\scshape}
\theorembodyfont{\normalfont}
\theoremsymbol{\ensuremath{_\blacktriangleleft}}

\theoremnumbering{arabic}
\theoremstyle{plain}
\usepackage{latexsym}
\theoremsymbol{\ensuremath{_\Box}}
\theorembodyfont{\itshape}
\theoremheaderfont{\normalfont\bfseries}
\theoremseparator{}

\theorembodyfont{\upshape}
\theoremprework{\bigskip\hrule}
\theorempostwork{\hrule\bigskip}

\usepackage[overload]{empheq}

\let\iftwocolumn\if@twocolumn
\g@addto@macro\@twocolumntrue{\let\iftwocolumn\if@twocolumn}
\g@addto@macro\@twocolumnfalse{\let\iftwocolumn\if@twocolumn}

\mathtoolsset{showonlyrefs=false,showmanualtags}
\let\underbrace\LaTeXunderbrace 
\let\overbrace\LaTeXoverbrace
\renewcommand{\eqref}[1]{\textup{(\refeq{#1})}} 
                                       
\newtagform{brackets}[\textbf]{(}{)}   
\usetagform{brackets}

\usepackage[Smaller]{cancel}

\PassOptionsToPackage{breaklinks,letterpaper,hyperindex=true,backref=false,bookmarksnumbered,bookmarksopen,linktocpage,colorlinks,linkcolor=BrickRed,citecolor=OliveGreen,urlcolor=Blue,pdfstartview=FitH}{hyperref}
\usepackage{hyperref}

\usepackage{graphicx,psfrag}
\graphicspath{{figure/}{image/}}

\usepackage{diagbox}

\usepackage[ruled,vlined,linesnumbered]{algorithm2e}
\usepackage{algpseudocode}
\usepackage{listings} 
\lstdefinelanguage{Maple}{
  morekeywords={proc,module,end, for,from,to,by,while,in,do,od
    ,if,elif,else,then,fi ,use,try,catch,finally}, sensitive,
  morecomment=[l]\#,
  morestring=[b]",morestring=[b]`}[keywords,comments,strings]
\DeclareMathAlphabet{\mathpzc}{OT1}{pzc}{m}{it}
\usepackage{upgreek} 

\usepackage{dsfont}

\def\multi@nostar#1#2{
  \expandafter\def\csname multi#1\endcsname##1{
    \if ##1.\let\next=\relax \else
    \def\next{\csname multi#1\endcsname}     
    
    \expandafter\newcommand\csname #1##1\endcsname{#2}
    \fi\next}}

\def\multi@star#1#2{
  \expandafter\def\csname #1\endcsname##1{#2}
  \multi@nostar{#1}{#2}
}

\newcommand{\multi}{
  \@ifstar \multi@star \multi@nostar}

\multi*{rm}{\mathrm{#1}}
\multi*{mc}{\mathcal{#1}}
\multi*{op}{\mathop {\operator@font #1}}

\multi*{ds}{\mathds{#1}}
\multi*{set}{\mathcal{#1}}
\multi*{rsfs}{\mathscr{#1}}
\multi*{pz}{\mathpzc{#1}}
\multi*{M}{\boldsymbol{#1}}
\multi*{R}{\mathsf{#1}}
\multi*{RM}{\M{\R{#1}}}
\multi*{bb}{\mathbb{#1}}
\multi*{td}{\tilde{#1}}
\multi*{tR}{\tilde{\mathsf{#1}}}
\multi*{trM}{\tilde{\M{\R{#1}}}}
\multi*{tset}{\tilde{\mathcal{#1}}}
\multi*{tM}{\tilde{\M{#1}}}
\multi*{baM}{\bar{\M{#1}}}
\multi*{ol}{\overline{#1}}

\multirm  ABCDEFGHIJKLMNOPQRSTUVWXYZabcdefghijklmnopqrstuvwxyz.
\multiol  ABCDEFGHIJKLMNOPQRSTUVWXYZabcdefghijklmnopqrstuvwxyz.
\multitR   ABCDEFGHIJKLMNOPQRSTUVWXYZabcdefghijklmnopqrstuvwxyz.
\multitd   ABCDEFGHIJKLMNOPQRSTUVWXYZabcdefghijklmnopqrstuvwxyz.
\multitset ABCDEFGHIJKLMNOPQRSTUVWXYZabcdefghijklmnopqrstuvwxyz.
\multitM   ABCDEFGHIJKLMNOPQRSTUVWXYZabcdefghijklmnopqrstuvwxyz.
\multibaM   ABCDEFGHIJKLMNOPQRSTUVWXYZabcdefghijklmnopqrstuvwxyz.
\multitrM   ABCDEFGHIJKLMNOPQRSTUVWXYZabcdefghijklmnopqrstuvwxyz.
\multimc   ABCDEFGHIJKLMNOPQRSTUVWXYZabcdefghijklmnopqrstuvwxyz.
\multiop   ABCDEFGHIJKLMNOPQRSTUVWXYZabcdefghijklmnopqrstuvwxyz.
\multids   ABCDEFGHIJKLMNOPQRSTUVWXYZabcdefghijklmnopqrstuvwxyz.
\multiset  ABCDEFGHIJKLMNOPQRSTUVWXYZabcdefghijklmnopqrstuvwxyz.
\multirsfs ABCDEFGHIJKLMNOPQRSTUVWXYZabcdefghijklmnopqrstuvwxyz.
\multipz   ABCDEFGHIJKLMNOPQRSTUVWXYZabcdefghijklmnopqrstuvwxyz.
\multiM    ABCDEFGHIJKLMNOPQRSTUVWXYZabcdefghijklmnopqrstuvwxyz.
\multiR    ABCDEFGHIJKL NO QRSTUVWXYZabcd fghijklmnopqrstuvwxyz.
\multibb   ABCDEFGHIJKLMNOPQRSTUVWXYZabcdefghijklmnopqrstuvwxyz.
\multiRM   ABCDEFGHIJKLMNOPQRSTUVWXYZabcdefghijklmnopqrstuvwxyz.

\newcommand{\dotleq}{\buildrel \textstyle  .\over {\smash{\lower
      .2ex\hbox{\ensuremath\leqslant}}\vphantom{=}}}
\newcommand{\dotgeq}{\buildrel \textstyle  .\over {\smash{\lower
      .2ex\hbox{\ensuremath\geqslant}}\vphantom{=}}}

\newcommand{\bM}{\begin{bmatrix}}
\newcommand{\eM}{\end{bmatrix}}
\newcommand{\bSM}{\left[\begin{smallmatrix}}
\newcommand{\eSM}{\end{smallmatrix}\right]}
\renewcommand*\env@matrix[1][*\c@MaxMatrixCols c]{
  \hskip -\arraycolsep
  \let\@ifnextchar\new@ifnextchar
  \array{#1}}

\newqsymbol{`N}{\mathbb{N}}
\newqsymbol{`R}{\mathbb{R}}
\newqsymbol{`P}{\mathbb{P}}
\newqsymbol{`Z}{\mathbb{Z}}

\newqsymbol{`|}{\mid}

\newqsymbol{`8}{\infty}
\newqsymbol{`1}{\left}
\newqsymbol{`2}{\right}
\newqsymbol{`6}{\partial}
\newqsymbol{`0}{\emptyset}
\newqsymbol{`-}{\leftrightarrow}
\newqsymbol{`<}{\langle}
\newqsymbol{`>}{\rangle}

\DeclarePairedDelimiter\abs{\lvert}{\rvert}

\DeclarePairedDelimiter\Set{\{}{\}}
\newcommand{\imod}[1]{\allowbreak\mkern10mu({\operator@font mod}\,\,#1)}

\newcommand{\threecols}[3]{
\hbox to \textwidth{
      \normalfont\rlap{\parbox[b]{\textwidth}{\raggedright#1\strut}}
        \hss\parbox[b]{\textwidth}{\centering#2\strut}\hss
        \llap{\parbox[b]{\textwidth}{\raggedleft#3\strut}}
    }
}

\newcommand{\reason}[2][\relax]{
  \ifthenelse{\equal{#1}{\relax}}{
    \left(\text{#2}\right)
  }{
    \left(\parbox{#1}{\raggedright #2}\right)
  }
}

\newcommand{\utag}[2]{\mathop{#2}\limits^{\text{(#1)}}}
\newcommand{\uref}[1]{(#1)}

\let\SavedDoubleVert\relax
\let\protect\relax
{\catcode`\|=\active
  \xdef\extendvert{\protect\expandafter\noexpand\csname extendvert \endcsname}
  \expandafter\gdef\csname extendvert \endcsname#1{\mskip-5mu \left.
      \ifx\SavedDoubleVert\relax \let\SavedDoubleVert\|\fi
     \:{\let\|\SetDoubleVert
       \mathcode`\|32768\let|\SetVert
     #1}\:\right.\mskip-5mu}
}
\def\SetVert{\@ifnextchar|{\|\@gobble}
    {\egroup\;\mid@vertical\;\bgroup}}
\def\SetDoubleVert{\egroup\;\mid@dblvertical\;\bgroup}

\begingroup
 \edef\@tempa{\meaning\middle}
 \edef\@tempb{\string\middle}
\expandafter \endgroup \ifx\@tempa\@tempb
 \def\mid@vertical{\middle|}
 \def\mid@dblvertical{\middle\SavedDoubleVert}
\else
 \def\mid@vertical{\mskip1mu\vrule\mskip1mu}
 \def\mid@dblvertical{\mskip1mu\vrule\mskip2.5mu\vrule\mskip1mu}
\fi

\makeatother

\usepackage{fouridx}

\usepackage{framed}
\usetikzlibrary{positioning,matrix,decorations.shapes}

\usepackage{paralist}

\usepackage{enumerate}

\usepackage[normalem]{ulem}

\numberwithin{equation}{section}
\makeatletter
\@addtoreset{equation}{section}
\renewcommand{\theequation}{\arabic{section}.\arabic{equation}}
\@addtoreset{Theorem}{section}
\renewcommand{\theTheorem}{\arabic{section}.\arabic{Theorem}}
\@addtoreset{Lemma}{section}
\renewcommand{\theLemma}{\arabic{section}.\arabic{Lemma}}
\@addtoreset{Corollary}{section}
\renewcommand{\theCorollary}{\arabic{section}.\arabic{Corollary}}
\@addtoreset{Example}{section}
\renewcommand{\theExample}{\arabic{section}.\arabic{Example}}
\@addtoreset{Remark}{section}
\renewcommand{\theRemark}{\arabic{section}.\arabic{Remark}}
\@addtoreset{Proposition}{section}
\renewcommand{\theProposition}{\arabic{section}.\arabic{Proposition}}
\@addtoreset{Definition}{section}
\renewcommand{\theDefinition}{\arabic{section}.\arabic{Definition}}
\@addtoreset{Subclaim}{Theorem}
\renewcommand{\theSubclaim}{\theTheorem\Alph{Subclaim}}
\makeatother

\newenvironment{ybox}{
  \setlength{\FrameSep}{1mm}
  \setlength{\FrameRule}{0mm}
  
  \MakeFramed {\FrameRestore}}
{\endMakeFramed}

\newenvironment{gbox}{
	  \setlength{\FrameSep}{1mm}
	  \setlength{\FrameRule}{0mm}
  
  \MakeFramed {\FrameRestore}}
{\endMakeFramed}

\newenvironment{bbox}{
	  \setlength{\FrameSep}{1mm}
	  \setlength{\FrameRule}{0mm}
  
  \MakeFramed {\FrameRestore}}
{\endMakeFramed}

\newcommand{\p}{\mathsf{p}}

\title{Info-Clustering:\\ A Mathematical Theory for Data Clustering}
\author{Chung Chan, Ali Al-Bashabsheh, Qiaoqiao Zhou, Tarik Kaced and Tie Liu
	\thanks{Preliminary results published in \cite{chan15allerton} and presented at the Claude Shannon Centenary Workshop in Hong Kong and the Claude Shannon's Centennial Day in Shanghai in 2016. To appear in the special issue of the IEEE Transactions on Molecular, Biological, and Multi-Scale Communications on Biological Applications of Information Theory.}
	\thanks{C.\ Chan (email: cchan@inc.cuhk.edu.hk, chungc@alum.mit.edu),
		A.\ Al-Bashabsheh, Q.\ Zhou are with the Institute of Network Coding at the
		Chinese University of Hong Kong, the Shenzhen Key Laboratory of
		Network Coding Key Technology and Application, China, and the
		Shenzhen Research Institute of the Chinese University of Hong
		Kong.
	}
	\thanks{T.\ Kaced is with with Universit\'e Paris-Est
		Cr\'eteil at the Algorithmic, Complexity and Logic Laboratory.}
	\thanks{T.\ Liu is with the Department of Electrical and Computer
		Engineering, Texas
		A\&M University, College Station, TX 77843 USA (email: tieliu@tamu.edu).}
	\thanks{The work described in this paper was supported by a grant from University Grants Committee of the Hong Kong Special Administrative Region, China (Project No. AoE/E-02/08), and supported partially by a grant from Shenzhen Science and Technology Innovation Committee (JSGG20160301170514984), the Chinese University of Hong Kong (Shenzhen), China.}
	\thanks{The work of C.\ Chan was supported in part by The Vice-Chancellor's One-off Discretionary Fund of The Chinese University of Hong Kong (Project Nos. VCF2014030 and VCF2015007), and a grant from the University
	Grants Committee of the Hong Kong Special Administrative Region,
	China (Project No. 14200714).}
	\thanks{The work of T. Liu was supported in part by the National
		Science Foundation under Grant CCF-13-20237. Part of the work was
		done while T. Liu was visiting the Institute of Network Coding at
		the Chinese University of Hong Kong.}
}

\begin{document}
	
	\IEEEoverridecommandlockouts
	\maketitle

\begin{abstract}
We formulate an info-clustering paradigm based on a  multivariate information measure, called multivariate mutual information, that naturally extends Shannon's mutual information between two random variables to the multivariate case involving more than two random variables. With proper model reductions, we show that the paradigm can be applied to study the human genome and connectome in a more meaningful way than the conventional algorithmic approach. Not only can info-clustering provide justifications and refinements to some existing techniques, but it also inspires new computationally feasible solutions.
\end{abstract} 


\section{Introduction}
\label{sec:introduction}
Clustering is the process of grouping similar objects together while separating dissimilar ones apart. This simple idea has a wide range of applications in different areas of scientific research. In \emph{bioinformatics}, clustering can identify co-expressed genes that work together for the same metabolic pathway~\cite{,butte00,sharan00}. In \emph{neuroscience}, clustering can also identify regions of neurons in the brain that are physically or functionally connected~\cite{deco15,hinne2015probabilistic,tononi1998functional}. Both the \emph{human genome} and the \emph{human connectome} are highly complex systems, with about $23,000$ protein-coding genes in the human genome~\cite{international2004finishing} and $16\times 10^{9}$ neurons composing the cerebral cortex of the human brain~\cite{azevedo2009equal}. Therefore, the ability to group similar genes or neurons together based on their interactions is very helpful, as it reduces the complex systems into smaller, and so, more manageable subsystems for further studies.

There are different techniques for capturing the detailed physical structure and functional interaction in a biological system. For the human genome, the expression levels of different genes in different individuals (or tissues) can be measured by the microarray analysis~\cite{brown1999exploring} or RNA-sequencing~\cite{morin2008profiling,wang2009rna}. For the physical connections of the neurons, called the physical connectome, electron microscopy (EM) has been used to map out the entire structural interconnections of the neurons in a small living creature called the nematode \emph{Caenorhabditis elegans} (\emph{C.\ elegans})~\cite{white1986structure,varshney2011structural,emmons2015beginning,jarrell2012connectome}. For the human brain of living subjects, EM does not apply, but a magnetic resonance imaging (MRI) technique called diffusion spectral imaging (DSI) can be used instead~\cite{wedeen2005mapping,wedeen2008diffusion,hagmann2007mapping}. The functional connectome of the neurons can also be studied by capturing the stimulation patterns of the neurons directly using electroencephalography (EEG) or indirectly using another MRI technique called functional MRI (fMRI). However, with the huge volume and variety of data available~\cite{hcp,ocp,mcp}, \emph{the main challenge is to automate the clustering process using a mathematical criterion that leads to meaningful, yet arithmetically simple to compute, clusters.} 

\begin{ybox}
We believe that the key to this challenge lies in \emph{a better understanding of what information is, and how we can measure mutual information quantitatively}. In this work, we propose a novel information-theoretic approach to clustering, called \emph{info-clustering}, and show that it applies to the study of the complex biological systems of the genes and neurons. The idea is to \emph{regard each object as a piece of information, and then group subsets of the objects together if their mutual information exceeds a certain threshold.} By varying the threshold value, a hierarchy of clusters can be obtained.
\end{ybox}

\subsection{Motivation: Human genome and connectome}

An application of info-clustering is in the study of the human genome. We know that the biological information of a human being is encoded entirely in its DNA sequences. A DNA sequence is further divided into segments called the genes. Some genes are protein-coding in the sense that they express themselves in the form of gene products such as enzymes, hormones and receptors. These proteins carry out important functions that sustain  different metabolic pathways. However, it is not entirely clear
\begin{enumerate}
	\item how the genes work together to sustain the metabolic pathways, and
	\item how do mutations of the genes cause a certain disease such as cancer.
\end{enumerate}

Clustering is a helpful first step in studying the metabolic pathways and disease pathology. This is because it helps identify smaller subsets of related genes that work closely together. More precisely, although different genes express differently in different people, or even in different tissues of the same person, genes that are co-regulated tend to have similar expression patterns~\cite{brown1999exploring,sorlie2003repeated}. Such co-expression of the genes means that there is mutual information among the genes. If we have a way to measure such information, then we can cluster the genes according to their mutual information.

Another application of info-clustering is in the study of the human brain. We know that the brain carries out important tasks such as perception, emotion, thought and memory. The way it works is that, the human brain consists of many cells called neurons. These neurons are physically wired together by fiber-like projections called axons. The neurons stimulate each other in some pathway circuitries to carry out the important brain functions. More precisely, when a neuron is excited by an external stimulation, it sends an electrical signal down the axon, which stimulates one or more target neurons through the synapses. 

The stimulation mechanism of the neurons has inspired a family of learning models in artificial intelligence called artificial neural networks. In machine learning, such artificial neural networks can be used by deep learning methods to perform complicated tasks such as image recognition~\cite{krizhevsky2012imagenet}. The performance of such methods often superior to alternative approaches. However, exactly why the technique works so well is not entirely clear. It remains a mystery as to
\begin{enumerate}
	\item how the stimulations of neurons lead to the complicated brain functions, and
	\item how the damages or anomalies in the brain lead to mental disorders such as schizophrenia, bipolar disorder, autism and attention deficit hyperactivity disorder.
\end{enumerate}

We believe information theory~\cite{yeung08,cover2012elements,csiszar2011information,gallager1968information,mackay2003information} lies close to the heart of these problems because the stimulation mechanism by electrical and chemical signals are simply transmission and processing of information. 
It was recently discovered~\cite{sporns2013network,van2013network,nigam2016rich} that the brain segregates into tightly connected regions, and there are important network hubs, called the rich-clubs, that connect between the different regions. Most neural signals pass through those network hubs, and therefore, damages to such network hubs can be detrimental. On the other hand, such information super highways were found to improve the performance of artificial neural networks~\cite{he2015deep}, because they allow many layers of neurons to communicate effectively with each other. Indeed, the formation of communities and the small-world topology~\cite{watts1998collective} are observed in social networks where people interact by communicating information. Since neurons also interact by transmitting information, we believe info-clustering can be applied to these information systems to discover or explain the communities with a large amount of intra-cluster communications as well as network hubs that support important inter-cluster communications.

\subsection{Contributions}
\label{sec:motivation}

In this section, we give a summary of the contributions of this work and a brief survey of previous
works pointing out, whenever possible, similarities and differences between info-clustering and existing approaches.
This brief survey is neither complete nor intended to present info-clustering as a
replacement of existing approaches, but rather to motivate info-clustering and help properly
position it relative to existing works.

Many clustering algorithms have been proposed, even for gene clustering~\cite{butte00,sharan00,thalamuthu2006evaluation}. However, the conventional approach has been typically of a heuristic nature with a primary focus on algorithmic simplicity~\cite{pelillo09}. Such an \emph{algorithmic approach} suffers several shortcomings, as it was already indicated by some researchers~\cite{von2012clustering,agarwal11}. For example, the well-known $k$-means clustering algorithm and self-organizing map require prior knowledge of the number of clusters, which is a well-known difficult task. 
For the $k$-means clustering algorithm, the similarity between objects is measured by the distance between the data points associated with the objects. This raises the concern that there are several different choices for defining the distance between two points or two clusters. Various mathematical criteria have been proposed. However, such criteria appear to be ``easy to fool" in the sense that there are examples for which the resulting clustering solution is obviously not the desired one~\cite{kannan2004clusterings}. \emph{The problem is that distance is fundamentally a pairwise measure, and there is no clear unique extension to the case involving more than two data points.}

There are clustering techniques that do not require any prior knowledge of the clusters, but their objective functions are often too difficult to compute. As a concrete example, correlation clustering~\cite{bansal04} specifies the similarity structure by a simple graph, with positive edges between similar nodes and negative edges between dissimilar nodes. The objective is to cluster the nodes in a way that minimizes the total number of pairs of similar nodes in different clusters and dissimilar nodes in the same cluster. Despite the conceptual simplicity in its formulation, the problem was shown to be NP-hard~\cite{bansal04}. This motivated the search for an approximation solution, such as the randomized $3$-approximation algorithm in~\cite{ailon2008aggregating}, which was also recently extended to a parallel version for clustering big data~\cite{pan15}. However, the obtained clusters are not reproducible since the randomization can result in very different-looking clusters. While there are indexes that evaluate the quality of the clusters, and algorithms that combine different clustering solutions together, a coherent theoretical ground is desired.

\begin{ybox}
	The problem of clustering is quite unique in the sense that it attempts to discover unknown patterns in the data. 
	Indeed,  \cite{von2012clustering} raised the question of whether clustering is more of an art than a science, because the existing methods of evaluating a clustering solution are not entirely justified. However, rather than declaring no satisfying solution to the problem, or jumping too quickly to a specific algorithm or dataset, we believe it is more important to \emph{lay a rigorous theoretical ground, upon which many meaningful and practical implementations can be developed}. Such a paradigm should be general enough to capture complicated similarity structures, and be able to reduce to computationally feasible algorithms under verifiable simplifying model assumptions.

\end{ybox}

Indeed, information theory has already been considered in some previous works on data clustering~\cite{kraskov09,Aghagolzadeh07,butte00,ver2014discovering,misra14,rosvall2008maps}. In particular, for gene clustering, the well-known Shannon's mutual information~\cite{shannon48} was used as a measure of similarity between two genes in the clustering algorithm by Mutual Information Relevance Networks (MIRN)~\cite{butte00}. The measure was reported to be less sensitive to outliers, among other benefits. 

Unfortunately, \emph{Shannon's mutual information only measures the amount of information mutual to two random variables} and so, its use for the multivariate case involving multiple random variables in \cite{butte00} was not properly justified. As an illustration of this, we give a concrete example where the clustering by MIRN fails to return the desired cluster. 

Many other information-theoretic frameworks make use of a proposed multivariate extension of Shannon's mutual information, called the total correlation~\cite{watanabe60}. Even with this choice of similarity measure, there have been very different approaches. For example, the hierarchical clustering by mutual information in \cite{kraskov09} made use of the grouping property of the total correlation for three random variables. The correlation explanation algorithm in \cite{ver2014discovering} used the conditional total correlation directly in the objective function to partition the random variables according to a latent tree model. In \cite{nemenman04}, the total correlation was further broken down into a sum of the so-called interaction multi-information. While these works consider information theory to be a promising framework for machine learning, a rigorous common theoretical ground is still missing. For instance, the clustering solutions in \cite{kraskov09} and \cite{nemenman04} have algorithmic characterizations which do not lead to a unique clustering solution. The approaches are mainly supported by experimental rather than theoretical results.

Instead of Shannon's mutual information or the total correlation, \emph{info-clustering makes use of a multivariate information measure called the multivariate mutual information (MMI) that can capture the higher-order correlation among multiple random variables.} The MMI originates from the divergence upper bound in \cite{csiszar04} for the capacity of the secret key agreement problem. Although the bound was shown by \cite{chan2008tightness} to be slack (in the case with helpers), \cite{chan2008tightness} also identified the rather general (no-helper) case when the bound is tight, and interpreted the corresponding expression as a measure of mutual dependency among multiple random variables. This established an alternative characterization of the secret key capacity that
 was formally studied as a measure of mutual information in \cite{chan15mi}, where many interpretations and properties of the measure were discovered to naturally extend those of Shannon's mutual information. The expression was therefore named and regarded as the same notion of mutual information as Shannon has defined in his seminal work~\cite{shannon48}, but extended to the multivariate case.  
 We pause here to make some important remarks on the MMI: 
\begin{ybox}
\begin{compactenum}
	\item The MMI has various concrete operational meanings. Indeed, it was shown in~\cite{chan10md} to be precisely the capacity of the secret key agreement problem in~\cite{csiszar04} and the max-flow min-cut characterization of network coding throughput~\cite{chan11isit}. It is also related to the source coding problem of communication for omniscience~\cite{csiszar04} and the problem's extension to successive omniscience~\cite{chan16so}. 
	\item Among other information-theoretic properties, the MMI satisfies the well-known data processing inequality, which has been used in \cite{chan15mi,chan16so,chan16isit,chan16itw} to derive new results or resolve some conjectures in other multiterminal information theory problems~\cite{mukherjee14,zhang15}.  
	\item The term MMI has also been used (though not very widely) to refer to McGill's multiple information. As we will explain in \S\ref{sec:MMI}, there is an issue with such an extension of Shannon's mutual information, causing it to be negative for the example shown in \figref{fig:Idiagram}. A correction of this extension will lead to the non-negative MMI we consider. The MMI was also called the minimum partition information in \cite{misra14}, but the name was based on the characterization of the MMI by partitions~\cite{chan10md}, which is only one of the many possible characterizations. e.g., an axiomatic formulation of the MMI is given in \cite{chan15mi} using the so-called mutually correlative property. 
	\item A more abstract mathematical form of the MMI for a submodular set function instead of multiple random variables appeared in the work~\cite{fujishige88} of Fujishige and the work~\cite{narayanan90} of Narayanan on the principal lattice of partitions of a submodular function. The MMI enriches the abstract mathematical structure with precise information-theoretic meaning by specializing the submodular function to the entropy function.
\end{compactenum}
\end{ybox}

The MMI has also been applied to clustering by \cite{misra14} and was shown to be superior to Shannon's mutual information under the proposed framework in \cite{misra14}.  However, unlike info-clustering, the work did not go deep into the information-theoretic interpretations of the MMI, and therefore, did not identify the clustering solution we found. Instead, it considered clustering as a universal communication problem, with a decoder that recovers patterns of the transmitted message as clusters. This idea is interesting although it is unclear whether this model assumption is fruitful or limiting, and whether the universal communication problem can lead to an efficient clustering solution. 

The theoretical underpinning of the MMI is a mathematical structure called the principal sequence of partitions (PSP)~\cite{narayanan90}. On the one hand, this structure enables the MMI and the clusters to be computed in strongly polynomial time~\cite{fujishige88,narayanan90} (see also \cite{nagano10}), and adds a new dimension to multi-terminal information theory~\cite{chan16so,chan16isit,mukherjee16,chan16itw,mukherjee14,mukherjee15,MKS16,zhang15,ding16,ding15}. On the other hand, the MMI enriches the abstract mathematical structure with information-theoretic meanings. 

There is also an existing clustering algorithm, called the minimum average cost (MAC) clustering~\cite{nagano10}, which builds implicitly upon the principal sequence of partitions to construct the clusters. However, the exact formulation is based on an abstract mathematical criterion that minimizes certain average of a submodular cost function, which differs from that of the principal sequence of partitions. We show by concrete examples that the MAC clustering is different from info-clustering in general. Instead of building our clustering solution on the abstract mathematical structure of the PSP, we \emph{start with a seemingly different but more meaningful formulation and eventually connect it to the PSP using the properties of the MMI}. We also prove the hierarchical structure of info-clustering separately based on a general property of the MMI, so that potentially other information measures satisfying such property can be applied. Building upon this abstract mathematical framework, a duality result was recently proved in \cite{chan16duality} relating the info-clustering problem with the feature selection problem. The info-clustering formulation was also extended slightly there to map to the more elaborate structure of the principal lattice of partitions (PLP) instead of just the PSP.


The info-clustering paradigm is general. Under some simplifying assumptions on the correlation structure, we show that the solution reduces to the clustering solution by MIRN~\cite{chan15allerton} for gene clustering. Another common model reduction is by assuming a jointly Gaussian distribution, as in the gene clustering method called the clustering identification by connectivity kernel (CLICK)~\cite{sharan00}. We  show that, under the jointly gaussian assumption, info-clustering reduces to a clustering solution that depends only on the covariance matrix through the spectra of its submatrices. This appears to be a new spectral clustering technique different from the spectral clustering algorithm in~\cite{shi2000normalized}, which was only used as an approximate solution to the NP-hard problem of finding the minimum normalized cut for image segmentation~\cite{shi2000normalized}.

Under the pairwise independent network (PIN) model~\cite{nitinawarat-ye10}, where the random variables have a graphical correlation structure, the MMI reduces to the partition connectivity for tree packing~\cite{nitinawarat10}, a well-known notion in combinatorial optimizations~\cite{schrijver02}.
The MMI was also shown to be equal to the maximum multicast throughput of an undirected network, giving it the usual connectivity notion of max-flow min-cut for graphs, which can be further extended to information flows over hypergraphs and, more generally, matroids~\cite{chan11isit,chan12ud,chan13isit,chan13itw}. We show that under the PIN model, the clustering solution corresponds to the PSP of graphs, and the idea is extended further to hypergraphs and more general channel models, following the usual extension of commodity flow to information flows in network coding. Because the physical interconnections among the neurons can be specified by a graph or a hypergraph, with edges being channels that transmit information, the graphical reduction of the info-clustering algorithm can potentially be applied to identify regions of tightly connected neurons in the brain with high intra-cluster communication rates. 

The infomap~\cite{rosvall2008maps} is another clustering algorithm applied to cluster the human connectome~\cite{hinne2015probabilistic}. The idea, like one of the interpretations of info-clustering, is to decompose the network by information flows. However, different from info-clustering, it uses a random walk over a graph as an analogy to information flows over a network. The clusters are obtained by optimizing a special two-stage source coding of the random walk. Unfortunately, the optimization is difficult, and can only be solved approximately. The two-stage source code is also far from the optimal source coding scheme that achieves the entropy rate~\cite{cover91}.

Another information-theoretic approach, called the integrated information theory (IIT), has also been proposed in neuroscience~\cite{balduzzi2008integrated} to study consciousness based on the structure and dynamics of the brain. A measure called the integrated information was defined to measure how integrated the subsystems are within a large system. Another information-theoretic measure is defined in \cite{deco15} to measure the segregation of a large system into separate subsystems. The motivation of such measures is the construction of a whole-brain computational model that can help explain some important features of the brain. However, despite similarity to the info-clustering paradigm, the proposed measures do not have clear operational meanings because some distribution, normalization factors and parameters are chosen in a rather ad-hoc manner. We will show that the info-clustering paradigm leads to a more meaningful measure of segregation and integration. Indeed, the info-clustering paradigm is not limited to biological systems. It can also apply to other information systems or social networks, like the measure of segregation proposed based on social interactions in~\cite{echenique2007measure}.

\begin{ybox}
In summary, info-clustering has the following advantages:
\begin{enumerate}
	\item The clustering procedure is driven by a new multivariate information measure called the MMI, which extends Shannon's mutual information between two random variables to the mutual information among multiple random variables. Like Shannon's mutual information~\cite{shannon48,shannon49}, the MMI has concrete operational meanings in various information-theoretic problems, including source coding, network communication and security.
	\item The clusters can be computed in strongly polynomial time due to the underlying mathematical structure called the PSP. However, unlike the related MAC clustering algorithm, info-clustering has a meaningful formulation not based directly upon the abstract mathematical structure.
	\item The clustering solution is unique and well-defined, unlike many other algorithmic formulations that may require an initial solution or an assumption on the number of clusters.
	\item Under the Markov tree model, info-clustering reduces to an existing gene clustering algorithm called the clustering by MIRN. This shows that info-clustering can apply to gene clustering and help justify existing clustering algorithms with the concrete operational meanings of info-clustering.
	\item For some non-Markov tree models, the clustering by MIRN fails to capture the higher-order statistical dependency among multiple random variables, while info-clustering succeeds to identify the correct clusters.
	\item Under a hypergraphical source model, info-clustering reduces to the PSP of hypergraphs. It gives the PSP a concrete operational meaning as clustering by network information flow. Such a model can be applied to cluster the neurons by their physical connections.
	\item Under the jointly Gaussian assumption, info-clustering reduces to a method of clustering by the covariance matrix. Compared to the existing spectral clustering method, it is a different algorithm that has concrete information-theoretic meaning. 
	\item A meaningful measure of integration and segregation can be derived in a more rigorous way than the integrated information theory (IIT), with applications beyond biological systems such as social networks.
\end{enumerate}
While there are many practical approximations and implementations possible for info-clustering, the focus of this paper is on the theoretical development and its potential biological applications in the study of human genome and connectome. 
\end{ybox}

\begin{gbox}
\noindent\textbf{Organization:}
The paper will be organized as follows. The info-clustering paradigm will be formulated in \S\ref{sec:HC} and characterized in \S\ref{sec:cPSP}, with the detailed clustering procedures implemented in Algorithm~\ref{algo:iteration}, \ref{algo:FP} and \ref{algo:PSP}. Its biological applications are through the model reductions in \S\ref{sec:reduction}. 

\noindent\textbf{Notations:}
Throughout this paper, unless otherwise specified, we use sans-serif upper-case letters (e.g., $\RZ$, $\RX$, etc.) to denote random variables
and calligraphic font upper-case letters (e.g., $\mathcal{C}$,
$\mathcal{F}$, etc.) to denote collections of sets.
For any collection of sets $\mathcal{F}$ whose elements are subsets of some finite set,
we use $\op{maximal} \mathcal{F}$ to denote the inclusion-wise maximal elements of
$\mathcal{F}$, i.e.,
\begin{align}
	\op{maximal} \mcF := \left\{ B \in \mcF \mid \not\exists B'\supsetneq B, B'\in \mcF \right\}.\label{eq:maximal}
\end{align}
\end{gbox}

\section{Hierarchical clustering formulation}
\label{sec:HC}

The info-clustering formulation in this paper separates into two main components. The first component is a formulation using a threshold constraint on the MMI. We show that the solution is hierarchical, and so, an iterative algorithm can be used to compute the clusters. We will make the proof general using only a simple property of the MMI rather than its detailed definition,
i.e., the results herein hold for any multivariate information measure that satisfies such a property. We use the two terms ``multivariate information" and ``multivariate mutual information (MMI)" for two distinct meanings, where the former refers to a \emph{general} information measure for multiple random variables as detailed in this section while the latter refers to a \emph{specific} information measure defined in the next section as \eqref{eq:mi}. The second component of the formulation is a refinement of the clustering solution based on further properties of the MMI as detailed in the next section. 
The reason for the two-step characterization is not only for theoretical elegance but also for practical implementations of info-clustering in subsequent work. The more general hierarchical solution developed in this section may allow the MMI to be approximated and estimated from data more efficiently with a tunable level of computational and sample complexity.



\subsection{Threshold constraint} 
\label{sec:cluster}

To cluster objects using information theory, we first associate each object we want to cluster, say $i$, with all the information that describes it. The information is represented by a random variable, say $\RZ_i$, which can be viewed as a file containing some measurements of the object~$i$. Then, we cluster all the objects based on the mutual information among the random variables $\RZ_i$'s. 

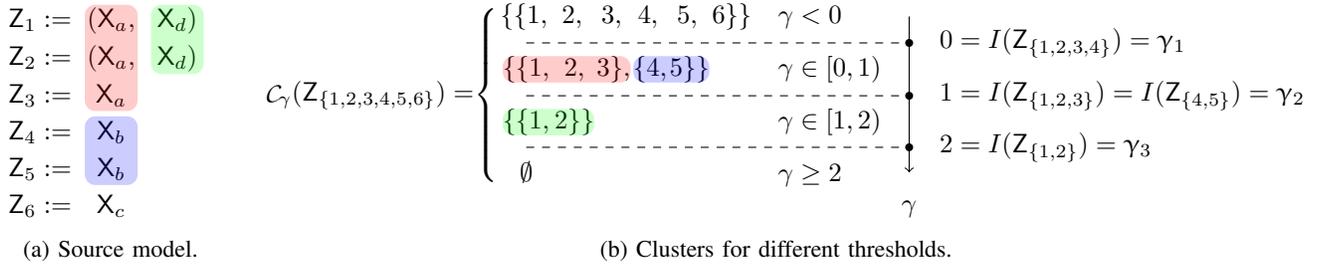
\begin{figure*}
	\centering
	\tikzstyle{cluster}=[opacity=0.2,fill=gray]
	\begin{subfigure}[b]{.15\textwidth}
		{\def\u{1em}
			\begin{tikzpicture}[inner sep=0,outer sep=0,x=\u,y=\u, every node/.style={rounded corners}]
			\matrix (Z) [matrix of math nodes, every cell/.style={anchor=base,minimum width=2*\u,minimum height=1.2*\u}, column 1/.style={anchor=base east}, row sep=0.2*\u,column sep=0.5*\u]
			{ && \\
				\RZ_1:=& (\RX_a, & \RX_d) \\
				\RZ_2:=& (\RX_a, & \RX_d) \\
				\RZ_3:=& \RX_a &  \\
				\RZ_4:=& \RX_b &  \\
				\RZ_5:=& \RX_b &  \\
				\RZ_6:=& \RX_c &  \\
			};
			\node[cluster,fill=red, fit=(Z-2-2) (Z-4-2)] {};
			\node[cluster,fill=blue, fit=(Z-5-2) (Z-6-2)] {};
			\node[cluster,fill=green, fit=(Z-2-3) (Z-3-3)] {};
			\end{tikzpicture}
		}
		\caption{Source model.}
		\label{fig:eg:Z}
	\end{subfigure}
	\hfil
	\begin{subfigure}[b]{.75\textwidth}
		{\def\u{1em}
			\tikzstyle{point}=[circle,minimum size=.3em,fill]
			\begin{tikzpicture}[inner sep=0,outer sep=0,x=\u,y=\u, every node/.style={rounded corners}]
			\matrix (C) [matrix of math nodes, nodes in empty cells, row sep=1*\u,column sep=0.1*\u,
			column 6/.style={column sep=1*\u},
			column 7/.style={anchor=base west,column sep=1*\u},
			left delimiter=\{]
			{
				\{\{1, & 2, & 3, & 4, & 5, & 6\}\} &  `g<0 & \\
				\{\{1, & 2, & 3\}, & \{4, & 5\}\} & & `g\in [0,1 )  \\
				\{\{1, & 2\}\} &  &  &  &  &  `g\in [ 1, 2 ) \\
				`0 &  &  &  &  &  &  `g\geq 2  \\
			};
                        \path (C.west) node [left] {$\pzC_{`g}(\RZ_{\Set{1,2,3,4,5,6}})=\kern1em$};
			\draw[->] (C-1-1-|C-1-8) to node (12) {} (C-2-1-|C-1-8) to node (23) {} (C-3-1-|C-1-8) to node (34) {} (C-4-1-|C-1-8) node[label={[label distance=\u]below:{$`g$}}] {};
			\foreach \x/\xtext in {12/$0=I(\RZ_{\Set{1,2,3,4}})=
				\upgamma_1$,23/$1=I(\RZ_{\Set{1,2,3}})=I(\RZ_{\Set{4,5}})=
				\upgamma_2$,34/$2=I(\RZ_{\Set{1,2}})=
				\upgamma_3$}
			\draw[dashed] (C-1-1|-\x) -- (\x) node [point,label={[label distance=\u]right:{\xtext}}] {};
			\node[cluster,fill=red, fit=(C-2-1) (C-2-3)] {};
			\node[cluster,fill=blue, fit=(C-2-4) (C-2-5)] {};
			\node[cluster,fill=green, fit=(C-3-1) (C-3-2)] {};
			\end{tikzpicture}
			\vspace{-1em}
		}
		\caption{Clusters for different thresholds.}
		\label{fig:eg:cluster}
	\end{subfigure}
	\caption[An example of clustering a set of random variables.]{An example of clustering a set of random variables based on their shared information. The statistical dependency of the random variables $\RZ_i$'s to be clustered are defined in terms of a set of independent uniformly random bits $\RX_j$'s.}
	\label{fig:eg}
\end{figure*}

As a motivating example, consider clustering the random variables $\RZ_i$'s defined in \figref{fig:eg:Z}. The random variables are correlated in the sense that they share some uniformly random independent bits $\RX_a,\RX_b,\RX_c$ and $\RX_d$.\footnote{Each random bit is uniformly random over $\Set{0,1}$, and the random bits are mutually independent.}  
It is desirable to group $\RZ_1,\RZ_2$ and $\RZ_3$ in a cluster because they share a common bit $\RX_a$, but it is desirable to group $\RZ_4$ and $\RZ_5$ in a different cluster because they share a different independent bit $\RX_b$. It is also desirable to group $\RZ_1$ and $\RZ_2$ as a smaller cluster (compared to $\Set{1,2,3}$) because they share the additional bit $\RX_d$ (in addition to the bit $\RX_a$).



More generally, let $V$ be a finite set of objects we want to cluster and
\begin{align*}
	\RZ_V:=(\RZ_i\mid i\in V)
\end{align*}
be the vector of random variables associated with the objects. For every subset $B\subseteq V$ of at least two random variables, we measure their shared information by some finite real number $I(\RZ_B)$, i.e.,
\begin{align*}
	I(\RZ_B)\in `R \kern1em \text{ for }B\subseteq V:\abs{B}\geq 2.
\end{align*}
This multivariate information quantity should depend on the joint distribution $P_{\RZ_B}$ of the random vector $\RZ_B$, but its precise definition will be postponed to \S\ref{sec:MMI} because \emph{the combinatorial structure of the following hierarchical clustering formulation of info-clustering does not depend on the particular choice of this measure, allowing potentially other measures to be used:}

\begin{ybox}
\begin{Definition}[Clusters]
	\label{def:cluster} 	
	\begin{subequations}
	For a \emph{threshold} $`g\in `R$, the set of \emph{clusters} is defined as
	\begin{align}
		\pzC_{`g}(\RZ_V):= \op{maximal}\Set{B\subseteq V \mid \abs{B}>1, I(\RZ_B)>`g}.\label{eq:cluster}
	\end{align}
	{The maximality~\eqref{eq:maximal} requirement ensures consistency among the clusters, i.e., a cluster $B$ with $I(\RZ_B)>`g$ does not separate apart any other highly correlated random variables in a larger set $B'\supsetneq B$ that also satisfies $I(\RZ_{B'})>`g$.\footnotemark}
 For notational convenience,
	\begin{align}
          \pzC(\RZ_V):= \bigcup_{`g\in `R} \pzC_{`g}(\RZ_V)\label{eq:clusters}
	\end{align}
	denotes the collection of all clusters at different thresholds.
	\end{subequations}
\end{Definition}
In words, given a threshold $`g$, we consider subsets $B$ of two or more elements from $V$, such that the random variables $\RZ_B$ indexed by the elements of $B$ have the multivariate information quantity $I(\RZ_B)$ \emph{strictly larger} than the threshold $`g$. Out of all such non-singleton subsets satisfying the \emph{threshold constraint}, we pick the inclusion-wise maximal subsets to be the clusters. To put it simply, the idea of clustering is to group together random variables, as many as possible, such that the group has increasingly larger amount of multivariate information. The desired level of multivariate information is specified by the threshold $`g$. 
\end{ybox}
\footnotetext{The maximality constraint can be relaxed slightly in \cite{chan16duality} to prove a stronger result on the duality between data clustering and feature selection. For simplicity, we limit the scope of this paper to the maximality constraint.}

The following is an illustration of the threshold-constraint formulation when applied to the motivating example above.

\begin{Example}
	Let $V:=\Set{1,\dots,6}$ and define $\RZ_V$ in terms of the independent uniformly random bits $\RX_a,\RX_b,\RX_c$ and $\RX_d$ as shown in \figref{fig:eg:Z}. For this example, the dependency structure is simple, and so, let us define the information measure as the number of shared bits:
	\begin{align}
		\label{eq:eg:I}
		I(\RZ_B)=
		\begin{cases}
			2 & B=\Set{1,2}\\
			1 & B\in \Set{\Set{1,2,3},\Set{4,5},\Set{2,3},\Set{1,3}}\\
			0 & \text{otherwise}.
		\end{cases}
	\end{align}
	For instance, $I(\RZ_{\Set{1,2,3}})=1=I(\RZ_{\Set{4,5}})$ because $\RZ_1$, $\RZ_2$, and $\RZ_3$ share the common bit $\RX_a$ while $\RZ_4$ and $\RZ_5$ share the common bit $\RX_b$. Similarly, $I(\RZ_{\Set{1,2}})=2$ because $\RZ_1$ and $\RZ_2$ share the common bit $\RX_d$ in addition to the bit $\RX_a$. Finally, $I(\RZ_V)=0$ because $\RZ_6$ is independent of all other random variables $\RZ_i$ for $i\neq 6$.
	
	If $`g=0$, then \eqref{eq:cluster} asks for the maximal subsets with shared information strictly larger than zero. By \eqref{eq:eg:I}, the sets whose shared information is larger than zero are $\Set{4,5}$, and any subset of two or more elements from $\Set{1,2,3}$. The maximal among such subsets are the clusters at threshold zero, i.e., $\pzC_0(\RZ_V)=\Set {\Set{1,2,3},\Set{4,5}}$. 
	If $`g=1$ instead, then \eqref{eq:cluster}  asks for the maximal subsets of random variables with more than $1$~bit of shared information. The only choice is the subset $\Set {1,2}$, and so it is the only cluster at threshold $1$, i.e., $\pzC_1(\RZ_V)=\Set {\Set {1,2}}$. For other values of $`g$:
	\begin{compactitem}
		\item For $`g<0$, the entire set $V$ is the only cluster because $I(\RZ_V)=0>`g$ and $V$ is the maximal set trivially.
		\item  For $`g\in [0,1)$, there are two clusters $\Set{1,2,3}$ and $\Set{4,5}$ because $I(\RZ_{\Set{1,2,3}})=I(\RZ_{\Set{4,5}})=1>`g$, and each of the sets is maximal.
		\item For $`g\in [1,2)$,  the set $\Set{1,2}$ is the only cluster because it is the only set with the shared information $I(\RZ_{\Set{1,2}})=2>`g$.
		\item There are no clusters for $`g\geq 2$ because no set of random variables has more than $2$~bits of shared information.
	\end{compactitem}
The complete clustering solution is illustrated in \figref{fig:eg:cluster}.
\end{Example}

\begin{bbox}
There are various important properties we can observe from the clustering solution in \figref{fig:eg:cluster}:
\begin{itemize}
	\item The set of clusters changes at a finite set of threshold values, namely the set of thresholds $\Set{0,1,2}$. The threshold values are the shared information of the clusters and each cluster occupies a contiguous interval between two of the threshold values. (E.g., the threshold value $2$ is the shared information of the cluster $\Set{1,2}$ that appears over the interval $`g\in [1,2)$.) In particular, the smallest threshold value $0$ is the shared information of the entire set of random variables, and the largest threshold value $2$ is the maximum shared information over all subsets that contain at least two random variables.
	\item For each threshold $`g$, the clusters are disjoint. For two different thresholds, two clusters are either disjoint or the larger-threshold cluster is a proper subset of the smaller-threshold cluster. (E.g., the cluster $\Set{1,2,3}$ at $`g=0.5$ does not intersect with the other cluster $\Set{4,5}$ at the same threshold, but it contains the cluster $\Set{1,2}$ that arises at the larger threshold $`g=1.5$.)
	\item There is an iterative relationship among the clusters: the cluster of a cluster of $\RZ_V$ is also a cluster of $\RZ_V$. For example, if we consider clustering the random variables from the cluster $\Set{1,2,3}$, then $\Set{1,2}$ is a cluster of $\Set{1,2,3}$. Note that $\Set{1,2}$ is also a cluster of $\RZ_V$.
\end{itemize}
\end{bbox}

In the next subsection, we will show that the above hierarchical structure holds more generally. 

\subsection{Hierarchical structure}
\label{sec:critical}

For convenience, we will use $`g^-$ ($`g^+$) to represent the value that is arbitrarily close to but strictly smaller (larger) than $`g$. More precisely, we write $\pzC_{`g^-}(\RZ_V)$ ($\pzC_{`g^+}(\RZ_V)$) for the limit of $\pzC_{t}(\RZ_V)$ as $t$ increases (decreases) to $`g$ from below (above). The limit exists because there is only a finite number of clusters, i.e.,
\begin{align}
	\abs*{\pzC(\RZ_V)} < \abs{\Set{B\subseteq V}} = 2^{\abs{V}} <`8,
\end{align}
even though the set $\pzC(\RZ_V)$ of all clusters~\eqref{eq:clusters} is a union over all real threshold values $`g$. Define the set of \emph{critical values} for clustering $\RZ_V$ as
\begin{align}
	\kern-1em \Upgamma(\RZ_V):=\Set{`g\in `R\mid \pzC_{`g^-}(\RZ_V)\neq \pzC_{`g^+}(\RZ_V)}.\label{eq:Upgamma}
\end{align}
This is the set of threshold values of interest, because the set of clusters changes at those values. The following theorem asserts that the set of critical values is also finite.

\begin{Theorem}[Discreteness]
	\label{thm:cluster}
	\begin{subequations}
		The set of critical values can be written as
		\begin{align*}
			\Upgamma(\RZ_V)=\Set{`g_i:1\leq i\leq N}
		\end{align*} 
		for some positive integer $N$
		with $`g_i<`g_{i+1}$
		for $1\leq i<N$. Furthermore, assuming  
		$\Set{\pzC_{`g_i}(\RZ_V):1\leq i\leq N}$ is the collection of the sets of clusters for
		the critical values, then the complete clustering~\eqref{eq:cluster} of
		$\RZ_V$ is given as
		\begin{align}
			\label{eq:cluster:solution}
			\pzC_{`g}(\RZ_V)=
			\begin{cases}
				\Set{V} & `g<`g_1\\
				\pzC_{`g_i}(\RZ_V) & `g\in [`g_i,`g_{i+1}), 1\leq i<N\\
				`0 & `g\geq`g_N.
			\end{cases}
		\end{align}
		Finally, the first and last critical values are
		\begin{align}
			`g_1 & =I(\RZ_V) \label{eq:upgamma1}\\
			`g_N & =\max_{B\subseteq V: \abs{B}>1} I(\RZ_B) \label{eq:maxMMI}
		\end{align}
		and $\pzC_{`g_{N-1}}(\RZ_V)$ is the set of all maximal $B$ achieving the
		maximum amount of multivariate information \eqref{eq:maxMMI}.
	\end{subequations}
\end{Theorem}

\begin{Proof}
	See Appendix~\ref{sec:A}.
\end{Proof}

\begin{Definition}[Critical values]
	\label{def:critical} 
	We will use $\upgamma_i(\RZ_V)$ and $\pzN(\RZ_V)$ to denote the $i$-th critical value and the number of critical values for $\RZ_V$ respectively. For simplicity and when there is no ambiguity, we may drop the dependency on $\RZ_V$ and write, e.g., $\pzC_{\upgamma_1(\RZ_V)}(\RZ_V)$ as $\pzC_{\upgamma_1}(\RZ_V)$.
\end{Definition}
Applying the definition of clusters~\eqref{eq:cluster} to an arbitrary subset $B \subseteq V$ with size at least two, the
definition above extends to any such subset, where in this case the simplified notation
$\mathcal{C}_{\upgamma_1}(\RZ_{B})$ will mean $\mathcal{C}_{\upgamma_{1}(\RZ_B)}(\RZ_B)$.

Next, we show that every cluster $B'$ of $\RZ_V$ can be obtained by computing the set of clusters of $\RZ_{B''}$ for some larger (previous) cluster $B''$ of $\RZ_V$: 

\begin{Theorem}[Iterative relation]
	\label{thm:iterate}
	For each $B'\in \pzC_{`g'}(\RZ_V)$ and $`g'\geq \upgamma_1(\RZ_V)$,
	\begin{align}
		\exists B'' \in \pzC_{`g''}(\RZ_V): `g''<`g', B'\in \pzC_{\upgamma_1}(\RZ_{B''}). \label{eq:iterate}
	\end{align}
	That is, we can obtain $B'$ by computing clusters that correspond to the first critical value of an earlier cluster $B''$.
\end{Theorem}

\begin{Proof}
	See Appendix~\ref{sec:B}.
\end{Proof}

According to Theorem~\ref{thm:iterate}, we can compute the complete solution to the clustering problem if we can compute the first set $\pzC_{\upgamma_1}(\RZ_{B})$ of clusters for all subsets $B\subseteq V:\abs{B}> 1$. 
However, without any additional properties of the multivariate information measure, it is unclear whether the iterative algorithm can be computed efficiently due to the following issues:
\begin{bbox}
\begin{enumerate}
	\item\label{item:issue1} While Theorem~\ref{thm:iterate} states that every cluster $B'$ is in the first set $\pzC_{\upgamma_1}(\RZ_{B''})$ of clusters of a larger cluster $B''$, the converse may not be true. That is, a cluster of $\RZ_{B''}$ in $\pzC_{\upgamma_1}(\RZ_{B''})$ may not be a cluster of $\RZ_V$, because the maximality in \eqref{eq:cluster} needs to be verified in addition.
	\item\label{item:issue2} The total number of clusters under a general multivariate information measure can be large. By~\eqref{eq:cluster}, $\pzC_{\upgamma_1}(\RZ_V)$ is an \emph{antichain} in the sense that a cluster in $\pzC_{\upgamma_1}(\RZ_V)$ cannot be a subset of another cluster in $\pzC_{\upgamma_1}(\RZ_V)$. Without any other restriction, the size of an antichain can be exponential in $\abs{V}$ by \emph{Sperner's theorem}~\cite{stanley12}. 
\end{enumerate}
\end{bbox}

To illustrate the issues above more clearly:
\begin{Example}
	\label{eg:IMm}
	Consider $V:=\Set{1,2,3,4}$ and let the multivariate information quantity be defined as follows: For $B\subseteq V:\abs{B}>1$,
	\begin{align*}
		I(\RZ_B)=
		\begin{cases}
			0 & B=V\\
			2 & B=\Set{2,3}\\
			3 & B=\Set{2,3,4}\\
			1 & \text{otherwise.}
		\end{cases}
	\end{align*}
	From this, it follows that the set of clusters is given by
	\begin{align*}
		\pzC_{`g}(\RZ_V) &=
		\begin{cases}
			\Set{\Set {1,2,3,4}} & `g<0\\
			\Set{B\subseteq V: \abs{B}=3} & `g\in [0,1)\\
			\Set{\Set{2,3,4}} & `g\in [1,3)\\
			`0 & `g\geq 3.
		\end{cases}
	\end{align*}
	Consider now the cluster $\Set{1,2,3}$ of $\RZ_V$. It is not hard to verify that $\Set{2,3}$ is a cluster (in the first set of clusters) of $\RZ_{\Set{1,2,3}}$. However, the set $\Set{2,3}$ is not a cluster of $\RZ_V$ because the proper superset $\Set{2,3,4}$ has a larger multivariate information of $3$.
\end{Example}

\begin{ybox}
Next, we show that both issues can be resolved if the following simple, but in our opinion fundamental, property is satisfied by the multivariate information measure:
\begin{align}
	I(\RZ_{B_1\cup B_2}) \geq \min\Set{I(\RZ_{B_1}),I(\RZ_{B_2})} \label{eq:I:lb}
\end{align}
for all $B_i\subseteq V:\abs{B_i}>1$, $i\in \Set{1,2}$ and $B_1\cap B_2\neq `0$. This property holds for the MMI we will consider in \eqref{eq:mi} (\cite[Corollary~5.1]{chan15mi}), but it also holds for some other multivariate information quantities.\footnotemark


\begin{Theorem}[Laminarity under \eqref{eq:I:lb}] 
	\label{thm:Laminarity}
	For any multivariate information measure that satisfies the property \eqref{eq:I:lb},
	the collection of all clusters $\pzC(\RZ_V)$ forms a \emph{laminar family}~\cite{schrijver02}, i.e.,
	\begin{align}
		B_1 \cap B_2  \in\Set{`0,B_1,B_2} \label{eq:laminar}
	\end{align}
	for all clusters $B_1,B_2\in \pzC(\RZ_V)$. In particular, for every $`g\in `R$, the set $\pzC_{`g}(\RZ_V)$ consists of disjoint clusters.
\end{Theorem}

\begin{Proof}
	Consider two clusters $B_1,B_2\in \pzC(\RZ_V)$ with $B_1\cap B_2\neq `0$. Without loss of generality, assume $I(\RZ_{B_1})\leq I(\RZ_{B_2})$. By \eqref{eq:I:lb}, $I(\RZ_{B_1\cup B_2})\geq I(\RZ_{B_1})$ and so $B_1\cup B_2=B_1$ or it would contradict the maximality of $B_1$. That is, $B_2\subseteq B_1$ and so we have \eqref{eq:laminar}.
\end{Proof}
Under the context of data clustering, the aforementioned laminarity is usually known as \emph{hierarchical clustering} or \emph{dendrogram}. By Theorem~\ref{thm:Laminarity}, any multivariate information measure that satisfies the property \eqref{eq:I:lb} will necessarily lead to a clustering solution that is guaranteed to be hierarchical. If we define a similarity relation $i\sim_{`g} j$ to mean that there exists $C\subseteq V$ containing both $i$ and $j$ such that $I(\RZ_C)>`g$, then it can be shown that $\sim_{`g}$ is an equivalence relation for any threshold $`g$. In particular, \eqref{eq:I:lb} implies that the relation is transitive. The set $\pzC_{`g}(\RZ_V)$ of clusters can be shown to be precisely the set of non-singleton equivalence classes, and so the clusters are disjoint.
\end{ybox}
\footnotetext{Watanabe's total correlation~\eqref{eq:JT}, Han's dual total correlation and the multivariate Wyner's common information described in~\cite{chan15mi} satisfy \eqref{eq:I:lb} because they are non-decreasing. However, their corresponding clustering solutions are also trivial due to the monotonicity. The normalized version of the total correlation considered in \cite[(6.2)]{chan15mi} satisfies \eqref{eq:I:lb}, as can be shown by \cite[Corollary~5.4]{chan15mi}). It is not monotonic and therefore gives non-trivial clustering solutions with the laminar structure by Theorem~\ref{thm:Laminarity}. McGill's multiple information and the multivariate G\'acs--K\"orner common information both fail to satisfy \eqref{eq:I:lb}.}

Our next result shows that under the condition \eqref{eq:I:lb}, the complete solution to the clustering problem can indeed be computed iteratively from the first set of clusters of a previous cluster.

\begin{Theorem}[Iterative relation under \eqref{eq:I:lb}]
	\label{thm:iterate2}
	For any multivariate information measure that satisfies the property \eqref{eq:I:lb}, 
	\begin{align}
		\pzC(\RZ_{V})=\Set{V}\cup \bigcup_{B'\in
			\pzC_{\upgamma_1}(\RZ_V)} 
		\pzC(\RZ_{B'}),\label{eq:it2}
	\end{align}
	where
	for any $B \subseteq V$, the set $\pzC(\RZ_B)$   is the collection of all clusters of
	$\RZ_{B}$ (similar to \eqref{eq:clusters}). 
\end{Theorem}

\begin{Proof}
	See Appendix~\ref{sec:C}.
\end{Proof}

\begin{Corollary}
	For any multivariate information measure that satisfies the property \eqref{eq:I:lb}, the total number of clusters
	\begin{align}
		\abs*{\pzC(\RZ_V)}\leq \abs{V}-1, \label{eq:num_clusters}
	\end{align}
	which is linear in the number of random variables to be clustered.
\end{Corollary}

\begin{Proof}
	Consider proving  \eqref{eq:num_clusters} by an induction on the size of $V$. In particular, consider the non-trivial case when $\pzC_{\upgamma_1}(\RZ_V)$ is non-empty. The base case $\abs{V}=2$ holds trivially. By \eqref{eq:it2},
	\begin{align*}
		\abs*{\pzC(\RZ_V)} &= 1+\sum_{B'\in \pzC_{\upgamma_1}(\RZ_V)} \abs*{\pzC(\RZ_{B'})}\\
		&\leq 1+\sum_{B'\in \pzC_{\upgamma_1}(\RZ_V)} (\abs{B'}-1) \leq \abs{V}-1,
	\end{align*}
	where the first inequality is by the inductive hypothesis and the last is because $\pzC_{\upgamma_1}(\RZ_V)$ consists of disjoint proper subsets of $V$ by Theorem~\ref{thm:Laminarity}.
\end{Proof}

Hence, given an algorithm that can compute the first critical value $\upgamma_1$ and the first set $\pzC_{\upgamma_1}$ of clusters of any given set of random variables, we can compute the entire clustering solution by applying the algorithm at most $\abs{V}$ times. The pseudocode is given in Algorithm~\ref{algo:iteration}. 

\begin{algorithm}
   	\caption{Hierarchical clustering by iteration.}
   	\label{algo:iteration}
   	\SetKwFunction{FirstClusters}{FirstClusters}
   	\SetKwProg{myfn}{function}{:}{end}
   	\DontPrintSemicolon
   	\SetAlgoLined
   	\KwData{Statistics of $\RZ_V$ sufficient for calculating \FirstClusters{$B$} for all $B\subseteq V:\abs{B}>1$.}
   	\KwResult{$\mcS$ is a list of $(I(\RZ_C),C)$ for $C\in \pzC(\RZ_V)$, which gives
   		 $\Upgamma(\RZ_V)=\Set{`g' \mid (`g',B)\in \mcS}$ and
   		 $\pzC_{`g}(\RZ_V)=\op{maximal}\Set{B\mid (`g',B)\in \mcS,`g'>`g}$.}
   	\BlankLine
   	$\mcS,\mcT\leftarrow$ empty queues;\;
   	enqueue $V$ to $\mcT$;\;
   	\While{$\mcT$ is non-empty}{
   		$B\leftarrow$ dequeue $\mcT$;\;
   		$(`g,\mcC)\leftarrow \FirstClusters(B)$;\;
   		enqueue $(`g,B)$ to $\mcS$;\;
   		enqueue all elements of $\mcC$ to $\mcT$;\;
   	}
 	\BlankLine
   	\myfn{\FirstClusters{$B$}}{
   		\KwRet $(\upgamma_1(\RZ_B),\pzC_{\upgamma_1}(\RZ_B))$;
   	}
\end{algorithm}

The algorithm computes the list $\mcS$ of all clusters $B\in \pzC(\RZ_V)$ and their associated values $I(\RZ_B)$. It calls the function \FirstClusters{$B$} iteratively to obtain the first critical value $`g$ and the first set $\mcC$ of clusters of every previously discovered cluster stored temporarily in $\mcT$. The newly discovered clusters in $\mcC$ are further added to $\mcT$.  

\section{Clustering by the MMI}
\label{sec:cPSP}

In this section, we focus on the clustering solution under the MMI measure in~\cite{chan15mi}. Although a general property, namely, property~\eqref{eq:I:lb}, suffices for a laminar hierarchical clustering solution  in \S\ref{sec:HC}, the resulting clusters may be trivial or meaningless if the multivariate information measure is not chosen properly. (For instance, any choice of $I(\RZ_B)$ that is non-decreasing in $B$ will satisfy \eqref{eq:I:lb} but will only produce the trivial cluster $\Set{V}$.) 
We will explain the meaning of the MMI measure precisely and show that the corresponding hierarchical clustering formulation is related to a non-trivial mathematical structure called the principal sequence of partitions (PSP)~\cite{narayanan90} of the entropy function. Consequently, the solution can be computed by some well-studied submodular function optimization techniques~\cite{narayanan90} that can run in strongly polynomial time. 

\subsection{Multivariate mutual information}
\label{sec:MMI}
Recall that for the simple example considered earlier in \figref{fig:eg:Z}, the mutual information of a given set was measured by the number of bits shared by the random variables in the set. 
For a general source model, \emph{Shannon's mutual information}~\cite{shannon48} provides a well-accepted measure in the bivariate case involving only two random variables:\footnote{For the bivariate case, there are also other measures of shared information such as the Wyner's common information~\cite{wyner75} and G\'acs--K\"orner common information~\cite{gac72}. Shannon's mutual information was also described by Shannon~\cite{shannon48} as the amount of information ``common" to two random variables. 
	However, Wyner's common information and G\'acs--K\"orner common information  are not as widely used as Shannon's mutual information measures. They measure more specific kind of shared information and also have their own multivariate extension in \cite{liu10,chan16itw,csiszar04}.} 
\begin{align}
	I(\RZ_1\wedge \RZ_2) := D`1(P_{\RZ_1,\RZ_2}\|P_{\RZ_1}P_{\RZ_2}`2),\label{eq:ShannonMI}
\end{align}
where $D`1(P_1\|P_2`2)$ denotes the Kullback--Liebler divergence~\cite{kullback1951information} between the distributions $P_1$ and $P_2$. The divergence on the R.H.S.\ of \eqref{eq:ShannonMI} can be interpreted as a statistical distance to independence, because it equals zero if and only if the two random variables are independent. 

A straightforward extension of Shannon's mutual information \eqref{eq:ShannonMI} to the multivariate case is Watanabe's total correlation~\cite{watanabe60}:
\begin{align}
  J_{\opT}(\RZ_V):= D`1(\extendvert{P_{\RZ_V}\|\prod\nolimits_{i\in V}P_{\RZ_i}}`2),\label{eq:JT}
\end{align}
which is equal to zero if and only if the random variables $\RZ_i$ for $i\in V$ are mutually independent. 
While the total correlation captures the mutual independence among the random variables, it fails to capture many other forms of independence relation. The work in \cite{chan15mi} aims at a more precise understanding of this and the formulation of the MMI, as follows, capable of capturing any form of independence that might exist among three or more random variables.

\begin{ybox}
Let $\Pi'(V)$ be the collection of all possible partitions of $V$ that splits $V$ into at least two nonempty disjoint subsets. (In other words, $\Pi'(V)$ is the collection of all set partitions of $V$ except the trivial partition $\Set{V}$.) For any partition $\mcP\in \Pi'(V)$ of $V$, the product distribution $\prod_{C\in \mcP} P_{\RZ_C}$ specifies an independence relation, namely, that the agglomerated random variables $\RZ_C$'s are mutually independent. A well-constructed measure needs to ensure that the mutual information is measured at zero as long as an independence relation exists among the random variables from $V$, not just when all the random variables from $V$ are mutually independent.

We now introduce the MMI measure from \cite{chan15mi}:
\begin{subequations}
	\label{eq:mi}
	\begin{align}
		I(\RZ_V) &:= \min_{\mcP\in \Pi'(V)} I_{\mcP}(\RZ_V) \kern1em
		\text{where} \label{eq:IS}\\
		I_{\mcP}(\RZ_V) &:= \frac1{\abs{\mcP}-1} D`1(\extendvert{P_{\RZ_V}\|\prod_{C\in \mcP} P_{\RZ_C}}`2).\label{eq:IP}
	\end{align}
\end{subequations}
Clearly, by the above definition, we have $I(\RZ_V)=0$ if and only if there exists an independence relation among the random variables from $V$. 
\begin{Example}
	\label{eg:I}
	For the example considered earlier in \figref{fig:eg:Z}, we have $I(\RZ_V)=0$ because of the independence relation
	\[P_{\RZ_{\Set{1,2,3,4,5,6}}}=P_{\RZ_{\Set{1,2,3}}} P_{\RZ_{\Set{4,5}}} P_{\RZ_6}\] 
	and so $I_{\mcP}(\RZ_V)=0$ with $\mcP=\Set{\Set{1,2,3},\Set{4,5},\Set{6}}$.
\end{Example}
The divergence expression~\eqref{eq:mi} of the MMI derives from a divergence upper bound~\cite[(26) in Example~4]{csiszar04} on the secrecy capacity for the multiterminal secret key agreement problem. The bound was derived in the general case with helpers, and was considered as a heuristically meaning upper bound to the LP characterization of the capacity in \cite{csiszar04}. The bound was shown to be tight for the case involving $2$ or $3$~users even involving helpers, but it was left open in \cite{csiszar04} whether the bound is tight beyond $3$ users.  \cite{chan2008tightness,chan10md} extended the brute-force search of \cite{csiszar04} and showed with the help of a computer program that the bound is tight for $4$ or $5$ users. However, a counter-example involving $6$ users with $3$ helpers was also discovered, showing that the bound is loose with the presence of helpers and therefore does not have the same meaning as the secrecy capacity. 
Nevertheless, it was identified and proved in \cite[Theorem~1]{chan2008tightness}\cite[Theorem~1.1]{chan10md}\cite[Theorem~2.1]{chan15mi} that the bound is tight in the no-helper case even under a general private source distribution, using only the well-known submodularity of entropy. This establishes the concrete operational meaning of the MMI as the secrecy capacity in the general no-helper case, much like the way Shannon's mutual information was shown to characterize the channel capacity in the seminal work~\cite{shannon48} of Shannon. A first attempt to interpret the MMI as a measure of mutual information among multiple random variables (and to explain the normalization factor of $\abs{\mcP}-1$ in \eqref{eq:mi}) appeared in \cite[Section~IV]{chan2008tightness}.
\end{ybox}

It is useful to compute $I_{\mcP}$ by rewriting the divergence in terms of Shannon's entropy or mutual information as follows:
\begin{subequations}
	\begin{align}
		D`1(\extendvert{P_{\RZ_V}\|\prod_{C\in \mcP} P_{\RZ_C}}`2)
		&= \sum_{C\in \mcP}H(\RZ_C) - H(\RZ_V) \label{eq:D:1}\\
		&= \sum_{i=1}^{k-1} I(\RZ_{C_1\cup \dots \cup C_i} \wedge \RZ_{C_{i+1}})\label{eq:D:2}
	\end{align}
\end{subequations}
where we used $C_1,\ldots,C_k$ to denote the blocks of the partition $\mcP$. The measure $I_{\mcP}$ is also written more explicitly in \cite{chan15mi} as
\begin{align*}
	I_{\mcP}(\RZ_V) &= I(\RZ_{C_1}\wedge \RZ_{C_2} \wedge \dots \wedge \RZ_{C_k}).
\end{align*}
Through \eqref{eq:D:2}, we can verify that the MMI measure defined as above is consistent with the measure of shared bits used in Example~\ref{eg:IMm} for the special source model in \figref{fig:eg:Z}.

\begin{figure*}
	\centering
	\begin{subfigure}[b]{.45\textwidth}
		\centering
		\includegraphics[width=2in]{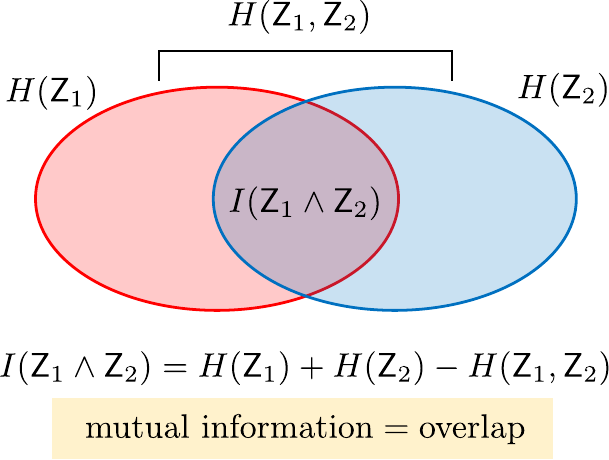}
		\caption{Venn diagram for Shannon's mutual information. (See \eqref{eq:venn}.)}\label{fig:venn}
	\end{subfigure}
	\begin{subfigure}[b]{.45\textwidth}
		\centering
		\includegraphics[width=2.8in]{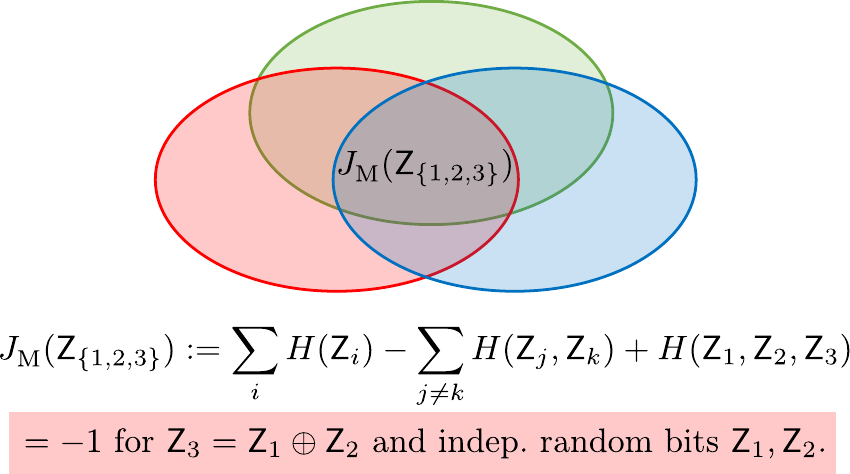}
		\caption{Information diagram for McGill's multiple information.}
		\label{fig:Idiagram}
	\end{subfigure}
	\caption{Interpretation of mutual information as the amount of overlap in randomness.}
	\label{fig:overlap}
\end{figure*}

\begin{figure*}
	\centering
	\begin{subfigure}[b]{.6\textwidth}
		\centering
		\kern2em \includegraphics[width=3.6in]{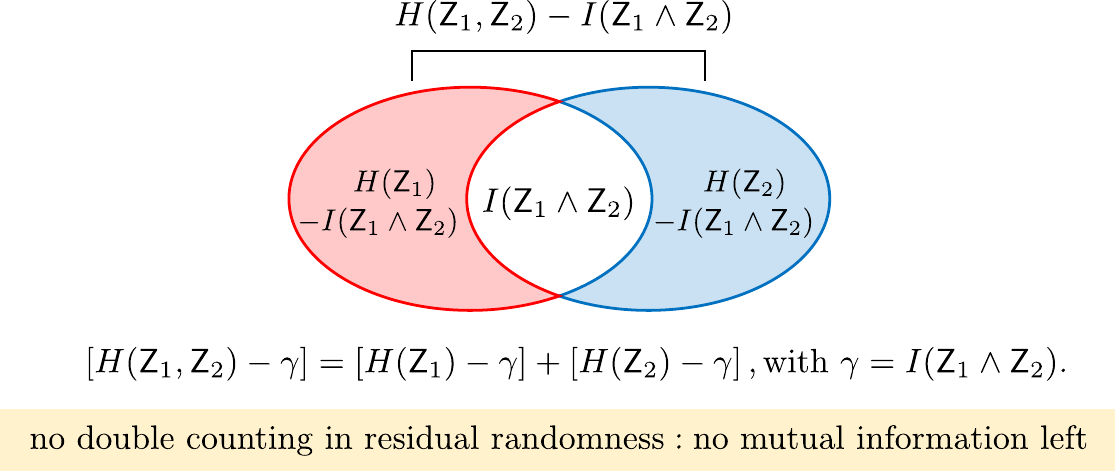}\kern2em
		\caption{Residual independence relation for Shannon's mutual information. (See \eqref{eq:RIR2}.)}
		\label{fig:RIR2}
	\end{subfigure}
	\begin{subfigure}[b]{.35\textwidth}
		\centering
		\kern2em \includegraphics[width=2in]{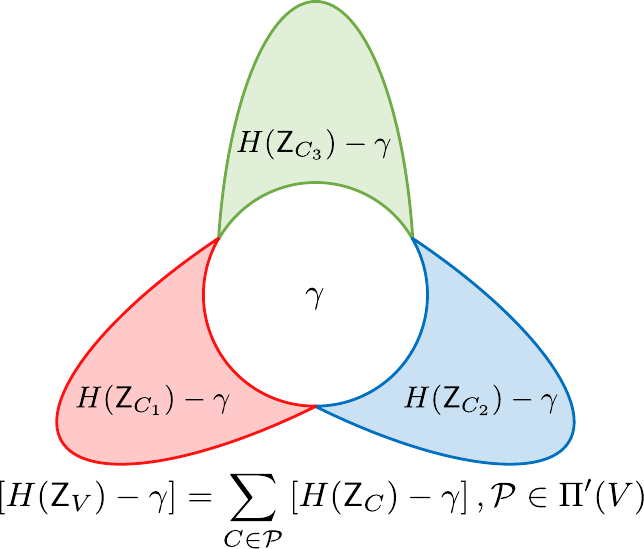} \kern2em
		\caption{Residual independence relation for the multivariate mutual information. (See \eqref{eq:RIR}.)}
		\label{fig:RIR}
	\end{subfigure}
	\caption{MMI as the smallest value satisfying the residual independence relation. 
		}
	\label{fig:RIRs}
\end{figure*}

\begin{Example}
	\label{eg:II}
	Consider the example in \figref{fig:eg:Z}. The values of $I_{\mcP}(\RZ_{\Set{1,2,3}})$ for different partitions $\mcP$ are
	\begin{align*}
			I(\RZ_1\wedge \RZ_{\Set{2,3}})&=2\\
			I(\RZ_2\wedge \RZ_{\Set{1,3}})&=2\\
			I(\RZ_3\wedge \RZ_{\Set{1,2}})&=1\\
			I(\RZ_1\wedge \RZ_2 \wedge \RZ_3) &= \tfrac{I(\RZ_1\wedge \RZ_2)+I(\RZ_{\Set{1,2}}\wedge \RZ_3)}{2}=\frac32,
	\end{align*}
	where the last term is obtained by applying \eqref{eq:D:2} 
	with $C_i=\Set{i}$. Hence, $I(\RZ_{\Set{1,2,3}})=1$ with $\Set{\Set{1,2},\Set{3}}$ being the unique optimal partition.
\end{Example}

At the first sight, the normalization factor $\abs{\mcP}-1$ on the R.H.S.\ of \eqref{eq:IP} may appear arbitrary. This factor is not included in other proposed information measures involving the divergence, such as the total correlation~\eqref{eq:JT} and the integrated information~\cite{balduzzi2008integrated}. However, it turns out that such a factor has an important information-theoretic meaning, which relates it to the non-trivial, but polynomial-time solvable, mathematical structure of the PSP. Indeed, the normalization factor is often overlooked in other proposed multivariate information measures based on the independence relations, such as the total correlation in \eqref{eq:JT}. This is because the factor only affects the measure when the independence relations do not hold, i.e., when the measure is non-zero. 

To help understand the reasoning behind such a normalization, we will introduce the residual independence relation~\cite{chan15mi} by
extending the well-known Venn diagram interpretation of Shannon's mutual information shown in \figref{fig:venn} according to the identity
\begin{align}
	I(\RZ_1\wedge \RZ_2) = H(\RZ_1)+H(\RZ_2)-H(\RZ_1,\RZ_2).\label{eq:venn}
\end{align} 

From the Venn diagram, the mutual information has the meaningful interpretation as the amount of overlap in the randomness of the individual random variables. This interpretation has been extended by \cite{yeung91} to the Information diagram, and the amount of overlap can be measured by the McGill's multiple information~\cite{mcgill54} using the inclusion-exclusion principle. (See \figref{fig:Idiagram}.) Unfortunately, the McGill's multiple information can be negative even for a very simple example involving three random variables~\cite{yeung91}, contradicting the basic intuition that mutual information should be non-negative. 

To ``fix" this problem, one may rewrite \eqref{eq:venn} equivalently as
\begin{multline}
	H(\RZ_1,\RZ_2)-I(\RZ_1\wedge \RZ_2) =\\ {\color{black}`1[H(\RZ_1)-I(\RZ_1\wedge \RZ_2)`2]} + {\color{black}`1[H(\RZ_2)-I(\RZ_1\wedge \RZ_2)`2]}.\label{eq:RIR2}
\end{multline}
Note that the L.H.S.\ is the total \emph{residual} randomness after removing the mutual information, and the equality states that the total residual randomness is equal to the sum of the individual residual randomness in each random variable, as illustrated in \figref{fig:RIR2}. The important interpretation of the equality is that:
\begin{bbox}
\noindent\emph{No double counting in the sum means precisely that there is no mutual information left in the residual randomness.}
\end{bbox}

The above idea can be extended to the multivariate case as follows. Consider a partition $\mcP\in \Pi'(V)$ and define the residual independence relation (RIR) as
\begin{align}
	H(\RZ_V)-`g = \sum_{C\in \mcP} `1[H(\RZ_C)-`g`2],\label{eq:RIR}
\end{align}
i.e., the total residual randomness after removing some real value $`g\in `R$ is equal to the sum of the individual residual randomness of the agglomerated random variables $\RZ_C$'s. This is illustrated in \figref{fig:RIR}. 

\begin{ybox}
We can now interpret the MMI~\eqref{eq:mi} as the \emph{smallest} $`g\in `R$ such that the RIR \eqref{eq:RIR} holds for some partition $\mcP\in \Pi'(V)$. To see this, a simple re-arrangement of the terms in \eqref{eq:RIR} gives
\begin{align*}
	`g&=\frac{1}{\abs{\mcP}-1}`1[\sum_{C\in \mcP} H(\RZ_C)-H(\RZ_V)`2]
	=I_{\mcP}(\RZ_V)
\end{align*}
where the last equality follows from \eqref{eq:D:1}. Minimizing over all possible $\mcP\in \Pi'(V)$ gives rise to the MMI, and the normalization factor $\abs{\mcP}-1$ appears naturally under the context of RIR due to the fact that any information mutual to the set of random variables should appear in every agglomerated random variable, and is therefore over-counted in the divergence. Note also that the residual randomness can be shown to be always non-negative as expected when the partition is the minimizing partition, and when the random variables are discrete.\footnotemark
\end{ybox}
\footnotetext{
\cite[Theorem~6.3]{chan15mi} also provides an axiomatic characterization of the MMI measure, but the RIR interpretation appears more intuitive; it seems to be more information-theoretically appealing than the mathematical form \eqref{eq:mi} as it explains the normalization factor not considered by previous multivariate information measures; and it serves as a fix to the Venn diagram extension of Shannon's mutual information. Furthermore, similar to the way Shannon's mutual information provides a theoretical limit to the communication rate~\cite{shannon48}, the MMI has various operational meanings~\cite{chan15mi}. For instance, it specifies the limit of the amount of common secret key (mutual information) extractable individually from different but correlated random sources.}

\subsection{Fundamental partition}

Back to the clustering problem, how do we compute the clusters efficiently using the MMI? Surprisingly, the optimal partition achieving the MMI~\eqref{eq:IS} gives us the desired clustering solution. 
The optimal partition achieving the MMI may not be unique, but it can be shown that the set of optimal partitions form a semi-lattice, which together with the partition $\Set {V}$ is referred to as the \emph{Dilworth truncation lattice}~\cite{narayanan90}. The lattice structure means that there exists a unique finest partition, which is called the \emph{fundamental partition}~\cite{chan15mi}. 
It turns out that the non-singleton elements of the fundamental partitions give us the desired clusters. 

\begin{figure}	
	\centering
	\includegraphics[width=3in]{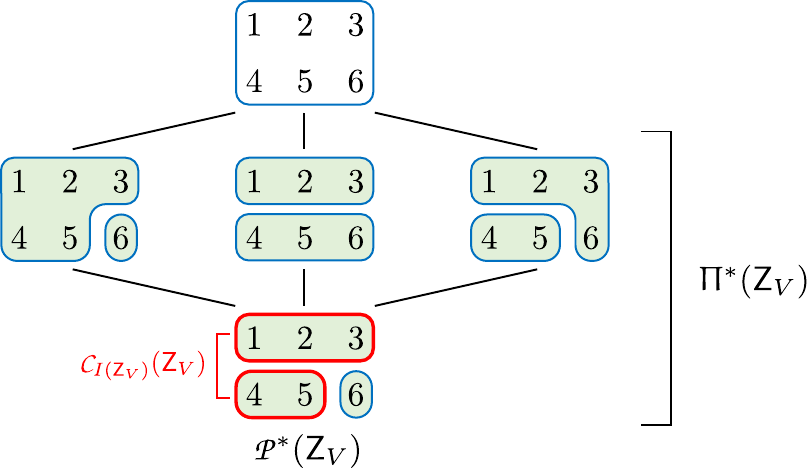}
	\caption[The Dilworth truncation lattice.]{Each partition in green is an optimal partition of the MMI~\eqref{eq:Pi*} (corresponding to an independence relation) for the source in \figref{fig:eg:Z}. The non-singleton elements of the fundamental partition give some of the clusters shown in \figref{fig:eg:cluster}, namely, the first set of clusters at the threshold $\upgamma_1$. (See Theorem~\ref{thm:CP}.)}\label{fig:DTL}
\end{figure}
\begin{bbox}
The proof is rather involved, but the idea can be illustrated using the previous example in \figref{fig:eg:Z}. As explained in Example~\ref{eg:I}, $I(\RZ_V)=0$ because of the independence relation $P_{\RZ_{V}} = P_{\RZ_{\Set {1,2,3}}} P_{\RZ_{\Set {4,5}}} P_{\RZ_6}$.
There are also other independence relations, which correspond to further merging of the agglomerated random variables as shown in \figref{fig:DTL}. Hence, the fundamental (finest optimal) partition is
\[ \pzP^*(\RZ_{V})=\Set{\underbrace{\Set{1,2,3},\Set{4,5}}_{\pzC_{I(\RZ_{V})}(\RZ_V)},\Set{6}}. \]
Note that, the non-singleton subsets in the fundamental partition, namely $\Set{1,2,3}$ and $\Set{4,5}$, are the clusters in $\pzC_{I(\RZ_V)}(\RZ_V)=\pzC_{0}(\RZ_V)$ in \figref{fig:eg:cluster}. 
If we take the cluster $\Set{1,2,3}$ and further compute its MMI, we have (from Example~\ref{eg:II}) $I(\RZ_{\Set{1,2,3}})=1$ with the unique optimal partition
\[ \pzP^*(\RZ_{\Set{1,2,3}})=\Set{\kern-2em\underbrace{\Set{1,2}}_{C_{I(\RZ_{\Set{1,2,3}})}(\RZ_{\Set{1,2,3}})}\kern-2em,\Set{3}}.\]
Note that the non-singleton subset $\Set {1,2}$ in the fundamental partition of the cluster $\Set {1,2,3}$ is a cluster of the entire set of random variables at the larger threshold of $I(\RZ_{\Set {1,2,3}})$.
In other words, clusters at larger thresholds can be obtained from the fundamental partition of some clusters at smaller thresholds. 
\end{bbox}


Below, we will show more generally that computing the fundamental partitions iteratively  gives the complete info-clustering solution.
Recall from Theorem~\ref{thm:cluster} that the first critical value $\upgamma_1(\RZ_V)$ is always given by the MMI $I(\RZ_V)$ of the entire set. To determine the corresponding first set $\pzC_{\upgamma_1}(\RZ_V)$ of clusters, denote the set of optimal partitions to \eqref{eq:IS} by
\begin{align}
	\Uppi^*(\RZ_V):=\Set{\mcP\in \Pi'(\RZ_V): I_{\mcP}(\RZ_V) = I(\RZ_V)}.\label{eq:Pi*}
\end{align}
In general, $\Uppi^*(\RZ_V)$ may contain more than one optimal partition, but the optimal partitions are related by the following partial order defined for any partitions $\mcP$ and $\mcP'$ as
\begin{align}
	\mcP \preceq \mcP' \kern1em \text{ iff }\kern1em \forall C\in \mcP,\; \exists C'\in \mcP': C\subseteq C'.\label{eq:<P}
\end{align}
We say that $\mcP$ is finer/smaller than $\mcP'$ whenever $\mcP \preceq \mcP'$, and we use $\prec$ to denote the strict inequality. The following result is known from \cite{chan15mi}.

\begin{Proposition}[\mbox{\cite[Theorems~5.2 and 5.3]{chan15mi}}]
	\label{prop:fundamental}
	There is a unique finest optimal partition, which we denote by $\pzP^*(\RZ_V)$ and refer to as the fundamental partition for $\RZ_V$. (Namely, $\pzP^*(\RZ_V)=\min \Uppi^*(\RZ_V)$, where the minimum is with respect to $\preceq$ in~\eqref{eq:<P}.) Furthermore, the fundamental partition $\pzP^*(\RZ_V)$ with the singletons removed, i.e., $\pzP^*(\RZ_V)`/\Set{\Set{i}\mid i\in V}$, is the set of maximal subsets $B\subseteq V$ with  $I(\RZ_B)>I(\RZ_V)$.
\end{Proposition}

Since we have $\upgamma_1(\RZ_V)=I(\RZ_V)$ by \eqref{eq:upgamma1}, and the non-singleton elements from $\pzP^*(\RZ_V)$ are, by Proposition~\ref{prop:fundamental}, the maximal subsets of random variables with mutual information strictly larger than $\upgamma_1(\RZ_V)$, we have by~\eqref{eq:cluster} that:

\begin{Theorem}
	\label{thm:CP}
	$\pzC_{\upgamma_1}(\RZ_V)=\pzP^*(\RZ_V)`/\Set{\Set{i}\mid i\in V}$ with the first critical value $\upgamma_1(\RZ_V)=I(\RZ_V)$.
\end{Theorem}

By Proposition~\ref{prop:fundamental}, the MMI measure defined in \eqref{eq:IS} is guaranteed to satisfy the key property \eqref{eq:I:lb} by \cite[Corollary~5.1]{chan15mi}. This can be argued as follows. Suppose to the contrary of \eqref{eq:I:lb} that both $I(\RZ_{B_1})$ and $I(\RZ_{B_2})$ are strictly larger than $I(\RZ_{B_1\cup B_2})$. Then, the non-singleton elements in the fundamental partition $\pzP^*(\RZ_{B_1\cup B_2})$ must consist of a superset of $B_1$ as well as a superset of $B_2$. However, the two supersets cannot be disjoint as $B_1\cap B_2\neq `0$, and they cannot be $B_1\cup B_2$ either, which is a contradiction.

The following result thus follows immediately from Theorems~\ref{thm:Laminarity} and \ref{thm:iterate2}.

\begin{Theorem}
	Under the MMI measure \eqref{eq:IS}, the clustering solution \eqref{eq:cluster} is guaranteed to be hierarchical. Furthermore, all clusters can be obtained by computing the fundamental partition iteratively for every previously obtained cluster.
\end{Theorem}

Given an algorithm that computes the fundamental partition exactly or approximately, we can compute the entire info-clustering solution following the iterative procedure in Algorithm~\ref{algo:iteration}. This is stated more precisely in Algorithm~\ref{algo:FP}. The complexity is again $\abs{V}$ times the complexity in calculating the fundamental partition by \eqref{eq:num_clusters}.\footnote{Indeed, we will see in the next section how the fundamental partition can be computed from the PSP of the entropy function. In fact, the additional factor of $\abs{V}$ can be saved by using the PSP rather than the iterative algorithm to compute the clusters. Nevertheless, the iterative algorithm is useful as it potentially allows us to compute the entire clustering solution approximately based on an approximate algorithm of computing the fundamental partition.}

\begin{algorithm}
	\caption{Clustering by fundamental partition.}
	\label{algo:FP}
	\SetKwFunction{FundamentalPartition}{FundamentalPartition}
	\SetKwProg{myfn}{function}{:}{end}
	\DontPrintSemicolon
	\SetAlgoLined
	\KwData{Statistics of $\RZ_V$ sufficient for calculating \FundamentalPartition{$B$} for all $B\subseteq V:\abs{B}>1$.}
	\KwResult{$\mcS$ is a list of $(I(\RZ_C),C)$ for $C\in \pzC(\RZ_V)$, which gives
		$\Upgamma(\RZ_V)=\Set{`g' \mid (`g',B)\in \mcS}$ and
		$\pzC_{`g}(\RZ_V)=\op{maximal}\Set{B\mid (`g',B)\in \mcS,`g'>`g}$.}
	\BlankLine
	$\mcS,\mcT\leftarrow$ empty queues;\;
	enqueue $V$ to $\mcT$;\;
	\While{$\mcT$ is non-empty}{
		$B\leftarrow$ dequeue $\mcT$;\;
		$(`g,\mcP)\leftarrow \FundamentalPartition(B)$;\;
		enqueue $(`g,B)$ to $\mcS$;\;
		enqueue all non-singleton elements of $\mcP$ to $\mcT$;\;
	}
 	\BlankLine
	\myfn{\FundamentalPartition{$B$}}{
		\KwRet $(I(\RZ_B),\pzP^*(\RZ_B))$;
	}
\end{algorithm}

Similar to Algorithm~\ref{algo:iteration}, the algorithm computes the list $\mcS$ of all clusters $B\in \pzC(\RZ_V)$ and their associated values $I(\RZ_B)$. It calls the function \FundamentalPartition{$B$} iteratively to obtain the first critical value $`g$ and the fundamental partition $\mcP$ of every previously discovered cluster stored temporarily in $\mcT$. The non-singleton elements of $\mcP$ are the desired clusters further added to $\mcT$.  

\subsection{Principal sequence of partitions of entropy function}
\label{sec:PSP}
As pointed out in \cite{chan11isit,milosavljevic11,chan15mi}, the MMI and the fundamental partition can both be computed in polynomial time assuming the entropies of arbitrary subsets of the random variables in hand are also computable in polynomial time. This result is based on the property that the (conditional) mutual information is non-negative, or equivalently, the entropy is submodular~\cite{fujishige78}. Hence, the iterative algorithm in the previous section can discover the info-clustering solution in polynomial time.

Quite surprisingly, based on the RIR~\eqref{eq:RIR} interpretation of the MMI in \S\ref{sec:MMI}, we find that the info-clustering solution can be mapped to the polynomial-time solvable mathematical structure of the principal sequence of partitions (PSP). The implication is that one can compute the general info-clustering solution more efficiently than the iterative algorithm, using techniques such as \cite{narayanan90,nagano10}. This understanding will also allow us to compare the info-clustering solution to the closely related approach of MAC clustering~\cite{nagano10}. The study of PSP from an information-theoretic perspective appears to be new, and we are beginning to discover more information-theoretic interpretations in other problems~\cite{chan16so,chan16itw,chan16isit}. 

\begin{ybox}
Define for $`g\in `R$ the residual entropy function~\cite{chan15mi}:
\begin{align}
	h_{`g}(B)&:= h(B)-`g \kern2em \text{for $B\subseteq V$,} \label{eq:residualH}
\end{align}
where $h(B) := H(\RZ_B)$ is the usual entropy function~\cite{yeung08}. $h_{`g}(B)$ measures the  residual randomness of $\RZ_B$ introduced in \S\ref{sec:MMI}. For notational simplicity, the dependency on $\RZ_B$ is implicit here. 
\end{ybox}

The entropy function is well-known\footnote{The submodularity follows directly from the non-negativity of the conditional mutual information $I(\RZ_{B_1}\wedge \RZ_{B_2}|\RZ_{B_1\cap B_2})\geq 0$~\cite{yeung08}.} to be submodular~\cite{fujishige78}, i.e., for all $B_1,B_2\subseteq V$,
\begin{align}
	h(B_1)+h(B_2)\geq h(B_1\cap B_2) + h(B_1\cup B_2),\label{eq:submodular}
\end{align}
and so, it is clear that the residual entropy function $h_{`g}$~\eqref{eq:residualH} is also submodular.
The Dilworth truncation of the submodular residual entropy function is defined as
\begin{subequations}
	\label{eq:DT}
	\begin{align}
		\hat{h}_{`g}(B)&:= \min_{\mcP\in \Pi(B)} h_{`g}[\mcP] \kern1em \text{for $B\subseteq V$, where}\label{eq:DT:1}\\
		h_{`g}[\mcP] &:= \sum_{C\in \mcP} h_{`g}(C)\label{eq:DT:2}
	\end{align}
	and $\Pi(B)$ is the collection of all partitions of $B$ into non-empty subsets. Note that the difference between $\Pi(B)$ and $\Pi'(B)$ is that $\Pi(B)$ includes the trivial partition $\{B\}$ as well, i.e., $\Pi'(B)=\Pi(B)\setminus\{\{B\}\}$.
\end{subequations}
The Dilworth truncation is itself a submodular set function, and can be calculated efficiently in strongly polynomial time~\cite{dunstan76} for any given set using Edmonds' greedy algorithm and the \emph{submodular function minimization (SFM)}. The running time is $O(\abs{V} \op{SFM}(\abs{V}))$, where $\op{SFM}(\abs{V})$ is the running time of the submodular function minimization over the ground set $V$. (See \cite{schrijver02} and \cite{chan15mi}.) $\op{SFM}(\abs{V})$ can be strongly polynomial assuming that the entropy function can be evaluated efficiently for every given subset of random variables.

\begin{ybox}
To characterize the info-clustering solution, we will focus on the Dilworth truncation evaluated at $V$:
\begin{align}
	\hat{h}_{`g}(V) = \min_{\mcP\in \Pi(V)} h_{`g}[\mcP]\label{eq:g2}
\end{align}
and think of it as a function of $`g$. More precisely,
it is a minimization of the function
\begin{align*}
	h_{`g}[\mcP]=\sum_{C\in \mcP} h_{`g}(C) = \sum_{C\in \mcP} H(\RZ_C) - `g\abs{\mcP},
\end{align*}
which is linear in $`g$ with
\begin{align*}
	\text{slope} &= -\abs{\mcP},&
	\text{y-intercept} &= \sum\nolimits_{C\in \mcP} H(\RZ_C).
\end{align*}
Since $\hat{h}_{`g}(V)$ is a minimization over a finite collection of linear curves, it must be
piecewise linear. More explicitly, for a given $\gamma$, let $\Pi^*$ be the set of partitions
attaining the minimization in (\ref{eq:g2}), then at $\gamma^{+}$ the Dilworth truncation is given as
the curve with the minimum slope among $\Pi^*$.
Thus, $\hat{h}_{`g}(V)$ is piecewise linear in $`g$ with slopes decreasing from $-1$ to $-\abs{V}$ and
taking only integer values, as shown in \figref{fig:P}. Since there is a finite number of partitions of $V$, the curve can be characterized by the set of
\emph{turning points} $p_i$'s where the slope changes. 
\begin{subequations}
	Denote the turning points as
	\begin{align}
		p_1=({`g}_1,y_1),\dots,p_N=({`g}_N,y_N) \label{eq:DT:turning_points}
	\end{align}
	for some positive integer $N$, and call
	\begin{align}
		{`g}_1< {`g}_2< \dots< {`g}_N \label{eq:DT:critical_values}
	\end{align}
\end{subequations}
the \emph{critical values for the Dilworth truncation} $\hat{h}_{`g}(V)$. 
\end{ybox}

\begin{Example}
	\label{eg:DT}
	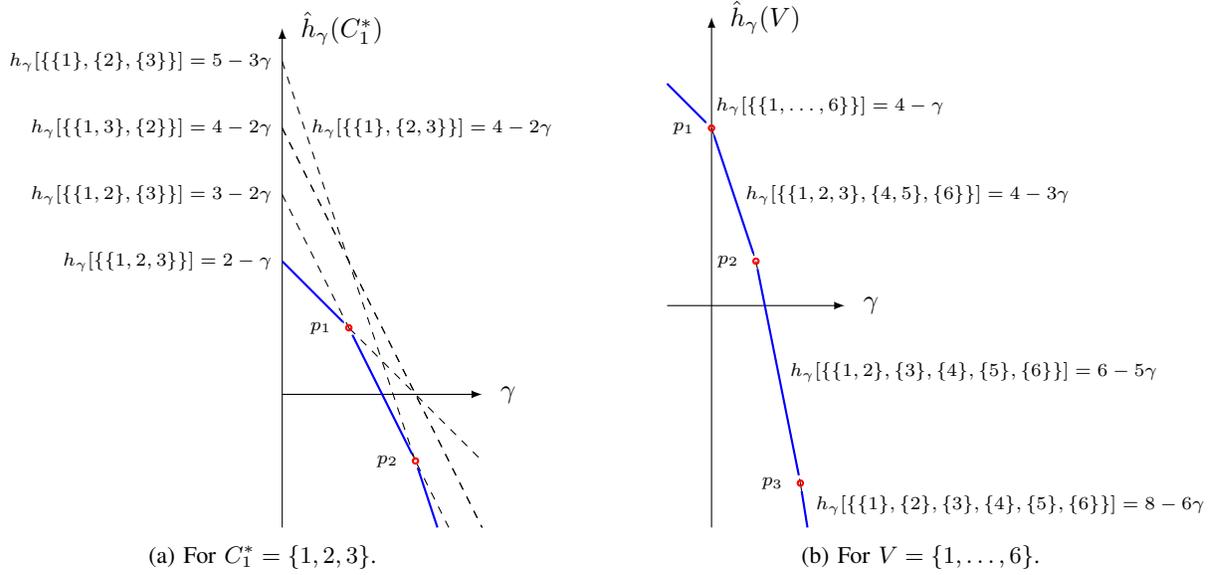
\begin{figure*}
		\centering
		\subcaptionbox{For $C^*_1=\Set{1,2,3}$.\label{fig:P:1}}{
			{\def\u{1.8}
				\tikzstyle{point}=[draw,circle,minimum size=.2em,inner sep=0, outer sep=.2em]
				\begin{tikzpicture}[x=1.4em,y=1.4em,>=latex]
				\draw[->] (0,-2*\u) -- (0,5.5*\u) node [label=right:$\hat{h}_{`g}(C^*_1)$] {};
				\draw[->] (0,0) -- (3*\u,0) node [label=right:$`g$] {};
				\foreach \i/\ya/\xa/\yb/\xb/\lp/\lb in {
					1/2/0/-1/3/left/{$h_{`g}[\Set{\Set{1,2,3}}]=2-`g$}, 
					2/3/0/-2/2.5/left/{$h_{`g}[\Set{\Set{1,2},\Set{3}}]=3-2`g$}, 
					3/4/0/-2/3/left/{$h_{`g}[\Set{\Set{1,3},\Set{2}}]=4-2`g$}, 
					5/5/0/-2/2.33/left/{$h_{`g}[\Set{\Set{1},\Set{2},\Set{3}}]=5-3`g$}, 
					4/4/0/-2/3/right/{$\kern1em h_{`g}[\Set{\Set{1},\Set{2,3}}]=4-2`g$}}
				\draw[dashed] (\xa*\u,\ya*\u)  node [inner sep=0,outer sep=0,label={[label distance=0em]\lp:{\scriptsize\lb}}] {} -- (\xb*\u,\yb*\u);
				\path (1*\u,1*\u) node (1) [point,red,thick,label=left:{\scriptsize$p_1$}] {};
				\path (2*\u,-1*\u) node (2) [point,red,thick,label=left:{\scriptsize$p_2$}] {};
				\draw[-,thick,blue] (0,2*\u)--(1)--(2)--(2.33*\u,-2*\u);
				\end{tikzpicture}}
		}
		\hfil
		\subcaptionbox{For $V=\Set{1,\dots,6}$.\label{fig:P:2}}{
			{\def\u{1.2}
				\tikzstyle{point}=[draw,circle,minimum size=.2em,inner sep=0, outer sep=.2em]
				\begin{tikzpicture}[x=1.4em,y=1.4em,>=latex]
				\draw[->] (0,-5*\u) -- (0,6.5*\u) node [label=right:$\hat{h}_{`g}(V)$] {};
				\draw[->] (-1*\u,0) -- (3*\u,0) node [label=right:$`g$] {};
				\foreach \i/\ya/\xa/\yb/\xb/\lp/\lb in {
					1/5/-1/4/0/right/{$\kern.8em h_{`g}[\Set{\Set{1,\dots,6}}]=4-`g$}, 
					2/4/0/1/1/right/{$h_{`g}[\Set{\Set{1,2,3},\Set{4,5},\Set{6}}]=4-3`g$},
					3/1/1/-4/2/right/{$h_{`g}[\Set{\Set{1,2},\Set{3},\Set{4},\Set{5},\Set{6}}]=6-5`g$},
					4/-4/2/-5/2.16/right/{$h_{`g}[\Set{\Set{1},\Set{2},\Set{3},\Set{4},\Set{5},\Set{6}}]=8-6`g$}    
				}
				\draw[dashed] (\xa*\u,\ya*\u)  to node [inner sep=0,outer sep=0,label={[label distance=0.1em]\lp:{\scriptsize\lb}}] {}  (\xb*\u,\yb*\u);
				\path (0*\u,4*\u) node (1) [point,red,thick,label=left:{\scriptsize$p_1$}] {};
				\path (1*\u,1*\u) node (2) [point,red,thick,label={[label distance=0em]left:{\scriptsize$p_2$}}] {};
				\path (2*\u,-4*\u) node (3) [point,red,thick,label=left:{\scriptsize$p_3$}] {};
				\draw[-,thick,blue] (-1*\u,5*\u)--(1)--(2)--(3)--(2.16*\u,-5*\u);
				\end{tikzpicture}}
		}
		\caption{Plots of Dilworth truncation for Example~\ref{eg:DT} as the minimum of the lines $h_{`g}[\mcP]$ over different partitions.}
		\label{fig:P}
	\end{figure*}
	Consider the example in \figref{fig:eg:Z}. To compute $\hat{h}_{`g}(C^*_1)$ for $C^*_1:=\Set{1,2,3}$, note that the values of $h_{`g}[\mcP]$ for different partitions $\mcP$ are
	\begin{align*} 
		\begin{cases}
			\underbrace{h(\Set{1,2,3})}_{2}-`g
			& \mcP=\Set{\Set{1,2,3}}\\
			\underbrace{h(\Set{1})+h(\Set{2,3})}_4-2`g
			& \mcP=\Set{\Set{1},\Set{2,3}}\\
			\underbrace{h(\Set{2})+h(\Set{1,3})}_4-2`g
			& \mcP=\Set{\Set{2},\Set{1,3}}\\
			\underbrace{h(\Set{3})+h(\Set{1,2})}_3-2`g
			& \mcP=\Set{\Set{3},\Set{1,2}}\\
			\underbrace{h(\Set{1})+h(\Set{2})+h(\Set{3}}_5)-3`g
			& \mcP=\Set{\Set{1},\Set{2},\Set{3}}.
		\end{cases}
	\end{align*}
	The minimum, $\hat{h}_{`g}(C^*_1)$, of the above lines is plotted in \figref{fig:P:1}. For $`g< 1=I(\RZ_{C^*_1})$, the minimum is achieved uniquely by $\mcP=\Set{C^*_1}=\Set{\Set{1,2,3}}$. For $`g\in (1,2)$, the minimum is achieved uniquely by $\mcP=\pzP^*(\RZ_{C^*_1})=\Set{\Set{3},\Set{1,2}}$. 
	For $`g> 2$, the minimum is achieved uniquely by the partition into singletons.
	
	Similarly, $\hat{h}_{`g}(V)$ can be plotted as the minimum of a set of lines in \figref{fig:P:2}. For $`g<0=I(\RZ_V)$, the minimum is achieved uniquely by $\mcP=\Set{V}=\Set{1,\dots,6}$. 
	For $`g\in (0,1)$, the minimum is achieved uniquely by $\mcP=\pzP^*(\RZ_{V})=\Set{\Set{1,2,3},\Set{4,5},\Set{6}}$. 
	For $`g\in (1,2)$, the minimum is achieved uniquely by $\mcP=\Set{\Set{1,2},\Set{3},\Set{4},\Set{5},\Set{6}}$. 
	For $`g> 2$, the minimum is achieved uniquely by the partition into singletons.
\end{Example}

The connection of the Dilworth truncation to the MMI is through the RIR, as shown in \cite[Theorem~5.1]{chan15mi}. When $`g$ is sufficiently small, $\hat{h}_{`g}(V)= h_{`g}(V)$ because
$h_{`g}[\mcP]$ has the largest slope of $-1$ when
$\mcP=\Set{V}$. More precisely,
\[
h_{`g}[\mcP]-h_{`g}[\Set{V}] =  D`1(\extendvert{P_{\RZ_V}\|\prod\nolimits_{C\in \mcP }P_{\RZ_C}}`2) + (1-\abs{\mcP})`g,
\]  
which will be positive, i.e., $h_{`g}[\mcP]>h_{`g}[\Set{V}]=h_{`g}(V)$, for $\abs{\mcP}>1$ (or $\mcP\neq \Set{V}$) and $`g$ is sufficiently small. Therefore, it follows from \eqref{eq:g2} that $p_1$ is the intersection
between $h_{`g}(V)$ and $\min_{\mcP\in \Pi'(V)}
h_{`g}[\mcP]$, and so $`g={`g}_1$ satisfies the equation
	\begin{align}
		h_{`g}(V) = \min_{\mcP\in \Pi'(V)} h_{`g}[\mcP], \label{eq:g}
	\end{align}
which translates directly to the RIR in \eqref{eq:RIR}. Hence, we have ${`g}_1=I(\RZ_V)$. Furthermore, $p_1$ lies on
$h_{`g}[\mcP]$ if and only if $\mcP\in
\Uppi^*(\RZ_V)\cup\Set{\Set{V}}$, since $\Uppi^*(\RZ_V)$ defined in \eqref{eq:Pi*} is the
set of solutions to the minimization in \eqref{eq:g}, as can be seen from the RIR interpretation of the MMI. Since
the fundamental partition $\pzP^*(\RZ_V)$ is the unique finest partition in $ \Uppi^*(\RZ_V)$,
$h_{`g}[\pzP^*(\RZ_V)]$ has the smallest slope and therefore
uniquely defines the line segment following $p_1$. 

\begin{bbox}
Note that, it is not clear a priori that the critical values~\eqref{eq:DT:critical_values} defined for the Dilworth truncation $\hat{h}_{`g}(V)$ are precisely the critical values in $\Upgamma(\RZ_V)$~\eqref{eq:Upgamma} defined for info-clustering~\eqref{eq:cluster}, even though the above result from \cite{chan15mi} shows that it is the case for the first critical value. 
We will show the stronger result that \emph{not only the two sets of critical values match, but that the line segments of the Dilworth truncation give the desired info-clustering solution.} E.g., from \figref{fig:P:2}, the critical values of the Dilworth truncation can be verified to be precisely the critical values for the info-clustering solution in \figref{fig:eg:cluster}. Furthermore, the sequence of partitions defining the line segments in \figref{fig:P:2} contains all the clusters in \figref{fig:eg:cluster} as its non-singleton elements. This sequence of partitions is the PSP, which will be defined more precisely below.
\end{bbox}


Let $\Uppi_i\subseteq \Pi(V)$ be the set of solutions to the minimization in
$\hat{h}_{`g_i}(V)$. The elements of $\Uppi_i$ form a lattice:
\begin{Proposition}[\mbox{\cite[Theorem~3.5]{narayanan90}}]
	\label{pro:DTL}
	The set of optimal solutions to the Dilworth truncation 
	$\hat{f}(V)$ of a submodular function $f:2^V\mapsto `R$ forms a lattice
	(with respect to the partial order in \eqref{eq:<P}) called the
	\emph{Dilworth truncation lattice}. 
\end{Proposition}
For instance, when we specialize the submodular function to the entropy
function $h$ for $\RZ_V$, the first critical value is $`g_1=I(\RZ_V)$ and the associated lattice of partitions
is $\Uppi_1=\Uppi^*(\RZ_V)\cup \Set{\Set{V}}$.

Let $\min \Uppi_i$ and $\max \Uppi_i$ be, respectively, the (unique) minimum and maximum
partitions in the lattice $\Uppi_i$. 
The following proposition asserts that, for all $i$, the extreme partitions $\min \Uppi_i$ and $\max\Uppi_{i+1}$ are equal. (In particular, the fundamental partition is $\pzP^*(\RZ_V)=\min\Uppi_1=\max\Uppi_2$.) Furthermore, the extreme partitions for different values of $i$ form a sequence of successively finer partitions, referred to as the PSP. 
\begin{Proposition}[\mbox{\cite[Theorem~3.7]{narayanan90}}]
	\label{pro:PSP}
	There is a unique sequence of partitions with respect to the partial order~\eqref{eq:<P}
	\begin{subequations}
		\label{eq:PSP}
		\begin{align}
			\pzP_0 \succ \pzP_1 \dots \succ \pzP_N \in \Pi(V), \label{eq:PSP1}
		\end{align}
		called the \emph{principal sequence of
			partitions (PSP)}, which satisfies
		\begin{align}
			\pzP_{i-1} =\max \Uppi_i\kern1em\text{and}\kern1em
			\pzP_{i} =\min \Uppi_i
			\label{eq:PSP2}
		\end{align}
	\end{subequations}
	for $i\in \Set{1,\dots,N}$. More explicitly, $\pzP_0=\max
	\Uppi_1=\Set{V}$,  
	\begin{align*}
	\min \Uppi_i =\pzP_i=\max
	\Uppi_{i+1}\kern2em\text {for $i\in \Set{1,\dots,N-1}$,}
	\end{align*}
	and $\pzP_N=\min \Uppi_N =\Set{\Set{i}\mid i\in V}$.
\end{Proposition}
Same the argument in \cite{nagano10}, the PSP is computable in strongly polynomial time in $O(\abs{V}^2 \op{SFM}(\abs{V}))$.
For completeness, we include a simple proof below. 
\begin{Proof}
	For $i\in \Set{1,\dots,N-1}$, the line segment of $\hat{h}_{`g}(V)$ for $`g\in (`g_i,`g_{i+1})$ is
	defined by $h_{`g}[\pzP_i]$ for some partition $\pzP_i\in \Uppi_{i}\cap
	\Uppi_{i+1}$ because it passes through both turning points $p_{i}$ and
	$p_{i+1}$. Since it has the smallest and largest slopes among all
	other lines through $p_i$ and $p_{i+1}$ respectively, $\pzP_i$ is
	the minimum in $\Uppi_i$ and maximum in $\Uppi_{i+1}$ as
	desired. $\pzP_0=\Set{V}$ and $\pzP_N=\Set{\Set{i}\mid i\in V}$ because
	$h_{`g}[\mcP]$ has the largest and smallest slopes when $\mcP$ is $\pzP_0$
	and $\pzP_N$ respectively.
\end{Proof}

The desired connection between info-clustering and the PSP of the entropy function follows from the main result below, which 
gives an interpretation to every
critical value of the Dilworth truncation using the PSP. 
\begin{Theorem}
	\label{thm:xi}
	The $i$-th critical value of $\hat{h}_{`g}(V)$~\eqref{eq:g2} is
	\begin{subequations}
		\label{eq:xi}
		\begin{align}
			`g_i &= \min_{\substack{\mcP\in \Pi(V):  \abs{\mcP}>\abs{\pzP_{i-1}}}}
			\frac{h[\mcP]-h[\pzP_{i-1}]}{\abs{\mcP}-\abs{\pzP_{i-1}}}\label{eq:xi1}\\
			&= \min_{C\in \pzP_{i-1}:\abs{C}>1} I(\RZ_{C}). \label{eq:xi2}
		\end{align}
	\end{subequations}
	The set of optimal solutions to \eqref{eq:xi1} is
	$\Uppi_i`/\Set{\pzP_{i-1}}$. The set of optimal solutions to \eqref{eq:xi2}, denoted as $\pzC^*_{i-1}$, 
	is equal to $\pzP_{i-1}`/\pzP_i$, or equivalently,
	\begin{align}
		\pzP_i &= `1(\pzP_{i-1}`/\pzC^*_{i-1} `2)
		\cup \bigcup_{C\in \pzC^*_{i-1}} \pzP^*(\RZ_{C}). \label{eq:C`gi}
	\end{align}
	Furthermore, with the product of set families $\mcF$ and $\mcG$ defined as $\mcF\times \mcG:=\Set{\Set{F,G}\mid F\in \mcF,G\in \mcG}$, we have
	\begin{align}
		\Uppi_i = \prod_{C\in \pzP_{i-1}`/\pzP_i} `1[\Uppi^*(\RZ_C)\cup\Set{\Set{C}}`2] \times \prod_{C\in
			\pzP_{i-1}\cap \pzP_i} \kern-1.1em \Set{\Set{C}} \label{eq:Pii}
	\end{align}
	which consists of refinements of $\pzP_{i-1}$ by successively
	partitioning one or more blocks $C\in \pzP_{i-1}`/\pzP_i$ according
	to $\Uppi^*(\RZ_C)$.
\end{Theorem}
\begin{Proof}
	See Appendix~\ref{sec:PSP_proof}
\end{Proof}
When $i=1$, \eqref{eq:xi1} reduces to $`g_1=I(\RZ_V)$ because
$\pzP_{i-1}=\Set{V}$ and
$h[\mcP]-h[\pzP_{i-1}]=D(P_{\RZ_V}\|\prod_{C\in \mcP}
P_{\RZ_C})$. 
\eqref{eq:Pii} reduces to
$\Uppi_1=\Uppi^*(\RZ_V)\cup\Set{\Set{V}}$ with
$\pzP_{i-1}`/\pzP_i=\Set{V}$ and $\pzP_{i-1}\cap \pzP_i=`0$.

For $i\geq 1$, \eqref{eq:xi2}  means that the other critical values can
be obtained simply by iteratively computing the MMI for the
non-singleton blocks of the fundamental partitions. \eqref{eq:C`gi} is essentially the iteration in \eqref{eq:it2} to obtain the clusters iteratively. Therefore, the critical values for the Dilworth truncation coincide with the critical values for the set of clusters, and the clusters are the non-singleton elements of the partitions in the PSP. This is summarized in the following corollary. (See Definition~\ref{def:critical}) for some of the notations.)
\begin{Corollary}
	\label{cor:main}
	For $1\leq i\leq \pzN(\RZ_V)$, we have $\pzC_{\upgamma_i}(\RZ_V)=\pzP_i(\RZ_V)`/\Set{\Set{j}:j\in V}$ with the critical value $\upgamma_i(\RZ_V)$ being the $i$-th critical value for $\hat{h}_{`g}(V)$.
\end{Corollary}

Since the info-clustering solution maps to the entire PSP of the entropy function, we can compute the clustering solution in strongly polynomial time as well. The algorithm is given in Algorithm~\ref{algo:PSP}, which is based on the algorithm of \cite{nagano10}. 

	 \begin{algorithm}
	 	\caption{Clustering by PSP for entropy function.}
	 	\label{algo:PSP}
	 	\DontPrintSemicolon
	 	\SetAlgoLined
	 	\SetKwFunction{split}{Split}
	 	\SetKwFunction{SFM}{SubmodularFnMin}
	 	\SetKwProg{myproc}{procedure}{:}{end}
	 	\SetKwProg{myfn}{function}{:}{end}
	 	\BlankLine
   	\KwData{Statistics of $\RZ_V$ sufficient for calculating the entropy function $h(B)$ for $B\subseteq V:=\Set{1,\dots,m}$.}
   	\KwResult{The array \DataSty{L} contains the values in $\Upgamma(\RZ_V)$. The array \DataSty{PSP} contains the PSP $\pzP_i$'s. More precisely, $\pzP_i$ is stored in \DataSty{PSP}$[\abs{\pzP_i}]$, and $\upgamma_i$ is stored in \DataSty{L}$[\abs{\pzP_{i-1}}]$. Hence, $\pzC_{`g}(\RZ_V)$ is the set the non-singleton values of \DataSty{PSP}$[s]$ where $s$ is the smallest index with \DataSty{L}$[s]$$>`g$. $\pzC_{`g}(\RZ_V)=`0$ if no such $s$ exists.}
	 	\BlankLine
	 	\DataSty{L}, \DataSty{PSP}$\leftarrow$ empty arrays of size $m$;\;
	 	$\mcQ\leftarrow\Set {V}$, $\mcP\leftarrow \{\{i\}\mid i \in V\}$;\;
	 	\DataSty{PSP}$[\abs{\mcQ}]\leftarrow\mcQ$;\;
	 	\split{$\mcQ,\mcP$};\;
	 	\BlankLine
	 	\myproc{\split{$\mcQ, \mcP$}}{
	 		$\gamma'\leftarrow \frac{1}{\abs{\mcP}-\abs{\mcQ}}(h[\mcP]-h[\mcQ])$;\;
	 		$h'\leftarrow\frac{1}{\abs{\mcP}-\abs{\mcQ}}(\abs{\mcP}h[\mcQ]-\abs{\mcQ}h[\mcP])$;\;
	 		\BlankLine
	 		$\mcP'\leftarrow\emptyset$, \DataSty{x}$\leftarrow$ all-zero array of size $m$;\label{ln:PSP:DT:1}\;
	 		\For{$l = 1$ \emph{\KwTo} $m$ }{
	 			$(`a,T)\leftarrow$\SFM{$B\mapsto h_{\gamma'}(\RZ_B)-\sum_{i\in B}\text{\DataSty{x}}[i]$,$l$};\;
	 			add $\alpha$ to \DataSty{x}$[l]$;\;
	 			\ForEach{$C$ in $\mcP'$}{
	 				\If{$C\cap T\neq `0$}{
		 				$T\leftarrow T\cup C$;\;
		 				remove $C$ from $\mcP'$;\;
	 				}
	 			}
	 			add $T$ to $\mcP'$;\;
	 		}\label{ln:PSP:DT:2}
	 		\BlankLine
	 		\eIf{ $h'= \sum_{i=1}^m \text{\DataSty{x}}[i]$\label{ln:PSP:check} }{
	 			\DataSty{L}$[\abs{\mcQ}]\leftarrow `g'$;\;
	 		}
	 		{
	 			\DataSty{PSP}$[\abs{\mcP'}]\leftarrow\mcP'$;\;
	 			\split{$\mcQ,\mcP'$};\;
	 			\split{$\mcP',\mcP$};\;
	 		}
	 	}
	 \BlankLine
	 \myfn{\SFM{$f$,$l$}}{
	 	$U\leftarrow \Set {1,\dots,l}$;\;
	 	\KwRet $(\min_{B\subseteq U:l\in B} f(B),\arg\min_{B\subseteq U:l\in B} f(B) )$\;
	 }
	 \end{algorithm}

Algorithm~\ref{algo:PSP} computes the sequence of critical values and the PSP, and stores them in the arrays \DataSty{L} and \DataSty{PSP} respectively. The desired clusters can then be obtained from the non-singleton subsets in the PSP. As in \cite{nagano10}, the procedure \split starts with two partitions $\mcQ\succ\mcP$ in the PSP, and then check if there is any other partition $\mcP'$ in the PSP with $\mcQ\succ \mcP' \succ \mcP$. To do so, it first computes the intersection point $(`g',h')$ of the two lines $h_{`g}[\mcQ]$ and $h_{`g}[\mcP]$, and then check whether $h'$ is equal to the Dilworth truncation $\hat{h}_{`g'}(V)$ (which is computed by lines~\ref{ln:PSP:DT:1}--\ref{ln:PSP:DT:2} and stored in $\sum_{i=1}^m \text{\DataSty{x}}[i]$). If they are equal (line~\ref{ln:PSP:check}), then $\mcQ$ and $\mcP$ are two consecutive partitions with no other partition between them in the PSP, and so $`g'$ is a critical value. Otherwise, the optimal partition $\mcP'$ achieving the Dilworth truncation must be a partition in the PSP satisfying $\mcQ\succ \mcP'\succ \mcP$. In this case, the procedure \split can be invoked in a recursive manner to further identify other partitions in the PSP that may lie between $\mcQ$ and $\mcP'$, and between $\mcP'$ and $\mcP$.

The complexity of the algorithm is mainly due to the computations of the Dilworth truncation (lines~\ref{ln:PSP:DT:1}--\ref{ln:PSP:DT:2}) by the submodular function minimization \SFM~\cite{schrijver02}. The number of such computations is at most $\abs{V}-1$, and each has a complexity of $O(\abs {V} \op{SFM}(\abs V))$. Therefore, the overall complexity is $O(\abs {V}^2 \op{SFM}(\abs V))$.\footnote{The fundamental partition is obtained as a special case since it is a partition in the PSP. Although we do not know of a faster exact algorithm to compute the fundamental partition for the general source model, faster approximation algorithms may be possible, in which case the iterative procedure in Algorithm~\ref{algo:FP} can be used to approximate the entire clustering solution.}

Indeed, \cite{nagano10} also proposed the MAC clustering algorithm that builds upon the algorithm for finding the PSP for a submodular cost function. Although we have shown that info-clustering is also intimately connected to the PSP of the entropy function, the two clustering approaches are different in two ways:
\begin{ybox}
	\begin{enumerate}
		\item Unlike info-clustering where the MMI is specified as a measure of mutual information under a meaningful hierarchical clustering formulation, the formulation of the MAC clustering does not specify how one should choose the submodular cost function for clustering. Hence, the mathematical criterion of MAC does not have a concrete operational meaning, that is, it is unclear in what sense are the elements in the same cluster are similar.
		\item Unlike info-clustering where the solution maps precisely to the \emph{entire} PSP of the entropy function, the solution of the MAC clustering is sensitive to shifts of the cost function by a constant, and is therefore not identical to the PSP of the submodular cost function.
	\end{enumerate}
	In Appendix~\ref{sec:mac}, we give detailed explanations with concrete examples differentiating the two algorithms.
\end{ybox}




\section{Model reductions}
\label{sec:reduction}

In this section, we show that info-clustering reduces to simpler clustering solutions under some special models. Model reduction is important for practical implementations because learning the entropy function from data, and even evaluating the entropy of an arbitrary distribution can take exponential time with respect to the number of random variables. 

In the following, we show that the clustering algorithm by mutual information relevance networks (MIRN)~\cite{butte00} is a special case when $\RZ_V$ forms a Markov tree. We also show that, if $\RZ_V$ is jointly Gaussian, the clustering solution will depend only on the covariance matrix, which may be estimated more easily from data. Finally, if $\RZ_V$ has a hypergraphical correlation, then info-clustering reduces to the procedure of computing the PSP for hypergraphs, which is useful in clustering the human connectome.

\subsection{Clustering by Chow--Liu tree approximation}
\label{sec:CL}


\begin{figure*}
	\fboxsep=1.5mm
	\fboxrule=0pt
	\centering
	\subcaptionbox{\label{fig:CL1}$`g<0$}{
		\fcolorbox{white}{green!8}{
		\includegraphics[height=1.4in]{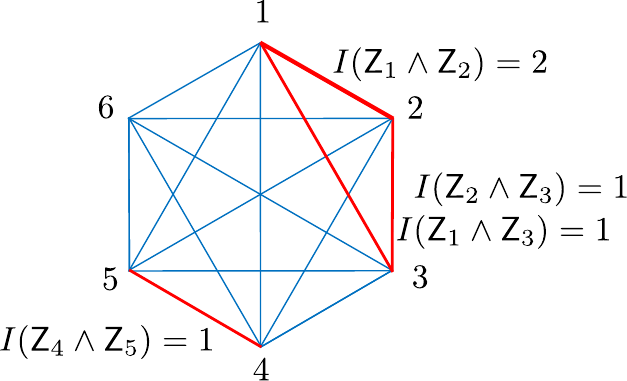}
		}
	} \hfill
	\subcaptionbox{\label{fig:CL2}$`g\in [0,1)$}{
		\fcolorbox{white}{green!8}{\includegraphics[height=1.4in]{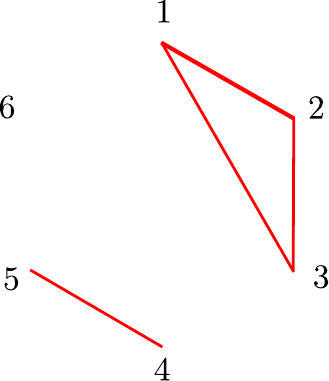}}
	} \hfill
	\subcaptionbox{\label{fig:CL3}$`g\in [1,2)$}{
		\fcolorbox{white}{green!8}{\includegraphics[height=1.4in]{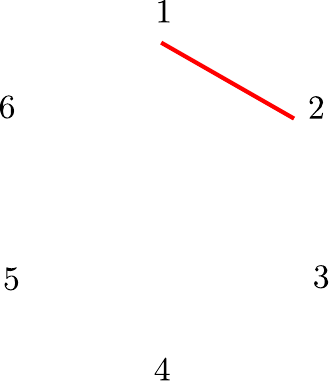}}
	} \hfill
	\subcaptionbox{\label{fig:CL4}$`g\geq 2$}{
		\fcolorbox{white}{green!8}{\includegraphics[height=1.4in]{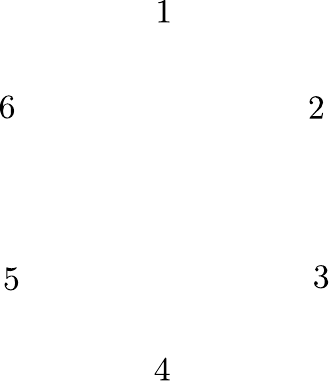}}
	}
	\caption{Clustering by MIRN~\eqref{eq:RN} for the source in \figref{fig:eg:Z}.}
	\label{fig:CL}
\end{figure*}
We first introduce the clustering by MIRN in~\cite{butte00} for gene clustering. This clustering algorithm first constructs a weighted complete graph, where the nodes represent the genes to be clustered. The weight of the edge between the nodes~$i$ and $j$ is equal to the Shannon's mutual information $I(\RZ_i\wedge \RZ_j)$, which may be estimated from measurements of the expression level $\RZ_i$ and $\RZ_j$ of the corresponding genes~$i$ and $j$ respectively. An example of such a graph is shown in \figref{fig:CL1} for the simple source model in \figref{fig:eg:Z}. In \figref{fig:CL1}, each blue edge has weight zero. Each red edge has weight one, except for the red edge between node~$1$ and $2$, which has weight two. 

Given a threshold $`g$, the algorithm filters the edges by removing all edges with weights no larger than $`g$. The clusters at threshold $`g$ are then defined as the non-singleton components of the resulting graphs. Such non-singleton components are called the MIRN. For instance, in the case of \figref{fig:CL}, the edge removal (or clustering) for different $`g$'s is as follows:
\begin{compactitem}
	\item When $`g<0$, we have the complete graph since the mutual information is non-negative. Consequently, we have the trivial cluster $V$. 
	\item For $`g\in [0,1)$, all the blue edges are removed since they have weight equal to $0$. Hence, we have the two clusters $\Set{1,2,3}$ and $\Set{4,5}$.
	\item For $`g\in [1,2)$, only the edge between nodes~$1$ and $2$ remains, and so, we have the cluster $\Set{1,2}$.
	\item For $`g\geq 2$, all edges are removed and so we have no clusters. 
\end{compactitem}

Note that the clusters we obtained from the above edge-filtering procedure are precisely the clusters we obtained by info-clustering in \figref{fig:eg:cluster}. We can show more generally that, info-clustering reduces to the clustering by MIRN when the random variables $\RZ_i$'s form a Markov tree. In this example, we indeed have a Markov chain structure, namely, \[\RZ_1"-"\RZ_2"-"\RZ_3"-"\RZ_4"-"\RZ_5"-"\RZ_6,\] which is a special case of the Markov tree. If the random variables do not form a Markov tree, then the MIRN solution turns out to correspond to applying info-clustering after approximating the correlation structure by a Markov tree. More precisely, for a set of random variables whose distribution does not necessarily factor according to a Markov tree, the clustering solution by MIRN corresponds to the solution resulting from applying info-clustering to any Markov tree obtained via the Chow--Liu tree approximation~\cite{chow68} of the distribution.\footnote{The preliminary result has been published in \cite{chan15allerton}.}
 
To explain the reduction above between info-clustering and clustering by MIRN, we first define the clustering by MIRN more formally using some graph-theoretic notations. For a simple graph $G$ with the vertex set $V$, we denote its edge set by $\mcE(G)\subseteq\Set{B\subseteq V:\abs{B}=2}$ (with the calligraphic font used for set families). For $i,j\in V$, we write $i\sim_G j$ to indicate that $j$ is reachable from $i$ via a path in $G$. Note that $\sim_G$ is an equivalence relation, and we denote the set of equivalence classes as:
\begin{align}
	\pzP(G):=\op{maximal}\Set{B\subseteq V\mid i\sim_G j, \forall i,j\in B}\in \Pi(V),\label{eq:PG}
\end{align}
where $\Pi(V)$ denotes the collection of all partitions of $V$ into non-empty disjoint sets. Each element in $\pzP(G)$ is the vertex set of a connected component of $G$, which will be considered as a cluster by MIRN as we describe below.

For any threshold $`g\in `R$, let $K_{`g}(\RZ_V)$, or simply $K_{`g}$, be a graph with vertex set $V$ and edge set
\begin{align}
	\mcE(K_{`g}):= \Set{\Set{i,j} \mid i,j \in V,i\neq j, I(\RZ_i\wedge \RZ_j)>`g}.
	\label{eq:Kg}
\end{align}
In words, we think of $K$ as a complete graph and associate each edge $\Set{i,j}$ with weight $I(\RZ_i\wedge \RZ_j)$. Then, the graph $K_{`g}$ can be obtained from $K$ by removing the ``light" edges, i.e., edges with weight no larger than $`g$.

\begin{Definition}[MIRN~\cite{butte00}]
	\label{def:RN}
	The non-singleton connected components of $K_{`g}(\RZ_V)$ are called the mutual information relevance networks (MIRN). The corresponding clusters are given by:
	\begin{align}
		\pzP(K_{`g})`/\Set{\Set{i}\mid i\in V},\label{eq:RN}
	\end{align}
	where $\pzP(K_{`g})$ is the partition of the vertices of $K_{`g}$ according to the connected components of $K_{`g}$.
\end{Definition}

Next, we introduce the Chow--Liu tree approximation under which \eqref{eq:RN} can be obtained from info-clustering. Consider a tree $T$ with vertex set $V$. A dependency-tree approximation~\cite{chow68} to $\RZ_V$, denoted as $\RZ_V^T$, can be written in terms of the marginal distributions $\RZ_B$ for $\abs{B}\leq 2$ as:
\begin{align}
	\kern-.5em P_{\RZ_V^T}(z_V) := `1(\prod_{i\in V} P_{\RZ_i}(z_i) `2) \prod_{\{i,j\}\in \mcE(T)}\kern-.5em \frac{P_{\RZ_i,\RZ_j}(z_i,z_j)}{P_{\RZ_i}(z_i)P_{\RZ_j}(z_j)},\kern-.5em
	\label{eq:ZT1}
\end{align}
for $z_V\in Z_V$.
Such a distribution forms a Markov tree or a Bayesian network (in which the in-degree of every vertex is at most one) with respect to $T$, i.e., we can relabel the indices in $V$ to $\Set{1,\dots,\abs {V}}$ such that
\begin{align}
	&P_{\RZ_V^T}(z_V) := \prod_{i\in V} P_{\RZ_i|\RZ_{\p_i}}(z_i|z_{\p_i})\kern1em\notag\\
	&\text{where $\p_1=`0$, $\p_i<i$, and $\Set{i,\p_i} \in \mcE(T)$ for $i>1$.}\label{eq:ZT2}
\end{align}

\begin{Definition}[Chow--Liu trees~\cite{chow68}]
	The set of Chow--Liu trees is defined as
	\begin{align}
		\pzT^*(\RZ_V)&:=  \arg\min_{T} D(P_{\RZ_V}\|P_{\RZ_V^T})\kern2em \text{where}\label{eq:T*}\\
		\kern-.5em D(P_{\RZ_V}\|P_{\RZ_V^T})&\utag{a}=D`1(\extendvert{P_{\RZ_V} \|\prod_{i\in V} P_{\RZ_i}}`2) - \sum_{e\in \mcE(T)} I(\RZ_e).\kern-.5em 
		\label{eq:D1}
	\end{align}
	Here, \uref{a} follows from \eqref{eq:ZT1}. For any $T\in \pzT^*(\RZ_V)$, $\RZ_V^T$ is called a Chow--Liu tree approximation to $\RZ_V$.
\end{Definition}

The celebrated Chow--Liu algorithm~\cite{chow68} computes a Chow--Liu tree as a maximum weight spanning tree since the minimization in \eqref{eq:T*} corresponds to maximizing the second term on the right-hand side of \eqref{eq:D1}, which is the total weight of the tree.

 The main result of this subsection is the following theorem on the equivalence between the clustering by MIRN and the clustering by MMI under the Chow--Liu tree approximation.
 \begin{Theorem}
 	\label{thm:TT}
 	The clustering of $\RZ_V$ by MMI~\eqref{eq:cluster} under the Chow--Liu tree approximation~\eqref{eq:T*} is
 	\begin{align}
 		\pzC_{`g}(\RZ_V^T) &= \pzP(K_{`g})`/\Set{\Set{i}\mid i\in V} 
 		\label{eq:ZT:cluster}
 	\end{align}
 	for any $`g\in `R$ and any $T\in \pzT^*(\RZ_V)$. Such a solution is identical to the clustering by MIRN~\eqref{eq:RN} and independent of the choice of $T\in \pzT^*(\RZ_V)$.
 \end{Theorem}
 \begin{Proof}
 	See Appendix~\ref{sec:D}.
 \end{Proof}
 
The proof of the equivalence makes use of the following result which evaluates the MMI for any dependency-tree distribution.
 \begin{Theorem}[MMI of dependency-tree distributions]
 	\label{thm:PP}
 	For any tree $T$ on the vertex set $V$, 
 	\begin{align}
 		I(\RZ_V^T) &= \min_{e \in \mcE(T)} I(\RZ_e) \label{eq:minI} \kern1em\text {and}\\
 		\pzP^*(\RZ_V^T) &= \pzP(T_{I(\RZ_V^T)}),\label{eq:T:P*}
 	\end{align}
 	where, as in \eqref{eq:Kg}, $T_{`g}$ denotes the tree $T$ with edges $e$ of weight $I(\RZ_e)\leq `g$ removed.
 \end{Theorem}
 \eqref{eq:minI} was discovered in \cite{csiszar04} to be the secrecy capacity for Markov trees but \eqref{eq:T:P*} is new.
  \begin{Proof}
  	See Appendix~\ref{sec:E}.
  \end{Proof}
In other words, the connected components of $T_{`g}$ for $`g=I(\RZ_V^T)$ characterize the fundamental partition for any dependency-tree distribution $\RZ_V^T$. 
 
 \begin{ybox}
The following theorem shows that $T_{`g}$ in fact characterizes the entire hierarchical clustering~\eqref{eq:cluster} of $\RZ_V^T$ for different values of $`g$:
 
 \begin{Theorem}[Clustering of dependency-tree distributions]
 	\label{thm:C}
 	For any tree $T$ on $V$ and any $`g\in `R$, 
 	\begin{align}
 		\pzC_{`g}(\RZ_V^T) &= \pzP(T_{`g})`/\Set{\Set{i}\mid i\in V}.\label{eq:C}
 	\end{align}
 	Furthermore, the critical value $`g_i$ in \eqref{eq:xi2} is the $i$-th smallest value in $\Set{ I(\RZ_e)\mid e\in \mcE(T) }$ and the partition $\pzP_i$ in \eqref{eq:C`gi} is  $\pzP(T_{`g_i})$, for $1\leq i \leq N$ and $\RZ_C^T$ in place of $\RZ_C$.
 \end{Theorem}
  \begin{Proof}
  	See Appendix~\ref{sec:F}.
  \end{Proof}
  
  However, the Chow--Liu tree approximation incurs a loss. Indeed, the clustering by MIRN fails to capture higher-order statistics beyond pairwise mutual information, because the algorithm only requires the knowledge of the pairwise mutual information. The following is a concrete example where the clustering by MIRN fails, while the general info-clustering without the Chow--Liu tree approximation succeeds.
  
  \begin{Example}
  	\label{eg:MIRN-fails}
  	Let $V=\Set{1,2,3,4}$,
  	\begin{align*}
  		\RZ_1:= \RX_a, \RZ_2:= \RX_b, \RZ_3:= \RX_a\oplus \RX_b \text{ and } \RZ_4:= \RX_c,
  	\end{align*}
  	where $\RX_j$'s are independent uniformly random bits.
  	It can be shown that $\RZ_i$'s are pairwise independent, and so their pairwise mutual information are all zero. The clustering by MIRN will construct a complete graph with zero weight on the edges. Hence, it will not return any cluster for threshold $`g>0$, because all the edges get removed in $K_{0}$~\eqref{eq:Kg}.
  	
  However, we know that $\RZ_1$, $\RZ_2$ and $\RZ_3$ share some mutual information, because $\RZ_3$ can be completely determined by $\RZ_1$ and $\RZ_2$. Indeed, it can be shown that
  \begin{align*}
  	I(\RZ_{\Set{1,2,3}}) &= \frac{\overbrace{H(\RZ_1)}^{=1}+\overbrace{H(\RZ_2)}^{=1}+\overbrace{H(\RZ_3)}^{=1}-\overbrace{H(\RZ_1,\RZ_2,\RZ_3)}^{=2}}{3-1} =\frac{1}{2},
  \end{align*}
  and so the random variables share non-negative mutual information. The general info-clustering algorithm will correctly find the cluster $\Set{1,2,3}$ at threshold $`g\in [0,\frac12)$.
  \end{Example}

\end{ybox}

\subsection{Clustering by covariance matrix}
\label{sec:CCM}
The Gaussian distribution is often used as a simplifying assumption because the distribution is completely characterized by its mean and covariance, both of which can be estimated quite efficiently from data. The measure of segregation in \cite{deco15}, for instance, is simply the differential entropy of a set of random variables assuming a jointly Gaussian distribution. The CLICK algorithm~\cite{sharan00} for gene clustering also makes certain assumption about the distribution being Gaussian, e.g., in the computation of the parameters and the threshold test. However, these assumptions are often mixed with other simplifications that make it rather difficult to tract the validity or the impact of the Gaussian assumption. For instance, the measure of segregation has a noise variance that is chosen in an ad-hoc manner to make the differential entropy in the desired range. For the CLICK algorithm, the clustering solution is defined as the end result of an algorithmic procedure, but only some of the steps are justified by the Gaussian assumption. In other words, the clustering solution does not appear to be uniquely defined from the mixture of algorithmic procedure and Gaussian assumption.

In contrast, we will derive a unique info-clustering solution assuming the random vector $\RZ_V$ is jointly Gaussian with zero mean and covariance matrix $\M\Sigma_V$.
It follows that any random subvector $\RZ_B$ for $B\subseteq V$ is also jointly Gaussian with zero mean and covariance matrix $\M \Sigma_B$ where $\M\Sigma_B$ is the submatrix of $\M\Sigma_V$ with the rows and columns indexed by elements outside $B$ removed. In the following, we use $\abs {\MA}$ to denote the determinant of any square matrix $\MA$.

\begin{Proposition}
	For the jointly Gaussian source $\RZ_V$ defined above,
	\begin{subequations}
		\begin{align}
			\kern-1em I(\RZ_V) &= \min_{\mcP\in \Pi'(V)} \frac{\sum_{C\in \mcP} \log \abs {\M\Sigma_C} - \log \abs {\M\Sigma_V} }{\abs{\mcP}-1} \label{eq:g:zi1}\\
			&= \min_{\mcP\in \Pi'(V)} \frac{\sum_{C\in \mcP} \sum_{i\in C}\log {\lambda^C_i} - \sum_{i\in V} \log \lambda^V_i }{\abs{\mcP}-1} \label{eq:g:zi2}\kern-1em \\
			&= \min_{\mcP\in \Pi'(V)} \frac{\log \abs {\M\Sigma_{\mcP}} - \log \abs {\M\Sigma_V} }{\abs{\mcP}-1} \label{eq:g:zi3}
		\end{align} 
		where $(\lambda^C_i\mid i\in C)$ is a vector of the eigenvalues of $\M\Sigma_C$ and $\M\Sigma_{\mcP}$ is the matrix $\M\Sigma_V$ but with the entry at row $i$ and column $j$ forced to $0$ if $i$ and $j$ belong to different blocks in $\mcP$.  
	\end{subequations}
\end{Proposition}

\begin{Proof}
	\eqref{eq:g:zi1} is obtained by substituting the following differential entropy into \eqref{eq:IS},
	\begin{align}
		H(\RZ_C) = \frac12`1(\abs {C} \log 2\pi e+ \log \abs {\M \Sigma_C}`2).
	\end{align}
	\eqref{eq:g:zi2} follows from the fact that the determinant~$\abs {\M\Sigma_C}$ is the product $ \prod_{i\in C}\lambda^C_i$ of the eigenvalues. \eqref{eq:g:zi3} is because, by possibly reordering the indices in $V$, $\M\Sigma_{\mcP}$ can be written as a block diagonal matrix with $\M\Sigma_C$ for $C\in \mcP$ being the blocks in the main diagonal. Hence, $ \abs {\M\Sigma_{\mcP}}=\prod_{C\in \mcP} \abs {\M\Sigma_C}$.
\end{Proof}

\begin{Proposition}
	\label{pro:gauss:psp}
	For the jointly Gaussian source $\RZ_V$ defined above, the clusters are the non-singleton subsets from the PSP of the submodular function $C\mapsto \log \abs {\M \Sigma_C}$.
\end{Proposition}

From \eqref{eq:g:zi2}, the clustering solution can be regarded as spectral clustering in the sense that it depends on the spectrum of the submatrices of the covariance matrix. However, it is a new clustering method different from the usual spectral clustering solution such as the one for approximately minimizing the normalized cuts~\cite{shi2000normalized}.

\subsection{Clustering by Network information flow}
\label{sec:CIF}

In order to make info-clustering applicable to the clustering of neurons based on their physical connectome, we need to convert the deterministic physical connections of neurons to a random source $\RZ_V$. We will show that this conversion is possible by reducing the info-clustering solution under the hypergraphical source model~\cite{chan10md,chan10phd}. To explain the idea, we start with the emulated source model.

\begin{Definition}[Emulated source~\mbox{\cite[Definition~2.1]{chan10md}}]
\label{def:esn}
For $i\in V$, let $\RZ_i=(\RX_i,\RY_i)$ such that,
\begin{align}
  	P_{\RX_V\RY_V}  &= \prod_{i\in V}  P_{\RX_i} P_{\RY_i|\RX_V}. 
  	\label{eq:esn}
\end{align}
The vector $\RZ_V$ is called an emulated source network.
\end{Definition} 
We can think of every $i\in V$ as a terminal that can send an input signal $\RX_i$ independently over a channel that returns the output signal
$\RY_i = f_i(\RX_V,\RN_i)$
to terminal $i$, where $f_i$ is deterministic and $\RN_i$'s are
independent channel noises that satisfy $P_{\RN_V|\RX_V}=\prod_{i\in V} P_{\RN_i}$. (Note that the observation $\RY_i$ of terminal~$i$ may depend on the input specified by other terminals.) Since $\RZ_i$ captures all the information in the input and output signals associated with terminal~$i$, the MMI among $\RZ_i$'s reflects the mutual information among the terminals, and so we can cluster the terminals accordingly. The MMI has the following special form:
\begin{Proposition}[\mbox{\cite[Proposition~2.1]{chan10md}}]
	\label{pro:esn}
	For the emulated source network in
	Definition~\ref{def:esn},
	\begin{align}
		I(\RZ_V)=\min_{\mcP\in \Pi'(V)} \frac1{\abs{\mcP}-1} \sum_{C\in \mcP}
		I(\RX_{V`/C}\wedge \RY_C|\RX_C),\label{eq:esn:I}
	\end{align}
	which is an achievable secret key rate under a multiterminal channel model~\cite[\S{II-B}]{chanzheng14}. 
\end{Proposition}

In network information theory, the mapping to the conditional mutual information, $C\mapsto I(\RX_{V`/C}\wedge \RY_C|\RX_C)$ in \eqref{eq:esn:I}, is a cut function (evaluated at the cut set $C$) that measures the total amount of information flow from the terminals in $V`/C$ to the terminals in $C$. Similar to the usual graphical cut function, this cut function is also submodular with respect to $C$. Therefore, the info-clustering algorithm will return the non-singleton subsets in the PSP of the cut function as the clusters.

A special case of interest is when the channel $P_{\RY_V|\RX_V}$ consists of a set of broadcast links among the subsets of the terminals. More precisely, consider a hypergraph with vertex set $V$, edge set $E$, and edge function $\phi:E\to 2^V`/\Set{`0}$. Each hyperedge $e\in E$ is regarded as a broadcast link with sender specified by $`r(e)\in \phi(e)$ and receivers being the terminals in $\phi(e)`/\Set{`r(e)}$. $`r$ is called the orientation of the edge $e$. The capacity of the broadcast link is specified by the non-negative weight $c(e)$. More precisely, the emulated source $\RZ_V$ is defined using
\begin{subequations}
\begin{align}
	\RX_i&:=(\RX_i^e\mid e\in E, i=`r(e))&& \text{for $i\in V$}\\
	\RY_j&:=(\RY_j^e\mid e\in E, j\neq `r(e)\in `\phi(e)) && \text{for $j\in V$},
\end{align}
and the input-output relationship of each broadcast link $e$:
\begin{align}
	\RY_j^e = \RX_{`r(e)}^e && \forall j\in \phi(e), \text{ and}\\
	\log \abs{X^e_{`r(e)}}=c(e)
\end{align} 
\end{subequations}
where the first equation says that the outputs of the broadcast link are equal to its input, and the second equation means that the capacity $c(e)$ is the log cardinality of the input alphabet set, which is the maximum amount of information that can be sent across the broadcast link. For instance, such a broadcast link can be used as a simple model for the physical connection between neurons because a neuron broadcasts signals to one or more neurons through the gap junctions and chemical synapses. The weight $c(e)$ can be obtained from the number of synapses. More elaborate models, such as the interference link in \cite{chan10md}, the ADT network in \cite{avestimehr09}, and the matroidal network link model~\cite{chan12ud,chan13isit,chan13itw}, can also be considered.

It is easy to argue that the MMI is maximized by the uniform input distribution, and the emulated source can be equivalently defined as follows without depending on the orientation $`r$:\footnote{The result of \cite{chan10md} is modified slightly to include edge weight $c$.}
\begin{Definition}[Broadcast Network~\mbox{\cite[Definition~2.4]{chan10md}}]
	\label{def:bc}
	A broadcast network with respect to the hypergraph $H:=(V,E,`f)$ is defined as
	\begin{align}
		\RZ_i:=\Set{\RZ^e:e\in E,i\in `f(e)} && \text{for $i\in V$},\label{eq:bc}
	\end{align}
	with $(\RZ^e:e\in E)$ uniformly distributed and $H(\RZ^e)=c(e)$.
\end{Definition}
The fact that the source model does not depend on the orientation $`r$ means that one needs not distinguish between directed and undirected links for info-clustering. For instance, even though the gap junction in neurons is undirected and the chemical synapses are directed, the direction does not affect the clustering. This is because each link, directed or not, leads to a piece of information shared \emph{symmetrically} among both the sender and the receivers.

The choice of the uniform input distribution can also be justified more rigorously. In the secret key agreement problem under the channel model~\cite[\S{VI-B}]{chanzheng14}, the uniform distribution on the input was shown to achieve the secrecy capacity, which is precisely the MMI $I(\RZ_V)$. Furthermore, the MMI can also be written in the form of a max-flow min-cut expression that characterizes the maximum multicast rate of network coding~\cite{chan11isit}.The MMI can be written in terms of the directed cut function for the hypergraph:
\begin{Proposition}[\mbox{\cite[Proposition~2.4]{chan10md}}]
	The MMI of the broadcast network~\eqref{eq:bc} is
	\begin{subequations}
		\label{eq:zi}
		\begin{align}
			I(\RZ_V) 
			&= \min_{\mcP\in \Pi'(V)} \frac{\sum_{C\in \mcP} \overbrace{\sum\nolimits_{e\in `d^-_{H^*}(C)} c(e)}^{c(`d^-_{H^*}(C)):=}}{\abs{\mcP}-1} \label{eq:zi1}\\
			&= \min_{\mcP\in \Pi'(V)} \frac{\sum_{e\in E} c(e)(\abs{`p_{\mcP}(`f(e))}-1)}{\abs{\mcP}-1}, \label{eq:zi2}
		\end{align}
	\end{subequations}
	where $H^*:=(V,E,`f,`r)$ is a hypergraph of $H$ with an arbitrary choice of the orientation $`r$ for each edge, 
	\begin{align}
		`d^-_{H^*}(C)&:=\Set{e\in E \mid `r(e)\in C^c \nsupseteq `f(e)}\\  
		`p_{\mcP}(`f(e))&:=\Set{C\cap `f(e) \mid C\in \mcP}`/\Set{`0}
	\end{align}
	are the set of in-coming edges into $C$ and the partition of $e$ respectively.
\end{Proposition}
Even though the MMI does not depend on the orientation $`r$, as shown in \eqref{eq:zi2}, it is informative to consider the alternative form in \eqref{eq:zi1} that is stated with an arbitrary choice of the orientation $`r$. In particular, from \eqref{eq:zi1}, we can deduce that:
\begin{Proposition}
	\label{pro:PSP:hyp}
	For the hypergraphical source $\RZ_V$ defined above, the clusters are the non-singleton subsets from the PSP of the submodular in-cut function $C\mapsto c(`d^-_{H^*}(C))$.
\end{Proposition}

Indeed, the physical connectome may be simplified as a graph instead of a hypergraph because the polyadic synapses that connect one neuron to multiple neurons are rare~\cite{white1986structure,varshney2011structural}. In the special case when the hypergraph is a graph $G=(V,E,`q)$ with $\abs{`q(e)}=2$, the broadcast network in \eqref{eq:bc} reduces to the graphical network called the pairwise independent network (PIN)~\cite{nitinawarat-ye10}. it is straightforward to show that the MMI in \eqref{eq:zi1} can be further written as the strength of the graph:
\begin{subequations}
\begin{align}
	I(\RZ_V)&= \min_{\mcP\in \Pi'(V)} \frac{\sum_{C\in \mcP} c(`d_{G}(C))}{2(\abs{\mcP}-1)} \label{eq:zi3}
\end{align}
where $C\mapsto c(`d_G(C))$ is the submodular undirected cut function with the edge cut
\begin{align}
	`d_G(C):=\Set{e\in E\mid `0\neq C\cap `q(e) \subsetneq C}.\label{eq:`d}
\end{align}
\end{subequations}
The factor of $2$ in the denominator of \eqref{eq:zi3} comes from the fact that an edge that crosses $\mcP$ overlap with two disjoint subsets in $\mcP$, so it is doubly counted in the numerator. Since the factor does not affect the PSP, we have the following result:
\begin{Proposition}
		\label{pro:PSP:g}
For the graphical source $\RZ_V$ defined above, the clusters are the non-singleton subsets from the PSP of the undirected cut function $C\mapsto c(`d_{G}(C))$.
\end{Proposition}

By the Tutte--Nash-Williams tree packing theorem, the strength of a graph has the meaningful interpretation as the maximum amount of fractional tree packings of the graph~\cite{nash1961edge,schrijver02}, which can also be extended to more general notion of partition connectivity for hypergraphs~\cite{frank03,bang-jensen01,chan10md}. It can be shown that the principal sequence for graphs correspond to successive packing of forests, with the first critical value being the strength of the graph and the last critical value being the fractional arboricity, defined as the maximum amount of forests one can fractionally pack in the graph~\cite{nash1964decomposition}.   

\section{Applications to biological datasets}
\label{sec:app}

In this section, we provide some discussions on how info-clustering can be used for the clustering of genes and neurons. For concreteness, we will describe some available datasets, and explain what one may potentially learn from them.

\subsection{Gene clusering}
As described in Section~\ref{sec:CL}, the clustering by MIRN~\cite{butte00} is a special case of info-clustering under the Chow--Liu tree approximation. Therefore, the experimental results in \cite{butte00} can be regarded as preliminary results of info-clustering, which may potentially be improved by considering higher-order correlation beyond pairwise mutual information as shown in Example~\ref{eg:MIRN-fails}.
   
The work in \cite{butte00} considered the dataset from \cite{eisen1998cluster}, which involves 2467 genes of a species of yeast called saccharomyces cerevisiae. The expression level of each gene was measured under 79 different conditions, including different stages of the cell cycle, temperatures, and time points. With $V$ denoting the set of all genes, the different expression levels of gene $i\in V$ were regarded as i.i.d.\ realizations of a random variable $\RZ_i$ that can be used for info-clustering. Towards this end, the pairwise mutual information $I(\RZ_i\wedge \RZ_j)$ between genes $i$ and $j$ was estimated using the empirical joint distribution of $\RZ_i$ and $\RZ_j$ after uniform quantization (since the expression levels are real-valued). We note that the empirical entropy after quantization can also be approximated without computing the empirical distribution~\cite{wu16}.

Similarly, the MMI beyond the pairwise mutual information can be estimated from the empirical distribution of the quantized expression levels. The idea is to compute the empirical entropies of subsets of random variables $\RZ_B$ after quantization, and use them in \eqref{eq:IS} to estimate the MMI. The MMI of the quantized random variables is shown to approach the MMI of the continuous random variables in \cite[Appendix~B]{chanzheng14}, and the details of the quantization can be found therein. However, computing the empirical joint distribution of a subset of random variables or estimating the joint entropy from the data samples takes exponential time with respect to the size of the subset~\cite{wu16}. This seems to suggest that some heuristics might be needed to tackle the problem of estimating the MMI. For example, an approach considered in \cite{kraskov09} was to use a file compression algorithm to return the file size after compressing the data associated with the subset of random variables. Alternatively, one may consider other model reduction techniques so that the simplifying assumption made is clear.

\subsection{Physical connectome}

As described in Section~\ref{sec:CIF}, info-clustering can be specialized to cluster graphical networks. When applied to the physical connectome of neurons, it can identify clusters of tightly connected neurons, as well as the important inter-cluster connections, the damage of which may cause detrimental effects. While the physical connectome may not represent the functional connectome, i.e., the task-specific stimulation patterns of the neurons, the resulting clusters may be studied in conjunction with the functional connectome to understand how neurons work together to manifest consciousness and to carry out brain functions~\cite{balduzzi2008integrated}. 

Instead of looking at the human connectome data, as an illustration, we will consider a small and nearly complete physical connectome dataset in \cite{varshney2011structural} for a small creature called nematode \emph{C.\ elegans}. A set of 279 neurons in the somatic nervous system was considered, and the dataset is in the form of an adjacency matrix, recording the total number of synaptic contacts (gap junction and chemical synapses) between every pairs of neurons.

The adjacency matrix defines the weighted graph for info-clustering. We can compute the cut function of the graph from the adjacency matrix, and then obtain the desired clusters from the PSP of the cut function as described in Section~\ref{sec:CIF}. It is worth pointing out that, owing to the similarity between MAC clustering algorithm of \cite{nagano10} and info-clustering as pointed out in Appendix~\ref{sec:mac}, we expect the performance of info-clustering to be close to that of the MAC clustering for cut functions, which was shown in \cite{nagano10} to be competitive with the existing leading algorithms for clustering graphical networks.

\section{Measures of integration and segregation}

Based on the info-clustering paradigm, we can derive some meaningful measures to describe the clustering solutions. For example, the MMI $I(\RZ_V)$ naturally measures how integrated the objects in $V$ are. This is because the more interaction among the objects in a system, the larger the mutual information they share. Such an argument is supported by the concrete operational meanings of $I(\RZ_V)$ as the secrecy capacity for the multiterminal secret key agreement problem, the multicast throughput for the network coding problem, and the partition connectivity for hypergraphical or graphical models. 

A measure called the integrated information was proposed in \cite[(2B)]{balduzzi2008integrated} to measure how integrated a system is. This measure may appear similar to the MMI in the sense that it is defined as the divergence from the joint distribution of the overall system to the product of the marginal distributions of some subsystems. (The subsystems are obtained by partitioning the system according to what is called the minimum information partition.) However, there are two fundamental differences between the two definitions:
\begin{enumerate}
\item In contrast to the fundamental partition $\pzP^*(\RZ_V)$, the minimum information partition is obtained by an additional normalization factor that forces the partition to be more balanced. However, this additional factor makes the problem intractable. This is similar to the normalized-cut minimization problem, which is NP-hard to solve. In general, the cluster size has nothing to do with the amount of information mutual to the elements in the cluster. Thus, this additional factor can steer the clustering procedure away from finding a small cluster that has high mutual information.
\item Unlike the MMI, the divergence expression in the integrated information with respect to the minimum information partition, say $\mcP$, is not normalized by the factor $\abs {\mcP}-1$. As we have described using the concept of residual independence relation, the factor is needed to account for the double counting in the mutual information in each subsystem. Therefore, the integrated information does not have the desired information-theoretic meaning.
\end{enumerate}
In addition, the integrated information is computed from the a posteriori probability that is marginalized using a uniform input distribution. However, the a posteriori distribution can be viewed as a channel, which can be handled by info-clustering as in the emulated source model in~\S\ref{sec:CIF}. Moreover, instead choosing a uniform distribution by assumption, we can justify such a choice as one that maximizes the MMI in the case of the hypergraphical model.

A measure of segregation was also proposed in \cite{deco15}. However, there are two issues of the formulation:
\begin{enumerate}
\item The measure assumes a jointly Gaussian distribution rather than a general source distribution, so it is unclear how the measure can capture a more general correlation structure. 
\item The measure is normalized using a noise variance, which is chosen in an ad-hoc manner without a concrete interpretation. The normalization is also done in a way different from the usual signal-to-noise ratio for the MIMO Gaussian channel~\cite{telatar1999capacity}.
\end{enumerate}
We believe that the measure of segregation is simply a dual to the measure of integration, i.e., the MMI can be used to measure segregation and there is no need to define another fundamental quantity. More precisely, we can measure the segregation of a cluster $C$ of $V$ as
\begin{align}
	1-\frac{I(\RZ_V)}{I(\RZ_C)}\in (0,1].\label{eq:segregation}
\end{align}
The index is non-negative because $I(\RZ_C)>I(\RZ_V)$ by the formulation~\eqref{eq:cluster} of clusters, and it is upper bounded by $1$ because of the non-negativity of the MMI. The index is large (the ratio $\frac{I(\RZ_V)}{I(\RZ_C)}$ is small) if the cluster $C$ is more integrated than the entire set $V$, that is to say, $C$ is more segregated from the rest of the nodes in $V`/C$. Depending on the application, one may further compute the average, minimum, or maximum segregation among a set of clusters to show how segregated the clusters are from each other.

\section{Conclusion}

In this work, we proposed a new information-theoretic approach to clustering biological systems. In particular, we formulated the info-clustering paradigm and showed how it can be applied to study the human genome and connectome. Compared to the conventional algorithmic approaches, info-clustering follows a bottom-up theoretical approach for clustering. Rather than justifying the algorithm purely by data, which was shown to have many issues, we believe that it is more important to lay a rigorous mathematical theory before algorithmic simplifications. In particular, the info-clustering is formulated in a meaningful way without requiring any prior knowledge of the number of clusters nor an initial solution to start the clustering algorithm. The solution is shown to be unique, with meaningful information-theoretic interpretations as well as an elegant mathematical structure for efficient computation.

More precisely, we formulated the clustering problem~\eqref{eq:cluster} using a threshold test on the MMI, and showed that the solution is hierarchical under a simple, but general, property~\eqref{eq:I:lb} of the MMI, which also holds for some other choices of multivariate mutual information measures. The clustering solution is characterized by a finite set of critical values and their corresponding finite sets of clusters~\eqref{eq:cluster:solution}. The formulation is different from the classical one in the sense that the set of clusters is not required a priori to form a partition. Instead, the set of all clusters is shown to be laminar~\eqref{eq:laminar} using the general property~\eqref{eq:I:lb} of the MMI. Consequently, the complete clustering solution can be computed iteratively in Algorithm~\ref{algo:iteration}.

Using the precise definition~\eqref{eq:IS} of the MMI, we further showed that the clustering solution maps to the PSP of the entropy function. More precisely, the set~\eqref{eq:Upgamma} of critical values for info-clustering is precisely the set of critical values~\eqref{eq:DT:critical_values} for the Dilworth truncation~\eqref{eq:DT} of the residual entropy function~\eqref{eq:residualH}. The corresponding set of clusters are the non-singleton subsets from the PSP~\eqref{eq:PSP} of the entropy function. This connection is non-trivial. It is based on the iterative relation~\eqref{eq:it2} among the clusters and the iterative relation~\eqref{eq:xi2} among the PSP.
This connection not only enriches the abstract mathematical structure of the PSP with the concrete operational meanings from information theory, but also provides a concrete clustering solution that can be computed from the PSP in strongly polynomial time.

Indeed, we showed that info-clustering reduces to simpler and more practical algorithms under some special source models. Unlike the approximation algorithms of many clustering formulations, which focuses mainly on algorithmic simplicity, the model reduction for info-clustering specifies precisely what kind of correlation structure is assumed in return for the algorithmic simplicity. Consequently, we can verify whether the simplifying model applies to the case of interest, and identify the weaknesses of the simplified algorithm.

In particular, we showed that under the Markov tree model info-clustering reduces to the gene clustering algorithm by MIRN~\eqref{eq:ZT:cluster}. If the correlation structure is not a Markov tree, the clustering by MIRN corresponds to the info-clustering algorithm under the Chow--Liu tree approximation. This shows that not only can info-clustering apply in practice to gene clustering, but it can also be used to justify existing techniques such as clustering by MIRN properly, with a concrete example showing how the Chow--Liu tree approximation may fail to capture the more complex multivariate correlation beyond the pairwise mutual information. 

We also considered the usual Gaussian assumption, which simplifies the info-clustering solution to a clustering algorithm by the covariance matrix, or more specifically, the eigenvalues of the submatrices of the covariance matrix (Proposition~\ref{pro:gauss:psp}.). This is a new spectral clustering technique that follows precisely from the info-clustering paradigm without any approximation. 

For the study of the human connectome, we also examined the specification of info-clustering to the hypergraphical model, which can capture the possibility of polyadic physical connections among neurons. In this case, the solution reduces to the PSP of hypergraphs and graphs (Proposition~\ref{pro:PSP:hyp} and \ref{pro:PSP:g}), which can be computed more efficiently than the PSP of the entropy function of a general source model. In addition to the algorithmic simplicity, the solution also has a meaningful interpretation as the network information flow: Clusters are simply subnetworks that support large information flows.

Finally, using the info-clustering paradigm, we also demonstrated how the MMI can be used as a measure of the integration of a cluster, which can further be used to measure how segregated a cluster is from the other objects or clusters~\eqref{eq:segregation}. The measures do not assume any particular source model or choice of parameters. Their values can be computed and justified from the info-clustering solution.

\clearpage

\appendices

\bgroup
\makeatletter
\renewcommand{\thesubsection}{\thesectiondis-\arabic{subsection}}
\renewcommand{\thesubsectiondis}{\arabic{subsection}.}
\@addtoreset{equation}{section}
\renewcommand{\theequation}{\thesection.\arabic{equation}}
\@addtoreset{Theorem}{section}
\renewcommand{\theTheorem}{\thesection.\arabic{Theorem}}
\@addtoreset{Lemma}{section}
\renewcommand{\theLemma}{\thesection.\arabic{Lemma}}
\@addtoreset{Corollary}{section}
\renewcommand{\theCorollary}{\thesection.\arabic{Corollary}}
\@addtoreset{Example}{section}
\renewcommand{\theExample}{\thesection.\arabic{Example}}
\@addtoreset{Remark}{section}
\renewcommand{\theRemark}{\thesection.\arabic{Remark}}
\@addtoreset{Proposition}{section}
\renewcommand{\theProposition}{\thesection.\arabic{Proposition}}
\@addtoreset{Definition}{section}
\renewcommand{\theDefinition}{\thesection.\arabic{Definition}}
\@addtoreset{Subclaim}{Theorem}
\renewcommand{\theSubclaim}{\theTheorem\Alph{Subclaim}}
\makeatother

\section{Proof of Theorems in \S\ref{sec:HC}}

\subsection{Proof of Theorem~\ref{thm:cluster}}
\label{sec:A}

The following is a necessary and sufficient condition for a set to be a cluster: 
\begin{Proposition}
	\label{pro:maxB}
	A non-empty non-singleton subset of $V$ is a cluster of $\RZ_V$ if and only if it cannot be enlarged without reducing multivariate information quantity, i.e., 
	\iftwocolumn
	\begin{multline}
		B \in  \pzC(\RZ_V) \iff \\
		I(\RZ_{B'})<I(\RZ_B) \kern1em \forall B'\subseteq V: B'\supsetneq B \label{eq:maxB}
	\end{multline}
	\else
	\begin{align}
		B \in \pzC(\RZ_V) \iff 
		I(\RZ_{B'})<I(\RZ_B) \kern1em \forall B'\subseteq V: B'\supsetneq B \label{eq:maxB}
	\end{align}
	\fi
	
	for $B\subseteq V:\abs{B}>1$.
\end{Proposition}

\begin{Proof}
	Suppose the R.H.S.\ of \eqref{eq:maxB} holds. Then, we have $B\in \pzC_{I(\RZ_B)^-}(\RZ_V)$ (and therefore the L.H.S. of \eqref{eq:maxB}) because $B$ is a maximal subset with multivariate information at least the threshold $I(\RZ_B)^-$.
	
	Suppose the R.H.S.\ of \eqref{eq:maxB} does not hold, i.e., there exists a proper superset $B'\supsetneq B$ with $I(\RZ_{B'})\geq I(\RZ_B)$. It follows that $I(\RZ_{B'})>`g$ whenever $I(\RZ_B)>`g$ and so $B$ cannot be maximal in \eqref{eq:cluster} for any threshold $`g\in `R$. Therefore, the L.H.S.\ of \eqref{eq:maxB} does not hold either.
\end{Proof}

\begin{Proposition}
	\label{pro:Upgamma}
	$\Upgamma(\RZ_V)=\Set{I(\RZ_B):B\in \pzC(\RZ_V)}$, consisting of the multivariate information quantities of the clusters.
\end{Proposition}
\begin{Proof}
	If $B\in \pzC_{`g}(\RZ_V)$ for some $`g\in `R$,  then $B\in \pzC_{I(\RZ_B)^-}(\RZ_V)`/\pzC_{I(\RZ_B)^+}(\RZ_V)$ and so $I(\RZ_B) \in \Upgamma(\RZ_V)$ by definition~\eqref{eq:Upgamma}.
	
 Consider any $`g\in \Upgamma(\RZ_V)$. Then, by \eqref{eq:Upgamma}, we have one of the following two cases:
	\begin{compactenum}
		\item There exists $B\in \pzC_{`g^-}(\RZ_V)`/\pzC_{`g^+}(\RZ_V)$, i.e., a cluster that disappears at $`g$. 
		We must have $`g^-<I(\RZ_B)\leq `g^+$ by \eqref{eq:cluster}, and so $`g=I(\RZ_B)$ as desired.
		\item There exists $B\in \pzC_{`g^+}(\RZ_V)`/\pzC_{`g^-}(\RZ_V)$, i.e., a cluster that appears at $`g$. By Proposition~\ref{pro:maxB}, this happens only if there is a larger cluster $B'\supsetneq B$ that disappears at $`g$, which reduces to the previous case.
	\end{compactenum}
\end{Proof}

We are now ready to prove Theorem~\ref{thm:cluster}. For $`g<I(\RZ_V)$, the set $\pzC_{`g}(\RZ_V)$ contains $V$ by definition~\eqref{eq:cluster}. Indeed, $V$ is the unique cluster because it is the largest subset of $V$. It follows that $`g_1=I(\RZ_V)$ and $\pzC_{`g}(\RZ_V)=\Set{V}$ for $`g<`g_1$. 

By the definition of critical values~\eqref{eq:Upgamma}, the cluster $\pzC_{`g}(\RZ_V)$ must remain unchanged for $`g$ between consecutive critical values. Therefore, $\pzC_{`g}(\RZ_V)=\pzC_{`g_i}(\RZ_V)$ for $`g\in [`g_i,`g_{i+1})$ and $1\leq i< N$.

When $`g\geq \max_{B\subseteq V:\abs{B}>1} I(\RZ_B)$, we have $\pzC_{`g}(\RZ_V)=`0$ because no solution $B$ to \eqref{eq:cluster} can have $I(\RZ_B)$ larger than the maximum~\eqref{eq:maxMMI}, which must therefore be the last critical value. Since the clusters in $\pzC_{`g_{N-1}}(\RZ_V)$ remain to be clusters for $`g\in [`g_{N-1},`g_N)$, they must achieve the maximum value of the multivariate information quantity.

\subsection{Proof of Theorem~\ref{thm:iterate}}
\label{sec:B}

\begin{figure}
	\centering
	\tikzstyle{cluster}=[fill=gray,opacity=0.3,draw=black]
	\tikzstyle{isclusterof}=[draw,thick,blue,<-]
	\tikzstyle{isnotclusterof}=[draw,thick,red,decorate,decoration={crosses,pre=moveto, pre length=1*\u,post=moveto,post length=1*\u,shape size=0.5*\u}]
	\begin{subfigure}[t]{.5\linewidth}
		\centering
		{\def\u{1em}
			\tikzstyle{point}=[circle,minimum size=.3em,fill]
			{\scriptsize \tikz{\draw[isclusterof,->] (0,0) --(4*\u,0) node [right] {is a cluster of};}}\\[1em]
			\begin{tikzpicture}[inner sep=0,outer sep=0,x=\u,y=\u, every node/.style={rounded corners}]
			\matrix (C) [matrix of math nodes, nodes={minimum width=1*\u,minimum height=1.5*\u}, 
			nodes in empty cells, row sep=1*\u,column sep=0.1*\u,
			column 6/.style={column sep=1*\u},  
			column 7/.style={anchor=base west,column sep=1*\u}]
			{
				&  &  &  &  & &  & \\
				&  &  &  &  & &  & \\
				&  &  &  &  & &  & \\
				&  &  &  &  & &  & \\
				&  &  &  &  & &  & \\
			};
			\draw[->] (C-1-1-|C-1-8) to node (12) {} (C-2-1-|C-1-8) to node (23) {} (C-3-1-|C-1-8) to node (34) {} (C-5-1-|C-1-8) node[label={[label distance=\u]below:{$`g$}}] {};
			\foreach \x/\xtext in {12/$`g''$,23/$`g'$}
			\draw[dashed] (C-1-1|-\x) -- (\x) node [point,label={[label distance=.1*\u]right:{\xtext}}] {};
			\node[cluster, fit=(C-1-1) (C-1-6),label={[label distance=.1\u]center:{$V$}}] (V) {};
			\node[cluster, fit=(C-2-1) (C-2-4),label={[label distance=.4\u]center:{$B''$}}] (B'') {};
			\node[cluster,rounded corners=.6em, fit=(C-3-3) (C-3-4),label={[label distance=.2\u]center:{$B'$}}] (B') {};
			\node[cluster, fit=(C-3-2) (C-3-4) (C-4-2) (C-4-4),label={[label distance=-1.1*\u]-90:{$B$}}] (B) {};
			\node[cluster, fit=(C-3-2) (C-3-5) (C-4-2) (C-5-5),label={[label distance=-1.1*\u]-90:{$C$}}] (C') {};
			\draw[isnotclusterof] (V.south east) to [out=-40,in=10]  ($(B.east)+(0,-1*\u)$);
			\draw[isclusterof] (V.south east) to [out=-40,in=10] ($(B.east)+(0,-1*\u)$);
			\draw[isclusterof] (V.east) to [out=-40,in=10] (C'.east);
			\draw[isclusterof] (V.south west) to [bend right=90] (B'.west);
			\draw[isclusterof] (B''.south east) to [out=-10,in=10] (B.east);
			\draw[isclusterof] (V.south) to [out=-10,in=10]  (B''.east); 
			\end{tikzpicture} 
		} 
		\caption{Theorem~\ref{thm:iterate}. Existence of $C$ contradicts $I(\RZ_{B''})$ is maximized.}
		\label{fig:iterate:1}
	\end{subfigure}
	\hfill
	\begin{subfigure}[t]{.45\linewidth}
		\centering
		{\def\u{1em}
			\tikzstyle{point}=[circle,minimum size=.3em,fill]
			{\scriptsize\tikz{
					\draw[isnotclusterof] (0,0)--(4*\u,0);
					\draw[isclusterof,->] (0,0) --(4*\u,0) node [right] {is not a cluster of};}}\\[1em]
			\begin{tikzpicture}[inner sep=0,outer sep=0,x=\u,y=\u, every node/.style={rounded corners}]
			\matrix (C) [matrix of math nodes, nodes={minimum width=1*\u,minimum height=1.5*\u}, 
			nodes in empty cells, row sep=1*\u,column sep=0.1*\u,
			column 6/.style={column sep=1*\u},  
			column 7/.style={anchor=base west,column sep=1*\u}]
			{
				&  &  &  &  & &  & \\
				&  &  &  &  & &  & \\
				&  &  &  &  & &  & \\
				&  &  &  &  & &  & \\
				&  &  &  &  & &  & \\
			};
			\draw[->] (C-1-1-|C-1-8) to node (12) {} (C-2-1-|C-1-8) to node (23) {} (C-3-1-|C-1-8) to node (34) {} (C-5-1-|C-1-8) node[label={[label distance=\u]below:{$`g$}}] {};
			\node[cluster, fit=(C-1-1) (C-1-6),label={[label distance=.1\u]center:{$V$}}] (V) {};
			\node[cluster, fit=(C-2-1) (C-2-4),label={[label distance=.4\u]center:{$B'$}}] (B') {};
			\node[cluster,rounded corners=.6em, fit=(C-3-3) (C-3-4),label={[label distance=.2\u]center:{$B''$}}] (B'') {};
			\node[cluster, fit=(C-3-2) (C-3-5) (C-4-2) (C-5-5),label={[label distance=.5\u]center:{$C$}}] (C') {};
			\draw[isnotclusterof] (V.south east) to [out=-40,in=10]  (B''.east);
			\draw[isclusterof] (V.south east) to [out=-40,in=10] (B''.east);
			\draw[isclusterof] (V.south west) to [bend right=45] (C'.west);
			\draw[isclusterof] (B'.south east) to [out=-10,in=10] (B''.east);
			\draw[isclusterof] (V.south) to [out=-10,in=10]  (B'.east); 
			\end{tikzpicture} 
		} 
		\caption{Theorem~\ref{thm:iterate2}. Existence of $C$ contradicts laminarity~\eqref{eq:laminar}.}
		\label{fig:iterate:2}
	\end{subfigure}
	\caption{Illustration of clusters in proofs by contradiction.}
	\label{fig:iterate}
\end{figure}
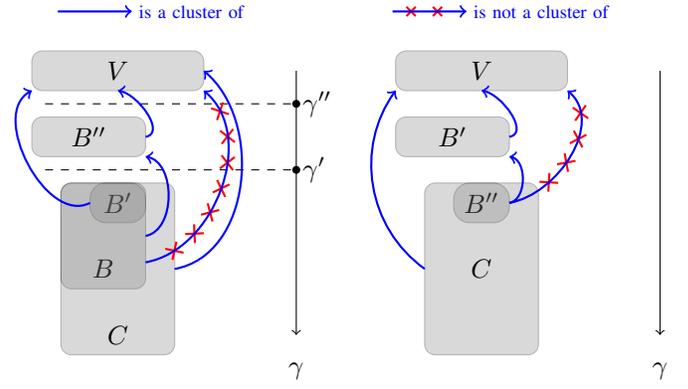
To help understand the proof, the readers may refer to \figref{fig:iterate:1} for a summary of the relationship among the clusters in the proof.
Let $B''$ be a cluster of $\RZ_V$ that is a proper superset of $B'$, i.e., we have
\begin{align}
	`g''\in `R, B''\supsetneq B',B''\in \pzC_{`g''}(\RZ_V). \label{eq:iterate:1}
\end{align}
Such a choice of $B''$ exists because $V$ is a feasible choice, but there can be multiple feasible choices. We choose any one that maximizes $I(\RZ_{B''})$. We will show that $B''$ and $`g''$  satisfy \eqref{eq:iterate}.

Note that $I(\RZ_{B''})>`g''$ and $I(\RZ_{B'})>`g'$ because $B''\in \pzC_{`g''}(\RZ_V)$ and $B'\in \pzC_{`g'}(\RZ_V)$ respectively. We also have  $I(\RZ_{B''})\leq`g'$ because, otherwise, $B''\supsetneq B'$ contradicts the maximality of $B'\in \pzC_{`g'}(\RZ_V)$. Altogether, we have 
\begin{align*}
	`g''\utag{a}<\upgamma_1(\RZ_{B''})\utag{b}=I(\RZ_{B''})\utag{c}\leq`g' \utag{d}<I(\RZ_{B'})
\end{align*}
where the equality~\uref{b} is by \eqref{eq:upgamma1}.

\uref{a}, \uref{b} and \uref{c} implies $`g''<`g'$ as desired by \eqref{eq:iterate}.
Furthermore, $I(\RZ_{B''})< I(\RZ_{B'})$ (from \uref{c} and \uref{d}) and the fact that $B'$ is a cluster of $\RZ_V$ implies that $B'$ is also a cluster of $\RZ_{B''}$. However, to establish \eqref{eq:iterate}, we need to show the stronger statement that $B'\in \pzC_{\upgamma_1}(\RZ_{B''})$. 

Now, $\pzC_{\upgamma_1}(\RZ_{B''})\neq `0$ because we at least have $B'\subseteq B''$ with $I(\RZ_{B'})> \upgamma_1(\RZ_{B''})$ (from \uref{b}, \uref{c} and \uref{d}). Therefore, we have $B'\in \pzC_{\upgamma_1}(\RZ_{B''})$ as desired by \eqref{eq:iterate} unless there exists 
\begin{align*}
	B\in  \pzC_{\upgamma_1}(\RZ_{B''}):B'\utag{e}\subsetneq B. 
\end{align*}
Suppose to the contrary that such a subset $B$ exists. Then, 
\begin{align*}
	\upgamma_1(\RZ_{B''})\utag{f}<I(\RZ_B)
\end{align*}
because $B\in  \pzC_{\upgamma_1}(\RZ_{B''})$. 
We will show that, regardless of whether $B$ is a cluster of $\RZ_V$ or not, there is a contradiction to the maximality of $I(\RZ_{B''})$ among all feasible subsets $B''$ satisfying \eqref{eq:iterate:1}.
\begin{compactenum}
	\item Suppose $B\in \pzC(\RZ_V)$. Then, \eqref{eq:iterate:1} holds with $B''$  replaced by $B$ and so $I(\RZ_B)>I(\RZ_{B''})$ (from \uref{b} and \uref{f}) contradicts the choice of $B''$.
	\item Suppose $B\not\in \pzC(\RZ_V)$. Then, there exists
	\begin{align*}
		C \in  \pzC(\RZ_V): B\utag{g}\subsetneq C,I(\RZ_C)\utag{h}\geq I(\RZ_B).
	\end{align*}
        This is the complete scenario shown in \figref{fig:iterate:1}.
	It follows that $C\supsetneq B'$ (from \uref{e} and \uref{g}) and $I(\RZ_C)>`g''$ (from \uref{a}, \uref{b}, \uref{f} and \uref{h}).
	Therefore, \eqref{eq:iterate:1} holds with $B''$ replaced by $C$,\footnote{We can also redefine $`g''$ to be the previous value of $I(\RZ_{B''})$, in which case \uref{a} need not be used to argue the contradiction.} but $I(\RZ_C)>I(\RZ_{B''})$ (from \uref{b}, \uref{f} and \uref{h}) contradicts the choice of $B''$.
\end{compactenum}

\subsection{Proof of Theorem~\ref{thm:iterate2}}
\label{sec:C}

It suffices to show that for any cluster $B'$ of $\RZ_V$, we have
\begin{align}
	\pzC(\RZ_{B'})=\Set*{B\in \pzC(\RZ_V):B\subseteq B'},\label{eq:it}
\end{align}
i.e., a cluster of $\RZ_{B'}$ must be a cluster of $\RZ_V$. 
Then, from \eqref{eq:it}, a simple induction on $`g$ over the finite set $\Upgamma(\RZ_V)$ will immediately lead to \eqref{eq:it2}.

To prove \eqref{eq:it}, consider any $B$ that is in the R.H.S.\ of \eqref{eq:it}. Then, by definition~\eqref{eq:cluster}, $B$ is a maximal subset of $V$ with $I(\RZ_B)>`g$ for some $`g\in `R$. Since $B\subseteq B'\subseteq V$, we also have that $B$ is a maximal subset of $B'$ with $I(\RZ_B)>`g$, i.e., $B$ is a cluster of $\RZ_{B'}$, belonging to the set in the L.H.S.\ of \eqref{eq:it}. Therefore, $\supseteq$ holds for \eqref{eq:it}. 

It remains to show the reverse inclusion $\subseteq$ for \eqref{eq:it}. To help understand the proof,  the readers may refer to \figref{fig:iterate:2} for a summary of the relationship among the clusters.
Suppose to the contrary that a cluster $B''$ of $\RZ_{B'}$ is not a cluster of $\RZ_V$. Note that $B''\neq B'$ because $B'$ is a cluster of $\RZ_V$ but $B''$ is not. Therefore, we have the strict inequality $I(\RZ_{B''})>I(\RZ_{B'})$. By \eqref{eq:maxB}, there exists
\begin{align*}
	C\in \pzC(\RZ_V): B''\subsetneq C, I(\RZ_C)\geq I(\RZ_{B''}).
\end{align*}
which implies that
\begin{align*}
	C\neq B'\kern1em\text{ and }\kern1em I(\RZ_C)\geq I(\RZ_{B''})>I(\RZ_{B'}).   
\end{align*}
We will show that $ C\cap B' \not\in\Set{`0,C,B'}$, contradicting laminarity~\eqref{eq:laminar}.
\begin{compactenum}
	\item $C\cap B'\neq `0$ because both $C$ and $B'$ contains the non-empty set $B''$.
	\item $C\cap B'\neq C$ or simply $C\nsubseteq B'$ because, if to the contrary that $C\subseteq B'$, then $B''\subsetneq C$ and $I(\RZ_C)\geq I(\RZ_{B''})$ assumed above contradicts the fact that $B''$ is a cluster of $\RZ_{B'}$.
	\item $C\cap B'\neq B'$ or simply $B'\nsubseteq C$ because, if to the contrary that $B'\subseteq C$, then $C\neq B'$ and $I(\RZ_C)> I(\RZ_{B'})$ derived above contradict the fact that $B'$ is a cluster of $\RZ_V$.
\end{compactenum}

\section{Proof of Theorem~\ref{thm:xi}}
\label{sec:PSP_proof}
	The line segment preceding the $p_i=(`g_i,y_i)$
	is $h_{`g}[\pzP_{i-1}]$ since $\pzP_{i-1}=\max \Uppi_i$ by
	\eqref{eq:PSP2}. The curve after $p_i$ has a strictly smaller slope
	than $-\abs{\pzP_{i-1}}$ by the definition of a turning point, and
	so $`g=`g_i$ is a solution to
	\begin{align*}
		h_{`g} [\pzP_{i-1}] = \min_{\substack{\mcP\in \Pi(V): \abs{\mcP}>\abs{\pzP_{i-1}}}} h_{`g}[\mcP]
	\end{align*}
	where the R.H.S.\ corresponds to $\hat{h}_{`g}(V)$ for $`g\geq
	`g_i$, with the set of optimal partitions at $`g=`g_i$ being
	$\Uppi_{i}`/\Set{\pzP_{i-1}}$. Rearranging the terms, it follows that
	\begin{align*}
		0 &= \min_{\substack{\mcP\in \Pi(V): \abs{\mcP}>\abs{\pzP_{i-1}}}} h_{`g_i}[\mcP] - h_{`g_i} [\pzP_{i-1}] \\
		&= \min_{\substack{\mcP\in \Pi(V): \abs{\mcP}>\abs{\pzP_{i-1}}}} h[\mcP] - h [\pzP_{i-1}] - `g_i(\abs{\mcP}-\abs{\pzP_{i-1}}) \\
		&= \min_{\substack{\mcP\in \Pi(V): \abs{\mcP}>\abs{\pzP_{i-1}}}} \frac{h[\mcP] - h [\pzP_{i-1}]}{\abs{\mcP}-\abs{\pzP_{i-1}}} - `g_i,
	\end{align*}
	which implies \eqref{eq:xi1}.
	The last expression is obtained from the previous by multiplying
	$\abs{\mcP}-\abs{\pzP_{i-1}}\geq 1$, which preserves both the
	minimum value of $0$ and the set of minimum solutions, namely
	$\Uppi_{i}`/\Set{\pzP_{i-1}}$. Since $\pzP_{i-1}=\max \Uppi_i$, every
	optimal solution is finer than $\pzP_{i-1}$, and so it does
	not lose optimality to impose $\mcP\prec \pzP_{i-1}$ in
	\eqref{eq:xi1}
	or equivalently,
	\begin{align*}
		\mcP=\bigcup\nolimits_{C\in \pzP_{i-1}} \mcP^C\kern1em \text{ for some  $\mcP^C \in \Pi(C)$.}
	\end{align*}
	Using the above, \eqref{eq:xi1} can be rewritten as
	\begin{align*}
          `g_i &= \min_{\mcP \prec \pzP_{i-1}} \frac{\sum_{C\in \pzP_{i-1}} `1[h[\mcP^C] - h(C)`2]}{\sum_{C\in \pzP_{i-1}} (\abs{\mcP^C}-1)} \\
		&\leq \min_{C\in \pzP_{i-1}:\abs{C}>1} \min_{\mcP^C\in \Pi'(C)} \frac{D(P_{\RZ_C}\|\prod_{C'\in \mcP^C} P_{\RZ_{C'}}) }{\abs{\mcP^C}-1} 
	\end{align*}
	The last expression is obtained
	by imposing $\mcP^C=\Set{{C}}$
        for all but one $C\in \pzP_{i-1}$ with $\abs{C}>1$, and substituting $h[\mcP^C] -
	h(C)=D(P_{\RZ_C}\|\prod_{C'\in \mcP^C} P_{\RZ_{C'}})$. It is equal to the R.H.S.\ of \eqref{eq:xi2} by the
	definition of $I$~\eqref{eq:IS}. (The existence of a block $C\in \pzP_{i-1}: |C|>1$ is guaranteed by
	Proposition~\ref{pro:PSP} since
	$\pzP_{i-1}$ is coarser than the partition into singletons, i.e.,
	$\pzP_{i-1} \succ \pzP_N$.)
        To show the reverse
	inequality, let $`g$ be the R.H.S.\ of \eqref{eq:xi2} and $\pzQ$ be
	the set of optimal solutions. Then, for all
	$C\in \pzP_{i-1}$ and $\mcP^C\in \Pi(C)$, we have $h[\mcP^C] -
	h(C)\geq`g (\abs{\mcP^C}-1)$ and so
	\begin{align*}
		`g_i &= \min_{\mcP \prec \pzP_{i-1}} \frac{\sum_{C\in \pzP_{i-1}} `1[h[\mcP^C] - h(C)`2]}{\sum_{C\in \pzP_{i-1}} (\abs{\mcP^C}-1)} \\
                &\geq \frac{\sum_{C\in \pzP_{i-1}} `g (\abs{\mcP^C}-1)}{\sum_{C\in
				\pzP_{i-1}} (\abs{\mcP^C}-1)}= `g 
	\end{align*}
	Equality happens if and only if, for all $C\in \pzP_{i-1}$, either we have
	$\mcP^C=\Set{C}$ or we have $C\in \pzQ$ and $\mcP^C\in
	\Uppi^*(\RZ_C)$. This implies $\pzQ=\pzP_{i-1}`/\pzP_i$ and therefore \eqref{eq:Pii}.

\section{Proof of Theorems in \S\ref{sec:CL}}

\subsection{Proof of Theorem~\ref{thm:TT}}
\label{sec:D}

The proof relies on Theorem~\ref{thm:C} proved in Appendix~\ref{sec:E}, which relies on Theorem~\ref{thm:PP} proved in Appendix~\ref{sec:F}.
First note that to prove Theorem~\ref{thm:TT}, it suffices to prove $\pzP(T_{`g})=\pzP(K_{`g})$, because this implies \eqref{eq:ZT:cluster} by  \eqref{eq:C}. In other words, we want to show that the vertex sets of the connected components of $T_{`g}$ are the same as those of $K_{`g}$, i.e., for any $i,j\in V$, we have $i\sim_{T_{`g}} j$ if and only if $i\sim_{K_{`g}} j$. The direct part (only if) is obvious, because $T_{`g}$ is a subgraph of $K_{`g}$. 
To prove the converse (if) part, we will use the following exchange property for spanning trees.

\begin{Lemma}[\mbox{\cite[Theorem~39.12]{schrijver02}}]\label{lem:exchange}
	Consider two spanning trees $T$ and $T'$ on the vertex set $V$. For any $e\in \mcE(T)`/\mcE(T')$, there exists $e'\in \mcE(T')`/\mcE(T)$ such that the graph $T$ with $e$ replaced by $e'$, denoted as $T-e+e'$, is a spanning tree.
\end{Lemma}

Now, suppose to the contrary that $i\not\sim_{T_{`g}} j$ but $i\sim_{K_{`g}} j$. Let $\mcE'$ be the set of edges in a path from $i$ to $j$ in $K_{`g}$. Let $e$ be an edge in the path from $i$ to $j$ in $T$ but with $I(\RZ_e)\leq `g$, and therefore not in $\mcE(K_{`g})$ nor $\mcE'$. Such an edge exists by the assumption $i\not\sim_{T_{`g}} j$. 

Let $G$ be the graph with edge set $\mcE(T)\cup \mcE'`/\Set{e}$. There exists a spanning tree $T'$ of $G$ since $G$ is connected, which follows from the facts that $T$ is spanning and $\mcE'$ connects the incident vertices of the removed edge $e$. Since $e\in \mcE(T)`/\mcE(T')$, we have by Lemma~\ref{lem:exchange} that there exists $e'\in \mcE(T')`/\mcE(T)$ such that $T-e+e'$ is a spanning tree. The tree $T-e+e'$ has a larger weight than $T$ because $I(\RZ_{e'})>`g\geq I(\RZ_e)$, as $e'\in \mcE(T')`/\mcE(T) \subseteq \mcE' \subseteq \mcE(K_{`g})$. This contradicts the maximality of $T\in T^*(\RZ_V)$. We have thus completed the proof of Theorem~\ref{thm:TT}.

\subsection{Proof of Theorem~\ref{thm:PP}}
\label{sec:E}

We remark that while the above arguments are purely of a graph-theoretical nature, in proving Theorems~\ref{thm:PP}, and subsequently Theorem~\ref{thm:C}, we rely on some information-theoretic properties of the MMI~\eqref{eq:IS}. The following is a lower bound on $I_{\mcP}$ specific to the dependency-tree distributions:
\begin{Lemma}
	\label{lem:IP}
	Consider the notation in \eqref{eq:ZT2}. For $\mcP\in \Pi'(V)$,
	\begin{align}
		I_{\mcP}(\RZ_V^T) &\geq \frac1{\abs{\mcE_{\mcP}}} \sum_{e\in \mcE_{\mcP}} I(\RZ_e)
		\label{eq:IP:t}
	\end{align}
	where $\mcE_{\mcP}:=\Set{\Set{\min C, \p_{\min C}}\colon 1\not\in C\in \mcP}$.
	Moreover, equality holds if we have 
	\begin{align}
		\p_i\in C\kern 2em \text {for all $C\in \mcP$ and $i\in C$ such that $i\neq \min C$}\label{eq:IP:eq}.
	\end{align}
	Note that \eqref{eq:IP:eq} simply means that the subgraph of $T$ induced on each $C\in \mcP$ is a subtree. 
\end{Lemma}

\begin{Proof}
	By \eqref{eq:D:1}, we can express $I_{\mcP}(\RZ_V^T)$~\eqref{eq:IP} in terms of the entropies as $\frac{\sum_{C\in \mcP} H(\RZ_C^T) - H(\RZ_V^T)}{\abs {\mcP}-1}
	$. It follows from the definition of $\mcE_{\mcP}$ that $\abs {\mcE_{\mcP}}=\abs {\mcP}-1$, and so
	\begin{align*}
		&\abs {\mcE_{\mcP}} I_{\mcP}(\RZ_V^T)\\
		&=\sum_{C\in \mcP} \underbrace{H(\RZ_C^T)}_{\text{(i)}} \kern.5em- \kern.5em \underbrace{H(\RZ_V^T)}_{\text{(ii)}}\\[-1.4em]
		&\utag{a}= \sum_{C\in \mcP} \overbrace{\sum_{i\in C} \underbrace{H(\RZ_i^T|\RZ^T_{\Set{j\in C\mid j<i}})}_{\text{(iii)}}}^{\text{(i)}'} - \overbrace{\sum_{i\in V} \underbrace{H(\RZ_i^T|\RZ^T_{\Set{j\in V\mid j< i}})}_{\text{(iv)}}}^{\text{(ii)}'}\\
		&\utag{b}\geq \sum_{C\in \mcP} `1[H(\RZ_{\min C})+\sum_{i\in C`/\Set{\min C}} H(\RZ_i^T|\RZ^T_{\p_i})`2] \\
		&\kern1em - \sum_{i\in V} H(\RZ_i^T|\RZ^T_{\p_i})\\
		&= \sum_{C\in \mcP} `1[H(\RZ_{\min C}) - H(\RZ_{\min C} | \RZ_{\p_{\min C}})`2]
	\end{align*}
	which is equal to $\sum_{e\in \mcE_{\mcP}} I(\RZ_e)$, completing the proof of \eqref{eq:IP:t}. To obtain~\uref{a}, we applied the chain rule $\text{(i)}=\text{(i)}'$ and $\text{(ii)}=\text{(ii)}'$. To obtain \uref{b}, we used  $\text{(iv)}=H(\RZ_i^T|\RZ^T_{\p_i})$ by the Markov relation~\eqref{eq:ZT2}. 
	$\text{(iii)}=H(\RZ_{\min C})$ when $i=\min C$, and  $\text{(iii)}\geq H(\RZ_i^T|\RZ^T_{\Set{j\in C\mid j<i}\cup\Set{\p_i}})=H(\RZ_i^T|\RZ^T_{\p_i})$ for $i>\min C$ by the Markov relation and the fact that conditioning reduces entropy. Equality holds if and only if \eqref{eq:IP:eq} holds, again due to the Markov relation~\eqref{eq:ZT2}. 
\end{Proof}

We are now ready to prove Theorem~\ref{thm:PP}. By \eqref{eq:IP:t} in Lemma~\ref{lem:IP},
\begin{align*}
	I_{\mcP}(\RZ_V^T)\geq \frac1{\abs{\mcE_{\mcP}}} \sum_{e\in \mcE_{\mcP}} I(\RZ_e) \utag{a}\geq \min_{e\in \mcE_{\mcP}} I(\RZ_e) \utag{b}\geq \min_{e\in \mcE(T)} I(\RZ_e)
\end{align*}
where \uref{a} is because the minimum edge weight on the right is no larger than the average on the left; \uref{b} is because $\mcE_{\mcP}\subseteq \mcE(T)$. The above implies $\geq$ for~\eqref{eq:minI} by the definition~\eqref{eq:IS} of MMI. 

To prove the reverse inequality, let
$\mcP:=\pzP(T_{`g})$ and $`g:=\min_{e\in \mcE(T)} I(\RZ_e)$. We shall argue that:
\begin{align*}
	I_{\mcP}(\RZ_V^T) &\utag{c}= \frac1{\abs{\mcE_{\mcP}}} \sum_{e\in \mcE_{\mcP}} I(\RZ_e) \utag{d}= \min_{e\in \mcE(T)} I(\RZ_e).
\end{align*}
\begin{itemize}
	\item \uref{c} is because the equality condition~\eqref{eq:IP:eq} holds. More precisely,  every $C\in \mcP$ is the vertex set of a connected component, and so, for all $i\in C$, we have $\p_i\in C$ unless $i=\min C$. 
	\item To argue \uref{d}, it suffices to show that $\mcE_{\mcP}=\mcE(T)`/\mcE(T_{`g})$, because $I(\RZ_e)=`g$ for all $e\in \mcE(T)`/\mcE(T_{`g})$.
	For any $C\in \mcP$, we have $\p_{\min C}\not\in C$ because $\p_{\min C}<\min C$ by \eqref{eq:ZT2}. Therefore, any edge $\Set{\min C,\p_{\min C}}\in \mcE_{\mcP}$ is also in $\mcE(T)`/\mcE(T_{`g})$. Conversely, consider any edge $\Set {i,\p_i}\in \mcE(T)`/\mcE(T_{`g})$ and $C\in \mcP:i\in C$ for some $i>1$ in $V$. Suppose to the contrary that $\Set{i,\p_i}\not\in \mcE_{\mcP}$. By the equality condition~\eqref{eq:IP:eq} proved earlier, we have $i=\min C$, contradicting $\Set{i,\p_i}\not\in \mcE_{\mcP}$.
\end{itemize}
Now that \eqref{eq:minI} is proved, we have $`g=I(\RZ_V^T)$ and $\mcP=\pzP(T_{I(\RZ_V^T)})$. To prove \eqref{eq:T:P*}, suppose to the contrary that $\pzP^*(\RZ_V^T)\neq \mcP$, i.e., there exists $\mcP'\in \Uppi^*(\RZ_V)$ with $\abs {\mcP'}>\abs {\mcP}$. We have
\begin{align*}
	\min_{e\in \mcE(T)} I(\RZ_e) \utag{e}= `g\utag{f}=I(\RZ_V^T)\utag{g}=I_{\mcP'}(\RZ_V^T),
\end{align*}
where \uref{e} is by the definition of $`g$; \uref{f} is by \eqref{eq:minI}; and \uref{g} is because $\mcP'\in \Uppi^*(\RZ_V)$. By \eqref{eq:IP:t}, every edge in $\mcE_{\mcP'}$ has weight $`g$, and so  
$\mcE_{\mcP'}\subseteq \mcE(T)`/\mcE(T_{`g})$, implying $\abs{\mcE_{\mcP'}}\leq \abs {\mcE(T)`/\mcE(T_{`g})}$. However,
\begin{align*}
	\abs {\mcE_{\mcP'}} &=\abs {\mcP'}-1 > \abs {\mcP}-1=\abs {\mcE_{\mcP}}=\abs {\mcE(T)`/\mcE(T_{`g})},
\end{align*}
which contradicts $\mcE_{\mcP'}\subseteq \mcE(T)`/\mcE(T_{`g})$. Note here that the last equality follows from the proof of \uref{d} above. This completes the proof of Theorem~\ref{thm:PP}

\subsection{Proof of Theorem~\ref{thm:C}}
\label{sec:F}

We shall prove by induction that, for $1\leq i\leq N$, $`g_i$ is the $i$-th smallest value of $I(\RZ_e)$ for $e\in \mcE(T)$, and $\pzP_i = \pzP(T_{`g_i})$ with $\pzP_i$ defined in \eqref{eq:C`gi} for $\RZ_C^T$ in place of $\RZ_C$. This will imply \eqref{eq:C} by Corollary~\ref{cor:main}. 

By \eqref{eq:xi2} and \eqref{eq:C`gi} with $i=1$, we have  $`g_1=I(\RZ_V^T)$ and $\pzP_1=\pzP^*(\RZ_V^T)$. This implies the base case under \eqref{eq:minI} and \eqref{eq:T:P*}, namely that, $`g_1$ is the smallest $I(\RZ_e)$ and $\pzP_1=\pzP(T_{`g_1})$. 

Let $T(C)$ be the subgraph of $T$ induced on the subset $C\subseteq V$ of vertices. By \eqref{eq:xi2}, for  $1<i\leq N$,
\begin{align*}
	`g_i &= \min_{C\in \pzP_{i-1}\colon\abs {C}>1} I(\RZ^T_C)
	\\
	&
	\utag{a}= \min_{C\in \pzP(T_{`g_{i-1}})} \min_{e\in \mcE(T(C))} I(\RZ_e) 
	\\
	&
	\utag{b}= \min_{e\in \mcE(T_{`g_{i-1}})} I(\RZ_e).
\end{align*} 
Here, \uref{a} is by the inductive hypothesis $\pzP_{i-1}=\pzP(T_{`g_{i-1}})$ as well as \eqref{eq:minI}  that $I(\RZ^T_C)= \min_{e\in \mcE(T(C))} I(\RZ_e)$. \uref{b} is because $\mcE(T_{`g})$ is the union of $\mcE(T(C))$ over $C\in \pzP(T_{`g})$ (for $`g=`g_{i-1}$). 

The above equalities implies that $`g_i$ is the $i$-th smallest value of $I(\RZ_e)$ for $e\in \mcE(T)$ because 	
the R.H.S.\ of \uref{b} is, by the inductive hypothesis that $`g_{i-1}$ is the $(i-1)$-st smallest value, and the fact that $T_{`g_{i-1}}$ contains all edges in $\mcE(T)$ with weights strictly larger than $`g_{i-1}$.

It remains to show $\pzP_i=\pzP(T_{`g_i})$. From \eqref{eq:C`gi}, we have
\begin{align*}
	\pzP_i &\utag{d}= `1(\pzP(T_{`g_{i-1}})`/\pzC^*_{i-1}`2) \cup \bigcup_{C\in \pzC^*_{i-1}} \pzP(T_{`g_{i}}(C))\\
	&= \bigcup_{C\in \pzP_{i-1}`/\pzC^*_{i-1}} \Set {C} 
	\cup \bigcup_{C\in \pzC^*_{i-1}} \pzP(T_{`g_{i}}(C))\\
	&
	\utag{e}= \bigcup_{C\in \pzP(T_{`g_{i-1}})} \pzP(T_{`g_i}(C))
	\utag{f}= \pzP(T_{`g_i}).
\end{align*} 
Here, \uref{d} is by applying \eqref{eq:T:P*} to \eqref{eq:C`gi}. \uref{e} is by rewriting $\pzP(T_{`g_{i-1}})`/\pzC^*_{i-1}$ as $\bigcup_{C\in \pzP_{i-1}`/\pzC^*_{i-1}} \Set {C}$ and then applying $\Set {C}=\pzP(T_{`g_{i}}(C))$ because $C\not\in \pzC^*_{i-1}$ means that every edge of $T_{`g_{i}}(C)$ has weight strictly larger than $`g_i$ by \eqref{eq:xi2}. Finally, \uref{f} follows from $\mcE(T_{`g_i})=\bigcup_{C\in \pzP(T_{`g_{i-1}})} \mcE(T_{`g_i}(C))$, which can be argued as follows. $\supseteq$ is obvious because $\mcE(T_{`g_i})\supseteq \mcE(T_{`g_i}(C))$. To prove the reverse inclusion, note that $`g_{i-1}<`g_i$ and so the edge in $T_{`g_i}$ must be in a connected component of $T_{`g_{i-1}}$, namely, a subtree $T_{`g_i}(C)$ induced on some $C\in \pzP(T_{`g_{i-1}})$. This completes the proof of Theorem~\ref{thm:C}.

\section{Clustering by minimum average cost}
\label{sec:mac}

The objective of \emph{minimum average cost (MAC) clustering} is to obtain a partition $\mcP$ of size $\abs{\mcP}>k$ for some threshold $k$ as the set of clusters, and the singleton elements in the partition are also regarded as clusters in satisfying the constraint $\abs{\mcP}>k$. To solve this problem using our clustering solution by multivariate mutual information, it is natural to use
\begin{align}
	\mcP = \pzP_i \kern1em\text{such that $\abs{\pzP_{i-1}} \leq k <
		\abs{\pzP_i}$.} \label{eq:dc:1}
\end{align}
where $\pzP_i$'s form the PSP of the entropy function in \eqref{eq:C`gi}. $\pzP_i$ is the coarsest partition from the PSP with more than $k$ parts. 
The clustering solution proposed by \cite{nagano10} obtains the partition by solving the following minimum average cost constraint instead:
\begin{subequations}
	\label{eq:dc:2}
	\begin{align}
		`g &=\min_{\substack{\mcP\in \Pi(V): \abs{\mcP}> k}}
		\frac{\sum_{C\in \mcP} f(C)}{\abs{\mcP}-k},\kern1em \text{or}\label{eq:dc:2:a}\\
		-k`g &= \min_{\substack{\mcP\in \Pi(V): \abs{\mcP}> k}} \sum_{C\in \mcP} `1[f(C)-`g`2]\label{eq:dc:2:b}
	\end{align}
\end{subequations}
where $f$ is a submodular function that needs to be chosen appropriately. The question of interest is, whether there is an obvious choice of $f$ in terms of the entropy function $h$ for which the two clustering solutions in \eqref{eq:dc:1} and \eqref{eq:dc:2} are the same. The similarity is more apparent by thinking of $\mcP$ in \eqref{eq:dc:1} as the solution to \eqref{eq:xi1}, namely,
\begin{align*}
  `g_i &= \min_{\substack{\mcP\in \Pi(V):  \abs{\mcP}>\abs{\pzP_{i-1}}}}
  \frac{h[\mcP]-h[\pzP_{i-1}]}{\abs{\mcP}-\abs{\pzP_{i-1}}},
\end{align*}
which is similar to \eqref{eq:dc:2:a} except for the numerator and $k$ in place of $\abs{\pzP_{i-1}}$. Note that the choice of $\pzP_{i-1}$ depends on $k$ according to \eqref{eq:dc:1}. In particular, $\abs{\pzP_{i-1}}\leq k$ but \emph{equality is not needed} so long as the solution $\pzP_i$ to the above minimization satisfies $\abs{\pzP_{i}}> k$ as required by \eqref{eq:dc:1}.

For graphical networks, \cite{nagano10} chooses $f$ to be the cut function of the graph, which is also the case for info-clustering by information flow in Section~\ref{sec:CIF}. The following is a concrete example that distinguishes info-clustering from MAC clustering. 
\begin{Example}
	\label{eg:MAC:cut}
	Consider a weighted graph $G$ with vertex set $V=\Set{1,2,3,4}$, edge set $E=\Set{e_{12},e_{23},e_{34}}$, and
	\begin{alignat*}{3}
	 `q(e_{12})&=\Set{1,2},&\kern1em `q(e_{23})&=\Set{2,3},&\kern1em `q(e_{34})&=\Set{3,4} \text{ and}\\
	 c(e_{12})&=2,& c(e_{23})&=3,& c(e_{34})&=4,
	\end{alignat*}
	where $`q$ and $c$ are the edge and weight functions as in Section~\ref{sec:CIF}. The PSP of the cut function $C\mapsto c(`d_G(C))$ (see \eqref{eq:`d}) can be shown to be 
	\begin{align*}
		\pzP_0 &= \Set{\Set{1,2,3,4}}\\
		\pzP_1 &= \Set{\Set{1},\Set{2,3,4}}\\
		\pzP_2 &= \Set{\Set{1},\Set{2},\Set{3,4}}\\
		\pzP_3 &= \Set{\Set{1},\Set{2},\Set{3},\Set{4}}.
	\end{align*}
	For $k=2$, info-clustering will return $\pzP_2$ according \eqref{eq:dc:1}, since $\pzP_2$ is the coarsest partition with more than $k$ parts. However, choosing $f$ to be the cut function, MAC clustering does not return the same solution because the average cost~\eqref{eq:dc:2:a} of $\pzP_3$ is strictly smaller than that of $\pzP_2$:
	\begin{align*}
		\frac{\sum_{C\in \pzP_2} f(C)}{\abs {\pzP_2}-k} &=\frac{c(e_{12})+c(e_{23})}{3-2}=5\\
		\frac{\sum_{C\in \pzP_3} f(C)}{\abs {\pzP_3}-k} &=\frac{c(e_{12})+c(e_{23})+c(e_{34})}{4-2}=4.5<5.
	\end{align*}
	Indeed, it can be shown that $\pzP_3$ achieves the minimum average cost among all other partitions of $V$, and so MAC clustering will return the less intuitive clustering by $\pzP_3$ instead of $\pzP_2$.
\end{Example}

Actually, $f$ was assumed to be non-negative in~\cite{nagano10}, because then, the constraint $\abs{\mcP}>k$ can be dropped from \eqref{eq:dc:2:b} without changing the solution. Doing so reduces \eqref{eq:dc:2:b} to computing the Dilworth truncation, which can be done efficiently and guaranteed to return a partition in the PSP, despite the possibility of returning one that is finer than required, as shown in the previous example (since cut function is non-negative). In the general case when $f$ can be negative, removing the constraint $\abs{\mcP}>k$ from \eqref{eq:dc:2:b} can potentially change the solution to something outside the PSP, and so it is unclear whether the clusters can be computed efficiently. In the following, we will further compare MAC clustering to info-clustering without assuming $f$ to non-negative.

It can be shown that constant scaling of $f$ does not change the solution to \eqref{eq:dc:2}, but constant shift does. To ensure submodularity, a reasonable choice of $f$ is $f=h_s$ for some appropriate constant shift $s$. We will show that there is a choice of $s$ such that the clustering solutions for $k=1$ are the same for \eqref{eq:dc:1} and \eqref{eq:dc:2}. However, there is no choice of $s$ for which the complete clustering solutions for different $k$ are the same.

More precisely, the fundamental partition $\pzP_1=\pzP^*(\RZ_V)$ in our clustering solution~\eqref{eq:dc:1} can be obtained from \eqref{eq:dc:2} with
\begin{align*}
	f(B):=h_{h(V)}(B)=-H(\RZ_{V`/B}|\RZ_B) \kern1em\text{for $B\subseteq V$}.
\end{align*}
To see this, rewrite \eqref{eq:dc:2:a} with the above choice and $k=1$:
\begin{align*}
	`g &=\min_{\substack{\mcP\in \Pi(V): \abs{\mcP}> 1}} \frac{\sum_{C\in \mcP} -H(\RZ_{V`/C}|\RZ_{C})}{\abs{\mcP}-1}\\
	&=\min_{\mcP\in \Pi'(V)} \frac{\sum_{C\in \mcP} H(\RZ_C)-H(\RZ_{V})}{\abs{\mcP}-1} -H(\RZ_V).
\end{align*}
The first term on the R.H.S.\ is $I(\RZ_V)$ by \eqref{eq:D:1} and \eqref{eq:mi}, and so the finest optimal partition is $\pzP^*(\RZ_V)$ as desired.

Note that we have allowed $f$ to be negative above $H(\RZ_B)<H(\RZ_V)$. However, it turns out that, even if we allow $f=h_s$ to be negative, there is no choice of $s$ for which the complete clustering solutions in \eqref{eq:dc:1} and \eqref{eq:dc:2} are the same:
\begin{Example}
	\label{eg:MAC}
	\begin{figure}
		\centering
		\subcaptionbox{For PSP~\eqref{eq:dc:1}.\label{fig:MAC:1}}{
			{\def\u{1.8}
				\tikzstyle{point}=[draw,circle,minimum size=.2em,inner sep=0, outer sep=.2em]
				\begin{tikzpicture}[x=1em,y=1em,>=latex]
				\draw[->] (0,0*\u) -- (0,9*\u) node [label=right:$\hat{h}_{`g}(V)$] {};
				\draw[->] (0,0) -- (5.5*\u,0) node [label=right:$`g$] {};
				\foreach \i/\ya/\xa/\yb/\xb/\lp/\lb in {
					1/5/0/0/5/left/{$h_{`g}[\Set{\Set{1,2,3,4}}]=5-`g$}, 
					2/5/0/0/2.5/left/{$\begin{aligned} \\[1em] h_{`g}[\Set{\Set{1,2,3},\Set{4}}]=5-2`g\end{aligned}$}, 
					3/7/0/0/2.33/left/{$\begin{aligned} & \\[0.7em] h_{`g}[\Set{\Set{1,2},\Set{3},\Set{4}}]&=7-3`g\\ =h_{`g}[\Set{\Set{1,3},\Set{2},\Set{4}}]&\\ =h_{`g}[\Set{\Set{2,3},\Set{1},\Set{4}}]& \end{aligned}$}, 
					4/8/0/0/2/left/{$\kern1em h_{`g}[\Set{\Set{1},\Set{2},\Set{3},\Set{4}}]=8-4`g$}}
				\draw[dashed] (\xa*\u,\ya*\u)  node [inner sep=0,outer sep=0,label={[label distance=0em]\lp:{\scriptsize\lb}}] {} -- (\xb*\u,\yb*\u);
				\path (1.5*\u,2*\u) node (2) [point,red,thick,label=left:{\scriptsize$p_2$}] {};
				\path (0*\u,5*\u) node (1) [point,red,thick,label=93:{\scriptsize$p_1$}] {};
				\draw[-,thick,blue] (0,5*\u)--(1)--(2)--(2*\u,0*\u);
				\end{tikzpicture}}
		}\\[1.5em]
		\subcaptionbox{For minimum average cost \eqref{eq:dc:2} with $k=1$, $s=3.5$.\label{fig:MAC:2}}{
			{\def\u{1.8}
				\tikzstyle{point}=[draw,circle,minimum size=.2em,inner sep=0, outer sep=.2em]
				\begin{tikzpicture}[x=1em,y=1em,>=latex]
				\draw[->] (3.5*\u,0*\u) -- (3.5*\u,9*\u) node [label=above:{\kern-1em $-`g,\displaystyle\min_{\abs{\mcP}>1}  \sum_{C\in \mcP} [f(C)-`g]$}] {};
				\draw[->] (0,0) -- (5.5*\u,0) node [label=right:$`g$] {};
				\foreach \i/\ya/\xa/\yb/\xb/\lp/\lb in {
					2/5/0/0/2.5/left/{$\begin{aligned} \\[1em] h_{s+`g}[\Set{\Set{1,2,3},\Set{4}}]\end{aligned}$}, 
					3/7/0/0/2.33/left/{$\begin{aligned} & \\[0.7em] h_{s+`g}[\Set{\Set{1,2},\Set{3},\Set{4}}] & \\
						=h_{s+`g}[\Set{\Set{1,3},\Set{2},\Set{4}}] & \\
						=h_{s+`g}[\Set{\Set{2,3},\Set{1},\Set{4}}] & \end{aligned}$}, 
					4/8/0/0/2/left/{$\kern1em h_{s+`g}[\Set{\Set{1},\Set{2},\Set{3},\Set{4}}]$}}
				\draw[dashed] (\xa*\u,\ya*\u)  node [inner sep=0,outer sep=0,label={[label distance=0em]\lp:{\scriptsize\lb}}] {} -- (\xb*\u,\yb*\u);
				\path (1.5*\u,2*\u) node (1) [point,red,thick,label=left:{\scriptsize}] {};
				\draw[-,thick,blue] (0,5*\u)--(1)--(2*\u,0*\u);
				\draw[red] (0*\u,3.5*\u)  to node [inner sep=0,outer sep=0,label={[label distance=0em]right:{\scriptsize$-`g$}}] {}  (3.5*\u,0*\u);
				\end{tikzpicture}}
		}\\[1.5em]
		\subcaptionbox{For minimum average cost~\eqref{eq:dc:2} with $k=2$, $s=3.5$.\label{fig:MAC:3}}{
			{\def\u{1.8}
				\tikzstyle{point}=[draw,circle,minimum size=.2em,inner sep=0, outer sep=.2em]
				\begin{tikzpicture}[x=1em,y=1em,>=latex]
				\draw[->] (3.5*\u,0*\u) -- (3.5*\u,9*\u) node [label=above:{\kern-1em $-2`g,\displaystyle\min_{\abs{\mcP}>2}  \sum_{C\in \mcP} [f(C)-`g]$}] {};
				\draw[->] (0,0) -- (5.5*\u,0) node [label=right:$`g$] {};
				\foreach \i/\ya/\xa/\yb/\xb/\lp/\lb in {
					3/7/0/0/2.33/left/{$\begin{aligned} & \\[0.7em] h_{s+`g}[\Set{\Set{1,2},\Set{3},\Set{4}}] & \\
						=h_{s+`g}[\Set{\Set{1,3},\Set{2},\Set{4}}] & \\
						=h_{s+`g}[\Set{\Set{2,3},\Set{1},\Set{4}}] & \end{aligned}$}, 
					4/8/0/0/2/left/{$\kern1em h_{s+`g}[\Set{\Set{1},\Set{2},\Set{3},\Set{4}}]$}}
				\draw[dashed] (\xa*\u,\ya*\u)  node [inner sep=0,outer sep=0,label={[label distance=0em]\lp:{\scriptsize\lb}}] {} -- (\xb*\u,\yb*\u);
				\path (1*\u,4*\u) node (2) [] {};
				\path (0*\u,7*\u) node (1) [point,red,thick,label=left:{\scriptsize}] {};
				\draw[-,thick,blue] (0,7*\u)--(2)--(2*\u,0*\u);
				\draw[red] (0*\u,7*\u)  to node [inner sep=0,outer sep=0,label={[label distance=0em]right:{\scriptsize$-2`g$}}] {}  (3.5*\u,0*\u);
				\end{tikzpicture}}
		}
		\caption{Computing the clustering solutions to Example~\ref{eg:MAC}.}
		\label{fig:MAC}
	\end{figure}
	Let $V=\Set{1,2,3,4}$ and
	\begin{align*}
		\RZ_1:=& (\RX_a, \hphantom{\RX_b,}\kern.2em\RX_c) & \RZ_4:=& (\RX_d, \RX_e),\\
		\RZ_2:=& (\RX_a, \RX_b\hphantom{\RX_b,}\kern.2em) \\
		\RZ_3:=& (\hphantom{\RX_a,}\kern.2em \RX_b, \RX_c)
	\end{align*}
	where $\RX_i$ are independent uniformly random bits.
	The Dilworth truncation $\hat{h}_{`g}(V)$ is plotted in \figref{fig:MAC:1}. Our clustering solution \eqref{eq:dc:1} consists of the PSP:
	\begin{align*}
		\pzP_1&=\Set{\Set{1,2,3},\Set{4}} && \text{for $k=1$}\\
		\pzP_2&=\Set{\Set{i}\mid i\in V} && \text{for $k>1$}
	\end{align*}
	The solution respects the symmetry of the correlation in $\RZ_{\Set{1,2,3}}$ and the independence between $\RZ_4$ and $\RZ_{\Set{1,2,3}}$.
	
	To compute the MAC clustering in \eqref{eq:dc:2}, rewrite \eqref{eq:dc:2:b} with $f=h_s$
	\begin{align*}
		-k`g 
		&= \min_{\substack{\mcP\in \Pi(V): \abs{\mcP}> k}} h_{s+`g}[\mcP].
	\end{align*}
	For $k=1$ and $s=3.5$, the L.H.S.\ and R.H.S.\ are plotted in \figref{fig:MAC:2}. Since the curve for the L.H.S.\ intersect the curve on the R.H.S.\ along the line segment $h_{s+`g}[\pzP_1]$, the partition $\pzP_1$ is an optimal solution to the R.H.S.. If $s<3.5$, then $-`g$ will intersect the line segment corresponding to $h_{s+`g}[\pzP_2]$ instead of $h_{s+`g}[\pzP_1]$. Therefore, in order to have $\pzP_1$ to be the solution, we must have $s\geq 3.5$. 
	For $k=2$ and $s=3.5$, the plot in \figref{fig:MAC:3} shows that the optimal partition to the R.H.S.\ is not a partition in the PSP. This is the case even for $s\geq 3.5$ because increasing $`g$ will only move the intersection point to the left further away from $h_{s+`g}[\pzP_2]$. The optimal partition, such as $\Set{\Set{1,2},\Set{3},\Set{4}}$, does not appear to respect the symmetry in the correlation among $\RZ_{\Set{1,2,3}}$.
\end{Example}




\egroup

\section*{Acknowledgment} 
%
%

The authors would like to thank Prof.\ Raymond W.\ Yeung, the Co-Director of the Institute of Network Coding (INC) at the Chinese University of Hong Kong, for his generous support of our research on information theory; Prof.\ Lav R.\ Varshney, Prof.\ Rosanna Y-Y.\ Chan, and Prof.\ Chen Change Loy for their suggestions of relevant works in machine learning and neuroscience; Dr.\ Javad B. Ebrahimi, Dr. Ravi K.\ Raman, and Dr. Ni Ding for their helpful discussions; Prof.\ Frank Kschischang, Prof.\ Devavrat Shah, and the colleagues at INC, whose comments have helped significantly improve the presentation of the paper. We would also like to thank the Associate Editor Prof. Peter Thomas and the reviewers for their detailed reading and insightful comments. 

The first author would like to thank Prof.\ Imre Csisz\'ar for the discussion on the divergence upper bound for secret key agreement and the issue of tightness, and Prof.\ Prakash Narayan for his recognitions of the contribution of this work. He would also like to thank his Ph.D. advisor, Prof.\ Lizhong Zheng, for leading him to the field of information theory. 




\bibliographystyle{IEEEtran}


\enlargethispage{-2.9in}

\begin{IEEEbiography}[{\includegraphics[width=1in,height=1.25in,clip,keepaspectratio]{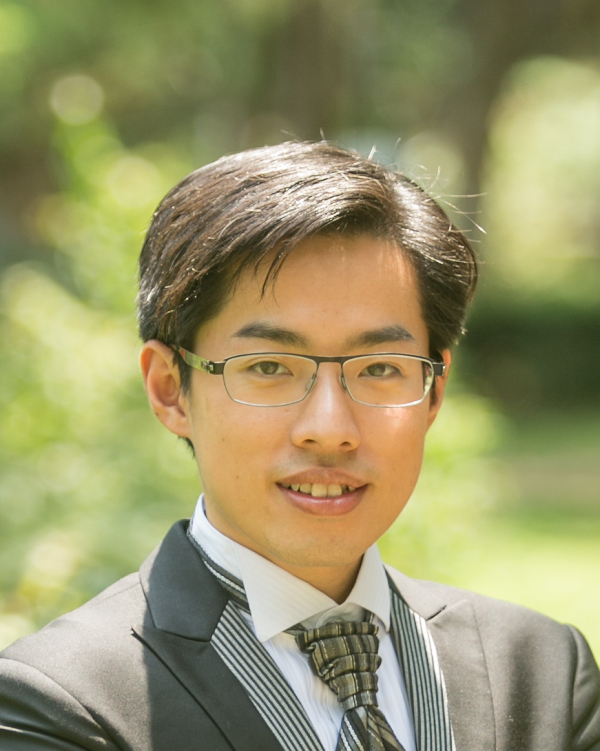}}]{Chung Chan} received the B.Sc., M.Eng. and Ph.D. from the EECS Department at MIT in 2004, 2005 and 2010 respectively. He is currently a Research Assistant Professor at the Institute of Network Coding, the Chinese University of Hong Kong. 	
	His research is in the area of information theory, with applications to network coding, multiple-terminal source coding and security problems that involve high-dimensional statistics. He is currently working on machine learning applications such as data clustering and feature selection. 
\end{IEEEbiography}
\vspace*{-2em}
\begin{IEEEbiography}[{\includegraphics[width=1in,height=1.25in,clip,keepaspectratio]{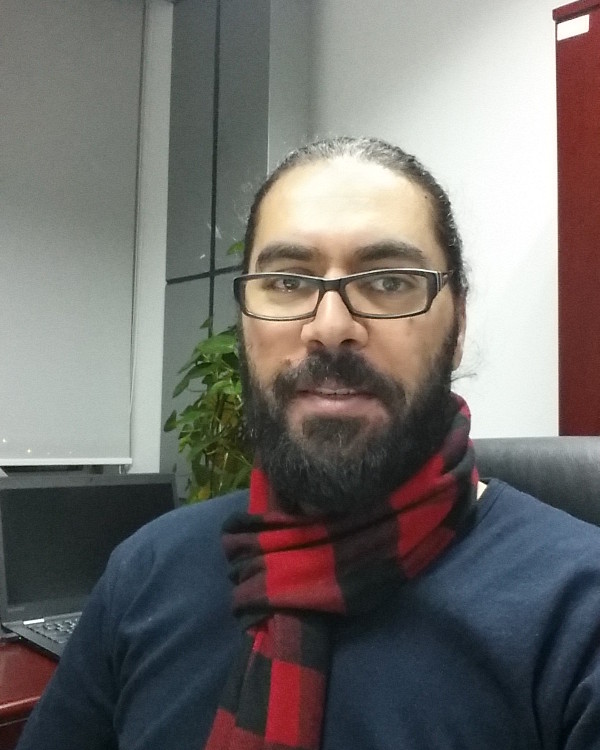}}]{Ali Al-Bashabsheh}
	received a B.Sc. (2001) and an M.Sc (2005) in electrical engineering from Jordan University of
	Science and Technology,
	an M.Sc. (2012) in mathematics from Carleton University,
	and a Ph.D. (2014) in electrical engineering from the University of Ottawa. 
	Since April 2014, he has been a postdoctoral fellow at the Institute of Network Coding
	at the Chinese University of Hong Kong. His research interests include graphical models, coding
	theory, and information theory.
\end{IEEEbiography}
\vspace*{-2em}
\begin{IEEEbiography}[{\includegraphics[width=1in,height=1.25in,clip,keepaspectratio]{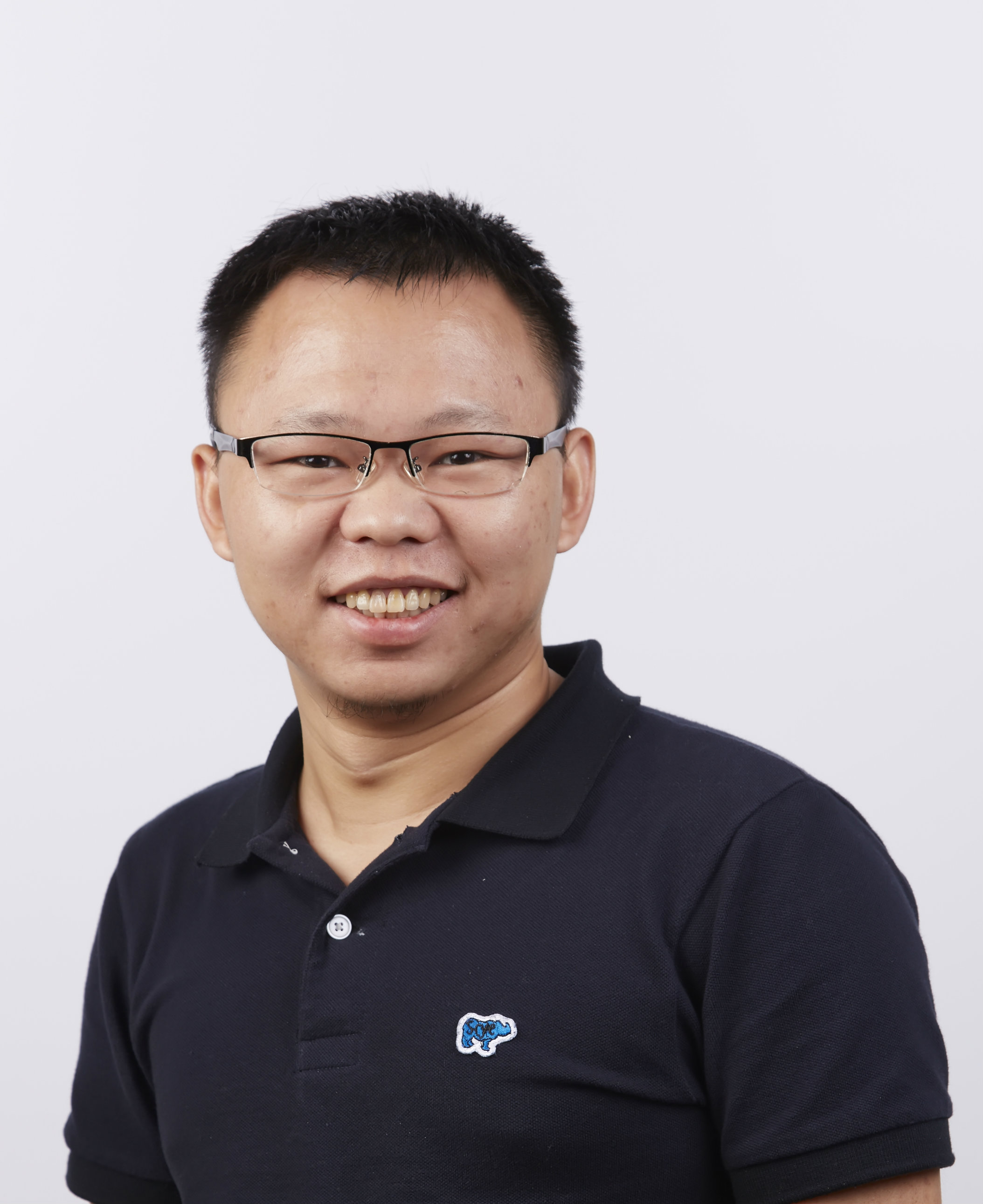}}]{Qiaoqiao Zhou} received his B.B.A. in business administration and M.S. in electrical engineering from Beijing University of Post and Telecommunication, China, in 2011 and 2014, respectively. From 2014 to 2015, he was a research assistant at the Institute of Network Coding, the Chinese University of Hong Kong. He is currently a Ph.D.\ candidate at the Department of Information Engineering, the Chinese University of Hong Kong. His research interests include information-theoretic security and machine learning. 
\end{IEEEbiography}
\vspace*{-2em}
\begin{IEEEbiography}[{\includegraphics[width=1in,height=1.25in,clip,keepaspectratio]{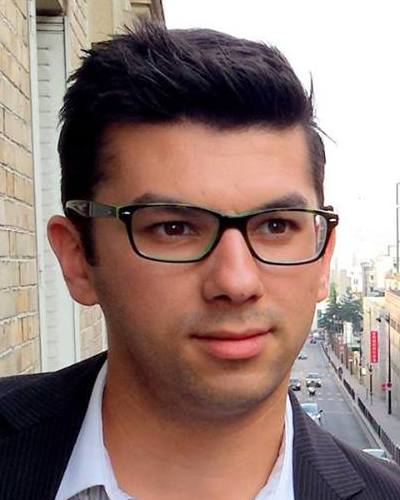}}]{Tarik Kaced}
	was born in France, he received his B.Sc. in Fundamental Computer Science from \'Ecole
	Normale Sup\'erieure de Lyon in 2007, and his M.Sc. from Universit\'e de
	Nice Sophia-Antipolis in 2009. He completed his Ph.D. degree in Computer Science in 2012
	at Universit\'e de Montpellier 2 in the ESCAPE team from LIRMM. 
	He has been a post-doctoral fellow at the
	Institute of Network Coding at The Chinese University of Hong Kong for two years.
	He was a post-doctoral fellow at in Universit\'e Paris-Est
	Cr\'eteil at the Algorithmic, Complexity and Logic Laboratory.
	
	His research interests include information theory, Kolmogorov complexity,
	matroid theory, computability, information inequalities, combinatorics, error correcting codes and secret sharing.
\end{IEEEbiography}
\vspace*{-2em}
\begin{IEEEbiography}[{\includegraphics[width=1in,height=1.25in,clip,keepaspectratio]{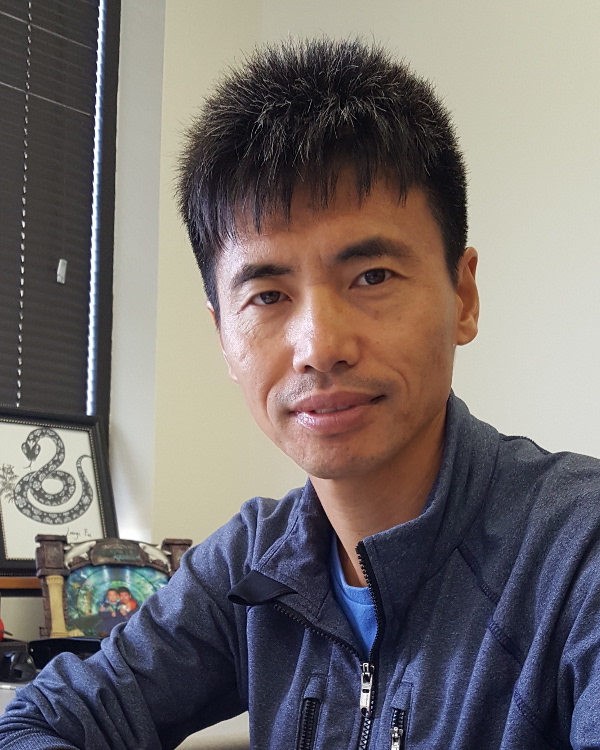}}]{Tie Liu} was born in Jilin, China in 1976. He received his B.S. (1998) and M.S. (2000) degrees, both in Electrical Engineering, from Tsinghua University, Beijing, China and a second M.S. degree in Mathematics (2004) and Ph.D. degree in Electrical and Computer Engineering (2006) from the University of Illinois at Urbana-Champaign. Since August 2006 he has been with Texas A\&M University, where he is currently an Associate Professor with the Department of Electrical and Computer Engineering. His primary research interest is in the area of information theory and statistical information processing.
	
	Dr. Liu received an M. E. Van Valkenburg Graduate Research Award (2006) from the University of Illinois at Urbana-Champaign and a Faculty Early Career Development (CAREER) Award (2009) from the National Science Foundation. He was a Technical Program Committee Co-Chair for the 2008 IEEE Global Communications Conference (GLOBECOM) and a General Co-Chair for the 2011 IEEE North American School of Information Theory. He currently serves as an Associate Editor for Shannon Theory for the IEEE Transactions on Information Theory.
\end{IEEEbiography}

\enlargethispage{-3.2in}

\end{document}